\documentclass[12pt]{article} 
\usepackage{epsfig,amsmath,amssymb,subfigure}
\usepackage{caption2}

   {\end{list}}%

\makeatletter
\renewcommand{\section}{\setcounter{equation}{0}\@startsection
  {section}%
  {1}%
  {0pt}%
  {-1\baselineskip}%
  {0.4\baselineskip}%
  {\bfseries\large}}%
\renewcommand{\subsection}{\@startsection
  {subsection}%
  {2}%
  {0pt}%
  {-0.75\baselineskip}%
  {0.2\baselineskip}%
  {\bfseries}}%
\renewcommand{\subsubsection}{\@startsection
  {subsubsection}%
  {3}%
  {0pt}%
  {-0.5\baselineskip}%
  {0.1\baselineskip}%
  {\sc}}%
\makeatother

\renewcommand{\theequation}{\thesection.\arabic{equation}}

\def\a{\alpha} 
\def\b{\beta}

\def\r{\rho}
\def\s{\sigma}

\def\Aslash{{A\mkern-11mu/}}

\def\Dirac{{D\mkern-12mu/}}

\def\pslash{{p\mkern-8mu/}{\!}}

\def\qslash{{q\mkern-8mu/}{\!}}

\def\Rslash{{R\mkern-11mu/}}
\def\Sslash{{S\mkern-11mu/}}

\def\idx{\int\!\! d^4\!x}
\def\iDx{\int\!\! d^D\!x}

\newcommand{\bea}{\begin{eqnarray}} 
\newcommand{\eea}{\end{eqnarray}}
\newcommand{\beann}{\begin{eqnarray*}} 
\newcommand{\eeann}{\end{eqnarray*}}
\newcommand{\ba}{\begin{array}}
\newcommand{\ea}{\end{array}}

\newcommand{\Tr}{\mathbf{Tr}}
\newcommand{\ST}{\star}

\def\psib{\bar{\psi}}
\def\Psib{\bar{\Psi}}
\def\g5{\gamma_{5}}

\def\pslash  {{p\mkern-7mu/}}
\def\Dcal{{\mathfrak{D}}}

\def\idx{\int\! d^{4}\!x\,}

\def\iDp{\int\! \frac{d^{D}\!p}{(2\pi)^{D}} \,\,}

 \def\psib{\bar{\psi}}
 \def\Psib{\bar{\Psi}}

 \def\Dirac{{D\mkern-12mu/}\,}

 \def\pslash  {{p\mkern-7mu/}}
\def\qslash  {{q\mkern-7mu/}}
 \def\Dcal{{\mathfrak{D}}}
 
 \def\hg {\hat{g}} 
 \def\hp {\hat{p}}
 \def\hq {\hat{q}} \def\hgamma {\hat{\gamma}}
 
 \def\Di {{\partial}_{\mu_1}}
 \def\Dii {{\partial}_{\mu_2}}
\def\Diii {{\partial}_{\mu_3}}
 \def\Div {{\partial}_{\mu_4}}
 \def\Ds {{\partial}_{\sigma}}
 \def\Dr {{\partial}_{\rho}}
 \def\Da {{\partial}_{\alpha}}
 \def\Db {{\partial}_{\beta}}
 \def\Dm {{\partial}_{\mu}}
 \def\Dn {{\partial}_{\nu}}
 \def\Di {{\partial}_{\mu_1}}

 \def\Dcs {{\Dcal}_{\sigma}}
 \def\Dcr {{\Dcal}_{\rho}}
 \def\Dca {{\Dcal}_{\alpha}}
 \def\Dcb {{\Dcal}_{\beta}}
 \def\Dcm {{\Dcal}_{\mu}}
 \def\Dcn {{\Dcal}_{\nu}}
 \def\ai {a_{\mu_1}}
 \def\aii {a_{\mu_2}}
 \def\aiii {a_{\mu_3}}
 \def\aiv {a_{\mu_4}}
 \def\as {a_{\sigma}}
 \def\ar {a_{\rho}}

 \def\am {a_{\mu}}
 
 \def\Ai {A_{\mu_1}}
 \def\Aii {A_{\mu_2}}
 \def\Aiii {A_{\mu_3}}
 \def\Aiv {A_{\mu_4}}
 
 \def\ar {a_{\rho}}
 \def\g {\gamma}

 \def\mi {{\mu_1}}
\def\mii {{\mu_2}}
 \def\miii {{\mu_3}}
\def\miv {{\mu_4}}
 \def\a {\alpha}
\def\b {\beta}
\def\r {\rho}
 \def\s {\sigma}

 \def\Tr{\text{Tr}}
\def\tr{\text{tr}}

\begin{document}
\begin{titlepage}
\rightline{FTI/UCM 70-2005}
\vglue 45pt

\begin{center}

{\Large \bf The $U(1)_{A}$ anomaly in noncommutative $SU(N)$ theories.}\\ 
\vskip 1.2 true cm 
{\rm C.P. Mart\'{\i}n}\footnote{E-mail: carmelo@elbereth.fis.ucm.es}
 and C. Tamarit\footnote{E-mail: ctamarit@fis.ucm.es}
\vskip 0.3 true cm
{\it Departamento de F\'{\i}sica Te\'orica I, 
Facultad de Ciencias F\'{\i}sicas\\ 
Universidad Complutense de Madrid,
 28040 Madrid, Spain}\\
\vskip 0.75 true cm

\flushright{\em Dedicated to Professor A. Galindo on his 70th birthday}

\vskip 0.25 true cm

{\leftskip=50pt \rightskip=50pt 
\noindent
We work out the one-loop $U(1)_A$ anomaly for noncommutative $SU(N)$ gauge 
theories up to second order in the noncommutative parameter $\theta^{\mu\nu}$.
We set $\theta^{0i}=0$ and conclude that there is no breaking of the 
classical $U(1)_A$ symmetry of the theory coming from the contributions that 
are either linear or quadratic in $\theta^{\mu\nu}$. Of course, the ordinary 
anomalous contributions will be still with us. We also show that the one-loop 
conservation of the nonsinglet currents holds at least up to second order 
in $\theta^{\mu\nu}$. We adapt our results to noncommutative gauge theories 
with $SO(N)$ and $U(1)$ gauge groups.  
\par }
\end{center}

\vspace{20pt}
\vfill
\end{titlepage}


\setcounter{page}{2}
\section{Introduction}

Some of the peculiar and beautiful properties of QCD in the low-energy regime 
can be explained with the help of the famous $U(1)_A$ anomaly equation. 
A conspicuous instance of this state of affairs is the occurrence of the 
interaction through instantons between left-handed quarks and right-handed 
antiquarks; a phenomenon which  is heralded by the existence of the $U(1)_A$ 
anomaly. That interaction process   
provided the solution given in ref.~\cite{'tHooft:1976up} to the so-called 
$U(1)_A$ problem. Other instances that show the importance of the $U(1)_A$ 
anomaly in particle physics can be found in ref.~\cite{Shifman:1988zk}.

Many are the pitfalls that one meets when constructing noncommutative gauge 
theories~\cite{Minwalla:1999px, Chepelev:2000hm, Gomis:2000zz,  Alvarez-Gaume:2003mb, Gayral:2004cu}. 
In particular, it is not easy to build noncommutative field theories for
$SU(N)$ gauge groups. Alas! The Moyal product of two local infinitesimal 
$SU(N)$ transformations is not a local infinitesimal $SU(N)$ 
transformation~\cite{Terashima:2000xq}.  Further, charges different from
$+1,0,-1$ do not fit in the standard noncommutative setup as developed for 
$U(N)$ groups~\cite{Douglas:2001ba, Szabo:2001kg,Chu:2005ev}. 
These problems were addressed and given a solution in 
refs.~\cite{Madore:2000en, Jurco:2001rq}, where the appropriate framework was 
developed: the framework is based on the concept of Seiberg-Witten map. 
Both the noncommutative Standard Model~\cite{Calmet:2001na} 
and the noncommutative generalizations~\cite{Aschieri:2002mc, Aschieri:2004ka} of the ordinary $SU(5)$ and $SO(10)$ grand unified theories have been 
constructed within this framework. These noncommutative generalizations of
ordinary theories are not 
renormalizable~\cite{Wulkenhaar:2001sq, Buric:2004ms},  
so that they must be formulated as effective quantum field theories. 
A nice feature of these theories is that their chiral matter content make 
them  free from gauge anomalies~\cite{Martin:2002nr, Brandt:2003fx}.
The study of the phenomenological consequences of the noncommutative Standard 
Model has just begun: 
see refs.~\cite{Behr:2002wx, Minkowski:2003jg, Melic:2005fm}. The reader is 
further referred to refs.~\cite{Chaichian:2001py, Khoze:2004zc} for other 
noncommutative models that generalize the ordinary standard model and are 
formulated within the standard noncommutative framework for $U(N)$ --not 
$SU(N)$-- groups. Now a point of terminology: by noncommutative $SU(N)$ gauge 
theories we shall mean field theories constructed, for  $SU(N)$ groups, 
within the framework in  refs.~\cite{Madore:2000en, Jurco:2001rq}.

The $U(1)_A$ anomaly and its consequences have been intensively studied for 
nocommutative $U(N)$ theories within the standard noncommutative setup, 
i.e., the Seiberg-Witten map is not used to define the noncommutative fields. 
The reader is referred to refs.~\cite{Gracia-Bondia:2000pz, Ardalan:2000cy,  
Intriligator:2001yu, Banerjee:2001un, Armoni:2002fh, Nishimura:2002hw, 
Ydri:2002nt, Nakajima:2003an} for  
further information. However, no such study has been carried out for 
noncommutative $SU(N)$ gauge theories as yet. The purpose of this paper 
is to remedy this situation and work out the anomaly equation for the $U(1)_A$ 
canonical Noether current up to  
second order in the noncommutative parameter $h$ --i.e., second order in $\theta^{\mu\nu}$-- and at the one-loop level. This is a highly non-trivial issue since already at first order in $h$ there are candidates to the $U(1)_A$ anomaly 
whose Wick rotated space-time volume integral does not vanish for a general field configuration with non-vanishing Pontriagin index. An instance of such candidates reads 
\begin{displaymath}
 \theta^{\rho\sigma}
\epsilon^{\mu_1\mu_2\mu_3\mu_4}\Tr\,
     [f_{\sigma \mu_1}f_{\mu_2\mu_3}f_{\rho\mu_4}].
\end{displaymath}
At second order in $h$ the situation worsens.

We shall also discuss the relationship, both at classical and quantum levels,  
between this canonical Noether current and other $U(1)_A$ currents that are 
the analogs of the  $U(1)_A$ canonical Noether currents  
--see refs.~\cite{Gracia-Bondia:2000pz, Ardalan:2000cy,  
Intriligator:2001yu, Banerjee:2001un, Armoni:2002fh}-- that 
occur in  noncommutative $U(N)$ gauge theories with fermions in the 
fundamental representation. These analogs, unlike the canonical Noether 
current of the noncommutative $SU(N)$ theory,  are local $\ST$-polynomials 
of the noncommutative fermion fields only. Barring a concrete instance, we 
shall not be able to give expressions for the $U(1)_A$ anomaly equation valid 
at any order in $h$ since the type of Feynman integrals to be computed 
depends on the order in $h$. This was not the case for chiral gauge anomalies  
--see ref.~\cite{Brandt:2003fx}--, since there the gauge current is of the 
planar kind and, thus, the one-loop Feynman integrals to be worked out are of 
the same type at any order in $h$.
We shall show besides that the nonsinglet chiral currents are conserved at 
the one-loop level and, this time, at any order in $h$. 

Our noncommutative $SU(N)$ theory will be massless and will have $N_f$ fermion 
flavours, all fermions carrying the same, but arbitrary, representation of 
$SU(N)$.  The generalization of our expressions to more general situations is 
achieved by summing over all representations carried by the fermions in the 
theory. The layout of this paper is as follows. The first section is devoted
to the study, at the classical level, of the chiral symmetries of the theory 
and the corresponding  conservation equations. In this section, we introduce 
as well several    
currents that are either conserved or covariantly conserved as a consequence 
of the rigid $U(1)_A$ symmetry of the action. In section two, we compute 
the would-be anomalous contributions to the classical conservation equations 
of these currents. In the third section, we discuss the conservation of the nonsinglet currents at the one-loop level. Then, it comes the section which 
contains a summary of the results obtained in this paper and where our 
conclusions are stated. In this last section we also adapt our results to 
$SO(10)$ and $U(1)$ noncommutative gauge theories. Finally, we  
include several Appendices that the  reader may find useful in reproducing the 
calculations presented in the sequel.

\section{Classical chiral symmetries and currents }

The classical action of the noncommutative $SU(N)$ gauge theory of a 
noncommutative gauge field, $A_{\mu}$, minimally coupled to  a
noncommutative Dirac fermion,$\Psi_{f}$, which we take 
to come in $N_f$ flavours, is given by
\begin{equation}
    S\,=\,\idx -\frac{1}{4g^2}\mathrm{Tr}\,F^{\mu\nu}\ST F_{\mu\nu}\,+\,
\sum_{f=1}^{N_f}\,\Psib_{f}\ST i\Dirac_{\ST}\Psi_{f}.
\label{ncaction}
\end{equation}
$F_{\mu\nu}$ denotes the field strength,
$F_{\mu\nu}=\partial_{\mu}A_\nu-\partial_{\nu}A_\mu-i[A_{\mu},A_{\nu}]_{\ST}$,   
and $\Dirac_{\ST}$ stands for the noncommutative Dirac operator, 
$\Dirac_{\ST}=\gamma^{\mu}(\partial_{\mu}-i\,A_{\mu}\ST\,)$. The symbol 
$\ST$ denotes the Weyl-Moyal product of functions: 
\begin{equation}
f\ST g(x)=\exp\Big(\frac{i}{2}\, h\,\theta^{\mu\nu}\frac{\partial}
{x^{\mu}}\frac{\partial}{y^{\nu}}\Big)f(x)g(y)\Big|_{y\rightarrow x}, 
\label{moyal}
\end{equation}
 and 
$[A_{\mu},A_{\nu}]_{\ST}= A_{\mu}\ST A_{\nu}- A_{\nu}\ST A_{\mu}$. We shall 
assume that time is commutative --i.e., that $\theta^{0\,i}=0,\; i=1,2,3$, 
in some reference system--, so that the concept of evolution is the 
ordinary one. Further, for this choice of $\theta^{\mu\nu}$ the action 
can be chosen to be at most quadratic in the first temporal derivative of 
the dynamical variables at any order in the expansion in $h$ 
--see the paragraph after the next-- and,   
thus, there is one conjugate momenta per ordinary field. This makes it 
possible to use simple Lagrangian and Hamiltonian methods to define the  
classical field theory and quantize it afterwards by using elementary and 
standard recipes. If time were not commutative the number of conjugate momenta 
grows with the order of the expansion in $h$ and then the Hamiltonian formalism 
has to be generalized in some way or another~\cite{Amorim:1999mr, Gomis:2000gy}. 
This generalization may affect the  quantization process in some nontrivial 
way and deserves to be analyzed separately, perhaps along the lines laid out 
in ref.~\cite{Amorim:1999mr}.

The noncommutative fields  $A_{\mu}$ and $\Psi_{f}$ are defined by the ordinary
fields (i.e., fields on Minkowski space-time) $a_{\mu}$ --the gauge field-- 
and $\psi_{f}$ --the Dirac fermion--  via the Seiberg-Witten map. We shall 
understand this map as a formal series expansion in $h$:
\begin{equation}
\begin{array}{l}
{A_\nu (x)\,=\,a_{\mu}(x)\,+\,\sum_{n=1}^{\infty}\,h^n\; 
A^{(n)}_{\mu}[\theta^{\rho\lambda},\partial_{\sigma},a_{\nu}](x),}\\
{\Psi_{f}(x)\,=\,\psi_{f}(x)\,+\,
\sum_{n=1}^{\infty}\,h^n\;\left (\,\mathrm{M}^{(n)}[\gamma^\rho,\theta^{\rho\lambda},a_{\nu},\partial_{\sigma}]\,\psi_{f}\,\right )(x),}\\
{\Psib_{f}(x)\,=\,\psib_{f}(x)\,+\,
\sum_{n=1}^{\infty}\,h^n\;\left (\,\bar{\mathrm{M}}^{(n)}[\gamma^\rho,\theta^{\rho\lambda},a_{\nu},\partial_{\sigma}]\,\psib_{f}\,\right )(x).}
\end{array}
\label{SWmap}
\end{equation}
Although the ordinary gauge field takes values on the Lie algebra, $su(N)$, of 
the group $SU(N)$, the noncommutative gauge field defined in eq.~(\ref{SWmap}) 
takes values on the enveloping algebra of $su(N)$. Both $\Psi_{f}(x)$ and 
$\psi_{f}(x)$ belong to the same vector space. Note that we made a restrictive, 
although natural, choice for the general structure of the Seiberg-Witten maps above: 
the map for the gauge fields does not depend on the matter fields and the map for 
the fermion fields is linear in the ordinary fermion. Also note that   
$A^{(n)}_{\mu}[\theta^{\rho\lambda},\partial_{\sigma},a_{\nu}](x)$, 
$\mathrm{M}^{(n)}[\theta^{\rho\lambda},\partial_{\sigma},a_{\nu}]$ and 
$\bar{\mathrm{M}}^{(n)}[\theta^{\rho\lambda},\partial_{\sigma},a_{\nu}]$
contains n-powers of $\theta^{\rho\lambda}$. On the other hand, $\mathrm{M}^{(n)}[\gamma^\rho,\theta^{\rho\lambda},\partial_{\sigma},a_{\nu}]$ and $\bar{\mathrm{M}}^{(n)}[\gamma^\rho,\theta^{\rho\lambda},\partial_{\sigma},a_{\nu}]$ are differential operators of finite order:
\begin{equation}
\begin{array}{l}
{\mathrm{M}^{(n)}[\gamma^\rho,\theta^{\rho\lambda},a_{\nu},\partial_{\sigma},]=
\mathrm{M}^{(n)}[\gamma^\rho,\theta^{\rho\lambda},a_{\nu}]_{0}\!+\!
\sum_{s=1}^{2n}\,\mathrm{M}^{(n)}
[\gamma^\rho,\theta^{\rho\lambda},a_{\nu}]_{\mu_1\cdots\mu_s}
\partial^{\mu_1}\cdots\partial^{\mu_n},}\\
{\bar{\mathrm{M}}^{(n)}[\gamma^\rho,\theta^{\rho\lambda},a_{\nu},\partial_{\sigma}]=
{\mathrm{M}^{(n)}}^{*}[\gamma^\rho,\theta^{\rho\lambda},a_{\nu}]_{0}\!+\!
\sum_{s=1}^{2n}{\mathrm{M}^{(n)}}^{*}
[\gamma^\rho,\theta^{\rho\lambda},a_{\nu}]_{\mu_1\cdots\mu_s}
\partial^{\mu_1}\cdots\partial^{\mu_n}.}\\
\end{array}
\label{Moperators}
\end{equation} 
The symbol $*$ in the previous equation stands for complex conjugation. 

Using the results in ref.~\cite{Cerchiai:2002ss}, it is not difficult to show 
that if  $\theta^{0i}=0$, $i=1,2,3$, the Seiberg-Witten map in 
eqs.~(\ref{SWmap}) and (\ref{Moperators}) can be appropriately chosen so that  
only the first temporal derivative, $\partial_{0} a_{i},\; i=1,2,3$, of the 
ordinary fields $a_{i}$ occurs in the map and that, besides, only 
$A^{(n)}_{0}[\theta^{\rho\lambda},\partial_{\sigma},a_{\nu}]$ depends on
$\partial_{0} a_{i}$; this dependence being linear. For this choice --or rather choices, see next paragraph-- of the Seiberg-Witten map the action  in 
eq.~(\ref{ncaction}) has a quadratic dependence on $\partial_{0} a_{i}$ and a 
linear dependence on $\partial_{0}\psi$ at any order in $h$. Hence, standard 
Hamiltonian and path integral methods can be used to quantize the theory. 
This is not so if time were noncommutative. 

The Seiberg-Witten map is not uniquely defined. 
There is an ambiguity to it \cite{Asakawa:2000bh,Goto:2000zj, Bichl:2001cq,
Jurco:2001rq, Brace:2001fj, Suo:2001ih,Calmet:2001na, Aschieri:2002mc, 
Barnich:2002pb}. At order $h$, we shall choose the form of the map that leads 
to the noncommutative Yang-Mills models, the noncommutative standard model 
and the noncommutative GUTS models of refs.~\cite{Jurco:2001rq,Calmet:2001na,Aschieri:2002mc}, respectively. Thus we shall take $A^{(1)}_{\mu}$ and 
$\mathrm{M}^{(1)}$ in eq.~(\ref{SWmap}) as given by 
\begin{equation}
\begin{array}{l} 
{A^{(1)}_{\mu}\,=\,
-\frac{1}{4}
    \theta^{\alpha\beta}\{a_\alpha ,\partial_\beta a_{\mu}
    +f_{\beta\mu}\},}\\
{\mathrm{M}^{(1)}\,=\,
-\frac{1}{2}\theta^{\alpha\beta}a_{\alpha}\partial_{\beta}+\frac{i}{4}
\theta^{\alpha\beta}a_\alpha a_\beta,}\\
\end{array}
\label{SWone}
\end{equation} 
where 
$f_{\mu\nu}(x)=\partial_{\mu}a_{\nu}-\partial_{\nu}a_{\mu}
-i[a_{\mu},a_{\nu}]$.

Several expressions --reflecting the ambiguity issue-- for the Seiberg-Witten 
map at order $h^2$ have been worked out  in several places 
\cite{Asakawa:2000bh, Jurco:2001rq, Goto:2000zj, Moller:2004qq}, 
but only in ref.~\cite{Moller:2004qq} has the action 
been computed at  second order in $h$. Here we shall partially follow 
ref.~\cite{Moller:2004qq} and choose the following forms for 
$A^{(2)}_{\mu}$ and $\mathrm{M}^{(2)}$ in eq.~(\ref{SWmap}):
\begin{equation}
\begin{array}{l} 
{A^{(2)}_{\mu}=
\frac{1}{32}
    \theta^{\alpha\beta}\theta^{\gamma\delta}\Big(
\{\{a_{\gamma},\partial_{\delta}a_{\alpha}\}-
\{f_{\gamma\alpha},a_{\delta}\},\partial_{\beta}a_{\mu}\}
-2i[\partial_{\gamma}a_{\alpha},\partial_{\delta}\partial_{\beta}a_{\mu}
+\partial_{\delta}f_{\beta\mu}]}\\
{\phantom{A^{(2)}_{\mu}=\frac{1}{32}\theta}
-\{a_{\alpha},
\{\partial_{\beta}f_{\gamma\mu},a_{\delta}\}+
\{f_{\gamma\mu},\partial_{\beta}a_{\delta}\}-
\{\partial_{\beta}a_{\gamma},\partial_{\delta}a_{\mu}\}-
\{a_{\gamma},\partial_{\delta}(\partial_{\beta}a_{\mu}+f_{\beta\mu})
+\Dcal_{\delta}f_{\beta\mu}\}\}}\\
{\phantom{A^{(2)}_{\mu}=\frac{1}{32}\theta}
-2\{a_{\alpha},\{f_{\beta\gamma},f_{\mu\delta}\}\}-
\{f_{\alpha\mu},\{a_{\gamma},\partial_{\delta}a_{\beta}\}
-\{f_{\gamma\beta},a_{\delta}\}\}
\Big),}\\
{\mathrm{M}^{(2)}=
-\frac{i}{8}\theta^{\alpha\beta}\theta^{\gamma\delta}\Big(
(\partial_{\gamma}a_{\alpha}+ia_{\gamma}a_{\alpha})\,
\partial_{\beta}\partial_{\delta}}\\
{\phantom{\mathrm{M}^{(2)}=-\frac{i}{8}\theta^{\alpha\beta}}
+i(-\partial_{\gamma}a_{\alpha}a_{\beta}+f_{\gamma\alpha}a_{\beta}
-a_{\beta}\partial_{\gamma}a_{\alpha}+2 a_{\beta}f_{\gamma\alpha}
-2i a_{\alpha}a_{\gamma}a_{\beta}+i a_{\alpha}a_{\beta}a_{\gamma})
\partial_{\delta}\;\Big)}\\
{\phantom{\mathrm{M}^{(2)}=}
-\frac{1}{32}\theta^{\alpha\beta}\theta^{\gamma\delta}\Big(
2(\partial_{\gamma}a_{\alpha}+ia_{\alpha}a_{\gamma})\,
\partial_{\delta}a_{\beta}-2i \partial_{\gamma}a_{\alpha}a_{\delta}a_{\beta}}\\
{\phantom{\mathrm{M}^{(2)}=-\frac{i}{8}\theta^{\alpha\beta}}
+i[[\partial_{\gamma}a_{\alpha},a_{\beta}],a_{\delta}]+
4i a_{\beta}f_{\gamma\alpha}a_{\delta}-a_{\gamma}a_{\delta}a_{\alpha}a_{\beta}
+2a_{\gamma}a_{\alpha}a_{\beta}a_{\delta}\Big)}\\ 
{\phantom{\mathrm{M}^{(2)}=}
-\frac{1}{64}\theta^{\alpha\beta}\theta^{\gamma\delta}\Big(
f_{\alpha\beta}f_{\gamma\delta}-4 f_{\gamma\alpha}f_{\delta\beta}\Big).}\\
\end{array}
\label{SWtwo}
\end{equation} 
Substituting eqs.~(\ref{SWone}) and (\ref{SWtwo}) in eq.~(\ref{ncaction}), one  obtains \cite{Moller:2004qq} the following expression for fermionic part of 
the action at second order in $h$:
\begin{equation}
S_{Fermi}\,=\,\sum_{f=1}^{N_f}\idx\;\psib_{f}\Big[i\Dirac\,+\, 
i \Rslash\,+\, i\Sslash\; \Big]\psi_{f}, 
\label{fermionaction}
\end{equation}
where
\begin{displaymath}
\begin{array}{l}
{\Dirac\,=\,\gamma^{\mu}\,(\partial_{\mu}-i A_{\mu}),}\\
{\Rslash\,=\,h\,\theta^{\alpha\beta}\,\big(
-\frac{1}{4}f_{\alpha\beta}\Dirac-\frac{1}{2}\,\gamma^{\rho}f_{\rho\alpha}\,
D_{\beta}\,\big),}\\
{\Sslash\,=\,h^2\,\gamma^{\mu}\,
\theta^{\alpha\beta}\theta^{\gamma\delta}\,\big(
\frac{1}{16}\{\Dcal_{\mu}f_{\gamma\alpha},f_{\delta\beta}\}-\frac{1}{64}\,
\{ \Dcal_{\mu}f_{\alpha\beta},f_{\gamma\delta}\}-
\frac{1}{8}f_{\alpha\gamma}f_{\delta\mu}D_{\beta}
-\frac{1}{4}f_{\alpha\mu}f_{\beta\gamma}D_{\delta} }\\
{\phantom{\Sslash\,=\,h^2\,\gamma^{\mu}\,
\theta^{\alpha\beta}\theta^{\gamma\delta}\,\big(}
-frac{1}{8}f_{\alpha\beta}f_{\gamma\mu}D_{\delta}+\frac{i}{8}\Dcal_{\alpha}f_{\beta\gamma}
D_{\delta}D_{\mu}+
\frac{i}{8}\Dcal_{\alpha}f_{\gamma\mu }D_{\beta}D_{\delta}\;\big).}\\
\end{array}
\end{displaymath}
The symbol $\Dcal_{\mu}$ will stand for $\partial_{\mu}-i\,[a_{\mu},
\phantom{a_{\mu}}]$ all along this paper.

 The action in eq.~(\ref{fermionaction}) is invariant under the group 
$SU(N_f)_{V}\times SU(N_f)_{A} \times U(1)_{V}\times  U(1)_{A}$ of the 
following rigid transformations:
\begin{equation}
\psi^{'}_{f'}\,=\,\big(e^{-i\alpha^{a}\rm{T}^a}\big)_{f^{'}f}\,\psi_{f},
\quad
\psi^{'}_{f'}\,=\,\big(e^{-i\alpha^{a}\rm{T}^a \gamma_5}\big)_{f^{'}f}
\psi_{f},\quad \psi^{'}_{f}\,=\,e^{-i\alpha}\,\psi_{f},\quad
\psi^{'}_{f}\,=\,e^{-i\alpha\gamma_5}\,\psi_{f}.
\label{rigidtrans}
\end{equation}
$\{T^a\}_a$ are the hermitian generators of $SU(N_f)$ in the fundamental 
representation and $\gamma_5=i\gamma^{0}\gamma^{1}\gamma^{2}\gamma^{3}$. 
According to the Noether theorem there exist   
currents which are classically conserved as a consequence of the symmetry.
That the currents associated to the vector-like transformations, 
$SU(N_f)_{V}\times U(1)_{V}$,
are conserved at the quantum level can be seen by using, for instance, 
dimensional regularization. The nonsinglet axial current which comes 
with $SU(N_f)_{A}$ is also conserved, at least at the one-loop level 
--see section 4. As for the singlet axial current attached to the 
$U(1)_{A}$ group, we shall show in the next section that it is not conserved 
at the quantum level. 

Promoting $\alpha$ in the $U(1)_{A}$ transformation in eq.~(\ref{rigidtrans}) 
to an infinitesimal space-time dependent parameter and working out 
the variation of $S_{Fermi}$ under such local transformation, one obtains
\begin{equation}
\delta S_{Fermi}\, =\,\idx\, \partial_{\mu}\,\alpha(x)\;
j^{\mu}_{5\,(cn)}(x),
\label{noethervar}
\end{equation}
where the Noether current $j^{(cn)\;\mu}_{5}$ is given by
\begin{equation}
\begin{array}{l}
{j^{(cn)\;\mu}_{5}(x)\,=\,\sum_{s=1}^{N_F} j^{(cn)\;\mu}_{5\,f}(x),}\\
{j^{(cn)\;\mu}_{5\, f}\,=\,\psib_{f}\gamma^{\mu}\gamma_5\psi_{f}
\,-\,h\,\psib_{f}\big(\theta^{\alpha\beta}
\frac{1}{4}f_{\alpha\beta}\gamma^{\mu}+\frac{1}{2}\theta^{\alpha\mu}
\gamma^{\rho}f_{\rho\alpha}\big)\gamma_5\,\psi_{f}
-h^2\,\theta^{\mu\gamma}\theta^{\alpha\beta}}\\
{\phantom{j^{\mu}_{5\,(cn)}}\psib_{f}\gamma^{\nu}\gamma_5
\big[\frac{i}{8}\Dcal_{\alpha}f_{\beta\gamma}\,D_{\nu}
+\frac{i}{8}\Dcal_{\gamma}f_{\alpha\nu}\,D_{\beta}+
\frac{i}{8}\Dcal_{\alpha}f_{\gamma\nu}\,D_{\beta}
-\frac{1}{8}f_{\gamma\alpha}f_{\beta\nu}
-\frac{1}{4}f_{\alpha\nu}f_{\beta\gamma}
-\frac{1}{8}f_{\alpha\beta}f_{\gamma\nu}\big]\psi_{f}}\\
{\phantom{j^{\mu}_{5\,(cn)\, f}\,=}\quad+\,h^2\,\frac{i}{8}\theta^{\mu\gamma}\theta^{\alpha\beta}\,
\big(\partial_{\nu}(\psib_{f}\gamma^{\nu}\gamma_5\Dcal_{\alpha}f_{\beta\gamma}
\psi_{f})+\partial_{\beta}(\psib_f \gamma^{\nu}\gamma_5\Dcal_{\alpha}
f_{\gamma\nu}\psi_{f})\big)}\\
{\phantom{j^{\mu}_{5\,(cn)\, f}\,=\,\psib_{f}\gamma^{\mu}\gamma_5}\quad\quad
+h^2\,\frac{i}{8}\theta^{\alpha\beta}\theta^{\gamma\delta}\psib_{f}
\gamma^{\mu}\gamma_5\Dcal_{\alpha}f_{\beta\gamma}D_{\delta}\psi_f.}\\
\end{array}
\label{swcurrent}
\end{equation}
As usual, we introduce the chiral charge which is defined by 
\begin{equation}
Q_{5}^{(cn)}(t)\,=\,\int\,d^{3}\vec{x}\;   j^{(cn)\;0}_{5}(t,\vec{x}).
\label{chargeSW}
\end{equation} 
This is a classically conserved quantity, whose properties upon quantization 
give us significant clues as to the dynamics of the quantum theory.

There is an ambiguity in the definition of the Noether current. Indeed, 
the current 
\begin{equation} 
{\tilde j}_{5}^{\mu}(x)\,=\,j^{(cn)\;\mu}_{5}(x)\,+\,{\cal Y}^{\mu}(x)
\label{ambi}  
\end {equation}
would also be a gauge invariant object that verifies eq.~(\ref{noethervar}) and would also yield the same chiral charge as $j^{(cn)\;\mu}_{5}(x)$, if 
${\cal Y}^{\mu}(x)$ were a gauge invariant quantity that satisfy 
\begin{equation}
a)\;\partial_{\mu}{\cal Y}^{\mu}(x)=0\;\quad\text{and}\;\quad b)\;\int\,d^{3}\vec{x}\;  {\cal Y}^{0}(t,\vec{x})\,=\,0.
\label{zerocalgy}
\end{equation} 
The current $j^{(cn)\;\mu}_{5}(x)$ is usually called the canonical Noether 
current since 
\begin{displaymath}
j^{(cn)\;\mu}_{5}(x)\,=\,
\sum_{f}\,\frac{\delta\,{\cal L}}{\delta\,(\partial_{\mu}\psi_{f})}\gamma_{5}
\psi(x)_{f},
\end{displaymath}
${\cal L}$ being the Lagrangian. Following  
ref.~\cite{Ramond:1989yd, Bailin:1986wt}, one 
may also relax a bit the constraints on ${\cal Y}^{\mu}$ and assume that 
$\partial_{\mu}{\cal Y}^{\mu}(x)=0$ holds only along the classical 
trajectories, while $b)$ in eq.~(\ref{zerocalgy}) holds for any field 
configuration, not only for those that are solutions to the equations 
of motion. Of course, this ${\tilde j}_{5}^{\mu}$ will not satisfy 
eq.~(\ref{noethervar}), but it will be  a conserved current such that its 
associated charge,$Q_{5}^{(cn)}$,  generates the action of the chiral 
transformations on the fields:
\begin{displaymath}
\{Q_{5}^{(cn)}(t),\psi(t,\vec{x})\}=-\gamma_{5}\,\psi((t,\vec{x}).
\end{displaymath}
$\{\phantom{a},\phantom{a}\}$ denotes the Poisson brackets. The latter 
current ${\tilde j}_{5}^{\mu}$ may be also called a Noether current.

In connection with the rigid (also called global) chiral 
symmetry $U(1)_A$, two currents have been introduced in noncommutative 
$U(N)$ gauge theories when defined without resorting to the 
Seiberg-Witten map. These currents are 
$\Psib_{s\,i}\ST(\gamma^{\mu}\gamma_5)_{st}\Psi_{t\,i}$ and 
$-\Psi_{s\, i}\ST\Psib_{t\, j}(\gamma^{\mu}\gamma_5)_{ts}$, where 
$\Psi_{t\,i}$ is a noncommutative Dirac fermion transforming under the 
 fundamental representation of $U(N)$. At the classical level, these currents 
are conserved and covariantly conserved, respectively, as a consequence of the 
rigid chiral invariance, $U(1)_A$, of the action. Further, unlike the 
current $j^{(cn)\;\mu}_{5}(x)$, they are local  
objects in the sense of noncommutative geometry, for they are 
$\ST$-polynomials of the noncommutative fields. For the theory defined by 
the action in eq.~(\ref{ncaction}), we have the following analogs of the 
previous currents
\begin{equation}
\begin{array}{l}
{j_{5}^{(np)\;\mu}\,=\,\sum_{f=1}^{N_f}\;j_{5\, f}^{(np)\;\mu}(x),\quad
j_{5\;ij}^{(p)\; \mu}\,=\,\sum_{f=1}^{N_f}\;j_{5\, f\;ij}^{(p)\;\mu}(x),}\\
 {j_{5\, f}^{(np)\;\mu}\,=\,
\Psib_{f\;s i}\ST(\gamma^{\mu}\gamma_5)_{st}\Psi_{f\; ti},\quad
(j_{5\,f}^{(p)\;\mu})_{ij}\,=\,-\Psi_{f\;s i}\ST\Psib_{f\;t j}(\gamma^{\mu}\gamma_5)_{ts}.}\\
\end{array}
\label{nccurrents}
\end{equation}
Now, $\Psi_{f\; t i}$ denotes a noncommutative Dirac fermion of our 
noncommutative $SU(N)$ theory. The reader may wonder why we should care about
a nongaugeinvariant current such as  $\sum_i\,(j_{5\,f}^{(p)\;\mu})_{ii}$. We
shall see in the next section that computing the quantum corrections to the
conservation equation of the chiral charge associated to it can be easily 
done at any order in $h$ and that, as we shall see below, this charge, 
even at the quantum level, is the same at any order in $h$ as the chiral 
charge of  $j_{5\, f}^{(np)\;\mu}$ and is also equal to the chiral 
charge of $j^{(cn)\;\mu}_{5}(x)$, at least at second order in $h$.

We shall show next that the currents in eq.~(\ref{nccurrents}) 
are conserved and covariantly conserved, respectively, 
at the  classical level and that this conservation comes from the  
invariance of the action under some type of 
transformations. To do so, we shall need the equation of motion for the 
ordinary fermion fields with action $S$ in eq.~(\ref{ncaction}), where 
the noncommutative fields are defined by the eq.~(\ref{SWmap}). 
Under arbitrary infinitesimal variations of $\psi_{f}$ and $\psib_{f}$, 
the action $S$ remains stationary if
\begin{displaymath}
\delta S\,=\,
\sum_{f=1}^{N_f}\idx\;\big[\delta\psib_{f}(1+\mathrm{M}^{\dagger})
i\Dirac_{\ST}\Psi_{f}+\Psib_{f}
i\Dirac_{\ST}(1+\mathrm{M})\delta\psi_{f}\big]\,=\,0.
\end{displaymath}
The symbol $\mathrm{M}^{\dagger}$ stands for the formal adjoint of 
$\mathrm{M}$. Taking into account that 
$(1+\mathrm{M})^{-1}=1+\sum_{n=1}^{\infty}\,(-1)^n\,\mathrm{M}^n$ and 
$(1+\mathrm{M}^{\dagger})^{-1}=1+
\sum_{n=1}^{\infty}\,(-1)^n\,(\mathrm{M}^{\dagger})^n$ formally exist as expansions in $h$, one easily shows that the previous equation is equivalent to
\begin{equation}
i\Dirac_{\ST} \Psi_{f}[\psi_{f}]\,=\,0,\quad 
i\overline{\Dirac_{\ST} \Psi_{f}[\psib_{f}]}\,=\,0.
\label{motion}
\end{equation}
These are the equations of motion for $\psi_{f}$ and $\psib_{f}$, whose left 
hand sides are to be understood as formal power expansions in $h$.
We use the notation $\overline{\Dirac_{\ST}\Psi_{f}}=
\partial_{\mu}\Psib_{f}[\psib_{f}]\gamma^{\mu}+i \Psib_{f}[\psib_{f}]\ST\Aslash$. Recall that 
the noncommutative spinors $\Psi_{f}$ and $\Psib_{f}$ depend on
the ordinary spinors $\psi_{f}$ and $\psib_{f}$ --see eq.~(\ref{SWmap}).

The equations of motion in eq.~(\ref{motion}) yield the following conservation
equations:
\begin{equation}
\partial_{\mu}\,j_{5}^{(np)\;\mu}(x)\,=\,0,\quad
\sum_{i}\,(\Dcal_{\mu}\,j_{5}^{(p)\;\mu})_{ii}(x)\,=\,0.
\label{conservation}
\end{equation}
Here $ \Dcal_{\mu\; ij}=\partial_{\mu}\delta_{ij}-i 
([A_{\mu},\phantom{A_{\mu}}]_{\ST})_{ij}$. The currents in the previous 
equation are defined in eq.~(\ref{nccurrents}). Note that the current 
$j_{5\;ij}^{(p)\;\mu}$ is covariantly 
conserved since it transforms covariantly under noncommutative gauge 
transformations. On the other hand, the current $j_{5}^{(np)\;\mu}$ is 
gauge invariant.

We shall show next that the conservation equations of  
eq.~(\ref{conservation}) are a consequence  
of the action in eq~(\ref{ncaction}) being chiral invariant under rigid  
transformations. Let us define the following
infinitesimal variations of $\Psib_f[\psi_f]$ and $\Psi_f[\psi_f]$:
\begin{equation}
\quad\quad \delta\Psi_f\,=\,-i\gamma_{5} \Psi_f[\psi_f]\ST\alpha,\quad  
\delta\Psib_f\,=\,-i\,\alpha\ST \Psib_f[\psib_f]\gamma_{5}.
\label{oddvariation}
\end{equation}
Here $\alpha$ is an infinitesimal arbitrary function of $x$. Note that for 
arbitrary $\alpha(x)$ neither $\delta\Psi_f$ nor  $\delta\Psib_f$ 
can be obtained by applying the Seiberg-Witten map in eq.~(\ref{SWmap}) to 
infinitesimal local variations of the corresponding ordinary fields, 
but this has no influence on our analysis. See however that 
if $\alpha(x)=\alpha=\rm{constant}$, then the variations in 
eq.~(\ref{oddvariation}) can be obtained by applying  
the Seiberg-Witten map of eq.~(\ref{SWmap}) to the rigid chiral transformations   
of eq.~(\ref{rigidtrans}). The variations of the previous equation induce the 
following change of the action in eq.~(\ref{ncaction}).
\begin{equation}
\delta S\,=\,\idx\,
\sum_{f=1}^{N_f}\big[ \alpha\ST \Psib_f[\psib_f]\gamma_{5}\ST \Dirac_{\ST}
\Psi_f[\psi_f]+\Psib_f[\psib_f]\ST\Dirac_{\ST}
(\gamma_{5} \Psi_f[\psi_f]\ST\alpha)\big].
\label{acvariation}
\end{equation}
Now, by partial integration one shows that 
\begin{displaymath}
\delta S\,=\,\idx\,
\sum_{f=1}^{N_f}\big[ \alpha\ST \Psib_f[\psib_f]\gamma_{5}\ST \Dirac_{\ST}
\Psi_f[\psi_f]-\overline{\Dirac_{\ST}\Psi_f[\psi_f]}\ST\
\gamma_{5} \Psi_f[\psi_f]\ST\alpha\big].
\end{displaymath}
Next, the r.h.s of equation eq.~({\ref{acvariation}) can be cast into the form
\begin{equation}
\delta S = \idx\!
\sum_{f=1}^{N_f}\big[ \alpha\ST \Psib_f[\psib_f]\gamma_{5}\ST \Dirac_{\ST}
\Psi_f[\psi_f]-\Psib_f[\psib_f]\gamma_{5}\ST(\Dirac_{\ST}
\Psi_f[\psi_f])\ST\alpha + \Psib_f[\psib_f]\ST\gamma^{\mu}\gamma_{5}
\Psi_f[\psi_f]\partial_{\mu}\alpha\big].
\label{variation}
\end{equation}
By setting $\alpha(x)=\alpha={\rm constant}$ in this equation, one easily 
shows that $S$ in eq.~(\ref{ncaction}) is invariant under the 
chiral transformations of eq.~(\ref{rigidtrans}). Finally, by combining 
eqs.~(\ref{acvariation}) and (\ref{variation}), and  choosing $\psi_f$ 
and $\psib_f$ to be solutions to the equation of motion 
--see eq.~(\ref{motion})--, one concludes that
\begin{displaymath}
  \idx\,\sum_{f=1}^{N_f}\big[ \Psib_f[\psib_f]\ST\gamma^{\mu}\gamma_{5}
\Psi_f[\psi_f]\partial_{\mu}\alpha\big]\,=\,0.
\end{displaymath}   
We have thus shown that the first identity in eq.~(\ref{conservation}) holds 
as a consequence of the invariance of the action under 
rigid chiral transformations. A similar analysis can be carried out for 
the transformations
\begin{displaymath}
\quad\quad \delta\Psi_f\,=\,-i\gamma_{5}\alpha\ST \Psi_f[\psi_f],\quad
\delta\Psib_f\,=\,-i\Psib_f[\psib_f]\ST\alpha\gamma_{5},
\end{displaymath}}
and explain the identity 
\begin{equation}
\sum_{i}\;(\Dcal_{\mu}\,j_{5}^{(p)\;\mu})_{ii}(x)\,=\,0
\label{tracedanom}
\end{equation}
as a by-product of the rigid chiral invariance of $S$ in eq.~(\ref{ncaction}).
Of course, one can use   the previous equation to introduce a new current, 
which is conserved, not covariantly conserved. 
Let $\ST_t$ denote the Moyal product obtained by changing $t$ for $h$ 
in eq.~(\ref{moyal}). Taking into account that 
\begin{displaymath}
\sum_{i}\;([A_{\mu},j_{5}^{(p)\;\mu}]_{\ST})_{ii}=
\partial_{\mu}\,\Big[\frac{1}{2}\;\theta^{\mu\beta}\,\sum_{i}
\,\int_{0}^{h}\,dt\;\big(\{A_{\nu},\partial_{\beta}j_{5}^{(p)\;\nu}\}_{\ST_t}\big)_{ii}\Big],
\end{displaymath}
one easily sees that 
\begin{displaymath}
\sum_{i}\;(\Dcal_{\mu}\,j_{5}^{(p)\;\mu})_{ii}(x)\,=\,
\partial_{\mu}\Big(\sum_{i}\,j_{5\;ii}^{(p)\;\mu}+\frac{1}{2}
\theta^{\mu\beta}\,\sum_{i}\,\int_{0}^{h}\,dt\;
\big(\{A_{\nu},\partial_{\beta}j_{5}^{(p)\;\nu}\}_{\ST_u}\big)_{ii}\Big).
\end{displaymath}
Then, one may introduce the current 
\begin{equation}
j_{5}^{(new)\;\mu}\,=\,\sum_{i}\,j_{5\;ii}^{(p)\;\mu}+\frac{1}{2}
\theta^{\mu\beta}\,\sum_{i}\,\int_{0}^{h}\,dt\;
\big(\{A_{\nu},\partial_{\beta}j_{5}^{(p)\;\nu}\}_{\ST_t}\big)_{ii},
\label{newcurrent}
\end{equation}
which is conserved if eq.~(\ref{tracedanom}) holds. Unfortunately, 
$j_{5}^{(new)\;\mu}$ is not gauge invariant, not even along the classical 
trajectories, so one would rather use the currents $j^{(cn)\;\mu}_{5}$ 
and $j_{5}^{(np)\;\mu}$ in eqs.~(\ref{swcurrent}) and (\ref{nccurrents}) to
analyse the properties of the theory. That  $j_{5}^{(new)\;\mu}$is not gauge invariant can be seen as follows. Let us express the r.h.s. of  
eq.~(\ref{newcurrent}) in terms of the ordinary fields by using the 
Seiberg-Witten map of eq.~(\ref{SWone}) and let us impose next the 
equation of motion of the fermion fields, then 
\begin{equation}
\begin{array}{l}
{j_{5}^{(new)\;\mu}\,=\,\sum_{f=1}^{N_f}\;j_{5\,f}^{(new)\;\mu},}\\
{j_{5\,f}^{(new)\;\mu}=\psib_{f}\,\gamma^{\mu}\gamma_5\,\psi_{f}
+\frac{i}{2}h\,\theta^{\alpha\beta}\,\overline{D_{\alpha}\psi}\gamma^{\mu}\gamma_5
D_{\beta}\psi}\\
{\phantom{j_{5\,f}^{(new)\;\mu}=\psib_{f}}
-ih\,\theta^{\alpha\beta}\,\partial_{\alpha}
\psib_f\,\gamma^{\mu}\gamma_5\partial_{\beta}\,\psi_f-ih\,\theta^{\mu\beta}\,
\partial_{\beta}\psib\,\gamma^{\nu}\gamma_5\,\partial_{\nu}\psi+ih\,
\theta^{\mu\beta}\,\partial_{\nu}\psib_f\,\gamma^{\nu}\gamma_5\,\partial_{\beta}\psi_f + o(h^2).}
\end{array}
\label{newexplicit}
\end{equation}
The previous expression is not gauge invariant. It can be seen that 
the current obtained from eq.~(\ref{newcurrent}) by using the most general 
Seiberg-Witten map differs from the current in eq.~(\ref{newexplicit}) in 
gauge invariant contributions. So changing the expression of the Seiberg-Witten map does not help in getting a gauge invariant  $j_{5}^{(new)\;\mu}$. And yet, 
for $\theta^{0i}=0$ and for fields that go  to zero fast enough as 
$\mid\vec{x}\mid\rightarrow\infty$, one can use  
$\sum_{i}\;(j_{5}^{(p)\;\mu})_{ii}(x)$ to define a conserved gauge  
invariant charge:
\begin{equation}
Q_5^{(p)}(t)\,=\,\int\,d^{3}\vec{x}\;   j^{(p)\;0}_{5}(t,\vec{x}).
\label{charge5p}
\end{equation}
Indeed, in this case
\begin{equation} 
Q_5^{(p)}(t)\,=\,Q_5^{(np)}(t), 
\label{pequalnp}
\end{equation}
with 
\begin{equation}
Q_5^{(np)}(t)\,=\,\int\,d^{3}\vec{x}\;   j^{(np)\;0}_{5}(t,\vec{x}),
\label{charge5np}
\end{equation}
To obtain eq.~(\ref{pequalnp}) we  have also assumed that the fermion fields
are already grassmann variables in the classical field theory. We have 
followed ref.~\cite{Bailin:1986wt} in making this assumption, for it  
positions us in the right place to start the quantization of the 
field theory. 

Let us now compute the difference $j^{(cn)\;\mu}_{5}-j_{5}^{(np)\;\mu}$ without
imposing the equation of motion. Taking into account that 
\begin{displaymath}
\begin{array}{l}
{j_{5\,f}^{(np)\;\mu}=
\psib_{f}\,\gamma^{\mu}\gamma_5\,\psi_{f}
+\frac{i}{2}h\,\theta^{\alpha\beta}\,\overline{D_{\alpha}\psi_f}\,\gamma^{\mu}\gamma_5\,
D_{\beta}\psi_f}\\
{\phantom{j_{5\,f}^{(np)\;\mu}=}
-\frac{1}{8}h^2\,\theta^{\alpha\beta}\theta^{\gamma\delta}\,
\overline{D_{\alpha}D_{\gamma}\psi_f}\,\gamma^{\mu}\gamma_5\,
  D_{\beta}D_{\delta}\,\psi_f-\frac{i}{4}h^2\,
\theta^{\alpha\beta}\theta^{\gamma\delta}\,\overline{D_{\alpha}\psi_f}
\,\gamma^{\mu}\gamma_5\,f_{\beta\gamma}D_{\delta}\,\psi_f}\\
{\phantom{j_{5\,f}^{(np)\;\mu}=}
-\frac{1}{32}h^2\,\theta^{\alpha\beta}\theta^{\gamma\delta}\,
\psib_f\,\gamma^{\mu}\gamma_5\,(f_{\alpha\beta}f_{\gamma\delta}-
4\,f_{\alpha\gamma}f_{\beta\delta})\,\psi_{f}\,+\,o(h^3),}\\
\end{array}
\end{displaymath}
one concludes that
\begin{equation}
{j_{5}^{(np)\,\mu}=j_{5}^{(cn)\,\mu}\,-\,{\cal Y}^{\mu},}\\
\label{diffnpcn}
\end{equation}
where
\begin{displaymath}
\begin{array}{l}
{{\cal Y}^{\mu}=\!-i\frac{h}{2}\Da(\theta^{\alpha\beta}\bar{\psi}\g^\mu\g^5 
D_\beta\psi)\!+\!i\frac{h}{2}\Da(\theta^{\mu\beta}\bar{\psi}\g^\alpha
\g^5 D_\beta\psi)\!+\!i\frac{h}{2}\theta^{\alpha\mu}\!
(\overline{D_\nu\psi}\g^\nu\g^5 D_\alpha\psi\!+\!\bar{\psi}\g^\nu\g^5 
D_\alpha D_\nu\psi)}\\
    {\phantom{{\cal Y}^{\mu}=}\!+\!h^2\theta^{\alpha\beta}\theta^{\rho\sigma}
\big[\frac{1}{8}
\Da\Dr(\bar{\psi}\g^\mu\g^5D_\beta D_\sigma\psi)\!-\!\frac{1}{8}
\Dr(\bar{\psi}\g^\mu\g^5D_\alpha\{D_\beta,D_\sigma\}\psi)\!
-\!\frac{1}{4}\Da(\bar{\psi}\g^{\mu}\g^5D_\beta D_\rho D_\sigma \psi)}\\
 {\phantom{{\cal Y}^{\mu}=}+\frac{1}{4}\Da(\bar{\psi}\g^\mu\g^5D_\rho D_\beta D_\sigma \psi)\big]+
   h^2  \theta^{\alpha\beta}\theta^{\mu\rho}\big[\frac{i}{8}
\Dn(\bar{\psi}\g^\nu\g^5\Dcal_\alpha f_{\beta\rho}\psi)+
\frac{i}{8}\Db(\bar{\psi}\g^\nu\g^5\Dcal_{\alpha}f_{\rho\nu}\psi)\big]}\\
{ \phantom{{\cal Y}^{\mu}=}-h^2\theta^{\alpha\beta}\theta^{\mu\rho}
\bar{\psi}\g^\nu\g^5\big[\frac{i}{8}\Dcal_\alpha f_{\beta\rho}D_\nu+
\frac{i}{8}\Dcal_\rho f_{\alpha\nu}D_\beta+\frac{i}{8}\Dcal_\alpha f_{\rho\nu}
D_\beta\big]\psi}\\
{\phantom{{\cal Y}^{\mu}=}+h^2\theta^{\alpha\beta}\theta^{\mu\rho}
\bar{\psi}\g^\nu\g^5\big[
+\frac{1}{8}f_{\rho\alpha}f_{\beta\nu}+\frac{1}{4}
f_{\alpha\nu}f_{\beta\rho}
    +\frac{1}{8}f_{\alpha\beta}f_{\rho\nu}\big]\psi+o(h^3).}\\
\end{array}
\end{displaymath}
Let us show next that $j_{5}^{(np)\,\mu}$ can also be interpreted as a Noether 
current, although not as the canonical current, if $\theta^{0i}=0$, $i=1,2,3$. 
First, one can prove by explicit computation that ${\cal Y}^{\mu}$ is 
conserved along the classical trajectories, which is not surprising since 
$\partial_{\mu}j_{5}^{(np)\,\mu}=0=\partial_{\mu}j_{5}^{(W)\,\mu}$. 
Secondly, if $\theta^{0i}=0$, $i=1,2,3$, then 
\begin{displaymath}
\begin{array}{l}
{{\cal Y}^{0}=\partial_{i}{\cal R}^{i},}\\
{{\cal R}^{i}=-i\frac{h}{2}\,\theta^{i j} \bar{\psi}\g^0 \g_5 D_j \psi}\\
{+h^2\,\theta^{i j}\theta^{i' j'}\big(+\frac{1}{8}
\partial_{i'} \bar{\psi}\g^0 \g_5 D_j D_{j'}\psi
-\frac{1}{8} \bar{\psi}\g^0 \g_5 D_{i'}\{D_{j'},D_j \}\psi
-\frac{1}{4}\bar{\psi}
\g^0 \g_5 D_j D_{i'} D_{j'} \psi }\\
{\phantom{+h^2\,\theta^{\alpha\beta}\theta^{\nu\rho}\big[}+\frac{1}{4}
\bar{\psi}
\g^o \g_5 D_{i'} D_j D_{j'}\psi \,\big).}\\
\end{array}
\end{displaymath}
Hence, 
\begin{displaymath}                
\int\,d^{3}\vec{x}\;\, {\cal Y}^{0}=\int\,d^{3}\vec{x}\;\,\partial_{i}
{\cal R}^{i}
=0,
\end{displaymath}
if the fields go to zero fast enough at spatial infinity. We thus conclude 
that, if time is commutative,  $j^{(cn)\;\mu}_{5}$ and $j^{(np)\;\mu}_{5}$ 
define the same charge, at least up to order $h^2$. Besides, for commutative 
time, we saw above that $j_{5}^{(p)\;\mu}$ and $j_{5}^{(np)\;\mu}$ 
yield the same chiral charge at any order in $h$. We thus come to the 
conclusion that for commutative time, and at least up to order $h^2$,  
$j^{(cn)\;\mu}_{5}$,  $j_{5}^{(p)\;\mu}$  and  $j_{5}^{(np)\;\mu}$ are 
such that  
\begin{displaymath}
Q_5^{(cn)}\,=\,Q_5^{(np)}\,=\, Q_5^{(p)}. 
\label{uniquecharge}
\end{displaymath}
$Q_5^{(cn)}$, $Q_5^{(np)}$ and $Q_5^{(p)}$ have been defined in 
eqs.~(\ref{chargeSW}),  (\ref{charge5np}) and (\ref{charge5p}), respectively. 
We shall take advantage of the previous equation to make a conjecture on 
the form of the anomalous equation satisfied by the quantum chiral charge 
at any order in $h$  --see section 5.

To close this section, we shall discuss the consequences of $Q_5^{(cn)}(t)$
being a constant of motion when we analyze the evolution of the fermionic 
degrees of freedom from $t=-\infty$ to  
$t=\infty$ in the background of a gauge field $a_{\mu}(x)$. With an eye on 
the quantization of the theory, we shall  introduce the  following boundary 
conditions for $a_{\mu}(t,\vec{x})$ in the temporal gauge  
$a_{0}(t,\vec{x})=0$: 
\begin{equation}
\begin{array}{l}
{a_{i}(t=\pm\infty,\vec{x})\,=\,i\,g_{\pm}(\vec{x})\,\partial_{i}\,g_{\pm}^{-1}(\vec{x}),}\\
{|a_{i}(t,\vec{x})|\leq \frac{c}{|\vec{x}|}\quad\text{as}\quad |\vec{x}|\rightarrow \infty,\quad i=1,2,3.}\\
\end{array}
\label{boundarycond}
\end{equation}
$g_{\pm}(\vec{x})$ is a element of $SU(N)$ for every $\vec{x}$ and 
$g_{\pm}(\mid\vec{x}\mid =\infty)=e$, $e$ being the identity of $SU(N)$. 
These boundary conditions arise naturally in the quantization of ordinary 
gauge theories when topologically nontrivial configurations
are to be taken into account~\cite{Jackiw:1976pf, Callan:1976je}. The boundary 
condition $g_{\pm}(\mid\vec{x}\mid =\infty)=e$ makes it possible the classification of the maps $g_{\pm}(\vec{x})$ in equivalence classes which are elements 
of the homotopy group $\Pi_{3}(SU(N))$. At $t=\pm\infty$ the ordinary gauge 
field yields pure gauge fields $a_{i}^{\pm}(\vec{x})$ with well-defined 
winding numbers, $n_{\pm}$, given by 
\begin{equation}
n_{\pm}\,=\, \frac{i}{24\pi^2}\;\int\,d^{3}\vec{x}\; \epsilon^{ijk}\;
Tr(a_{i}^{\pm}a_{j}^{\pm}a_{k}^{\pm}).
\label{npm}
\end{equation}
The reader should note that by keeping the same boundary conditions for the 
ordinary fields $a_{\mu}$ in the noncommutative theory as in the corresponding 
ordinary gauge theory, we are assuming that the space of noncommutative fields 
is obtained by applying the Seiberg-Witten map --understood as an expansion in 
powers of $h$-- to the space of gauge fields of ordinary gauge 
theory. At least for $U(N)$ groups, this approach misses \cite{Kraus:2001xt} 
some topologically nontrivial noncommutatative gauge configurations 
\cite{Nekrasov:1998ss} and it is not known whether it is possible to modify 
the boundary conditions for the ordinary fields so as to iron this problem out.
Here, we shall be discussing the evolution of the fermionic degrees of freedom 
given by the action in eq.~(\ref{ncaction}) in any noncommutative gauge field 
background which is obtained by applying the $\theta$-expanded Seiber-Witten 
map to a given ordinary field belonging to the space of gauge fields 
of ordinary gauge theory. For $SU(N)$ groups, this is interesting on its own, 
but, as with $U(N)$ groups, it might not be the end of the story.

From eqs.~(\ref{swcurrent}), (\ref{chargeSW}) and (\ref{boundarycond}), 
we conclude that, up to second order in $h$, we have
\begin{displaymath}
Q_5^{(cn)}(t=\pm\infty)\,=\,\sum_{f}\;\int\,d^{3}\vec{x}\;
\psi^{\dagger}_{f}(t=\pm\infty,\,\vec{x})\gamma_5\,\psi_{f}
(t=\pm\infty,\,\vec{x}).
\end{displaymath}
Recall that $Q_5^{(cn)}(t)$ is gauge invariant object, so that the choice of
gauge has no influence on its value. Here we have chosen the gauge $a_0(x)=0$.
In the quantum field theory, the r.h.s of the previous equation yields the 
difference between the fermion number,$n_{R}^{\pm}$,  of asymptotic 
right-handed fermions and the fermion number,$n_{L}^{\pm}$, of asymptotic 
left-handed fermions.   
Hence, if $Q_5^{(cn)}(t)$ were conserved upon second quantization, the
following equation would hold in the quantum field theory:
\begin{equation}
0\,=\,Q_5^{(cn)}(t=\infty)- Q_5^{(cn)}(t=-\infty) \,=\,(n_{R}^{+}-n_{R}^{-})-
(n_{L}^{+}-n_{L}^{-}).
\label{asymptoticchiral}
\end{equation}
We saw above --see discussion below eq.~(\ref{rigidtrans})-- that the  vector 
$U(1)_{V}$ symmetry of the classical theory survives renormalization. So, in 
the quantum theory we have 
\begin{equation}
0\,=\,Q^{(cn)}(t=\infty)- Q^{(cn)}(t=-\infty) \,=\,
(n_{R}^{+}-n_{R}^{-})+
(n_{L}^{+}-n_{L}^{-}).
\label{asymptoticbarion}
\end{equation}
The reader should notice that $Q^{(cn)}(t)$ can be obtained from 
$Q_5^{(cn)}(t)$ by stripping the latter of its $\gamma_{5}$ matrix. Now, by 
combining eqs.~(\ref{asymptoticchiral}) and (\ref{asymptoticbarion}), we would 
reach the conclusion that in the presence of a  
background field satisfying the boundary conditions in 
eq.~(\ref{boundarycond}), if we prepare a scattering experiment where we have 
$n_{R}$ right-handed fermions at $t=-\infty$, there will come out  $n_{R}$ 
right-handed fermions at  $t=+\infty$. The same analysis could  be carried 
out  independently for left-handed fermions, reaching an analogous conclusion. 
The conclusions just discussed are a consequence of the fact that in the 
massless classical action right-handed fermions are not coupled left-handed 
fermions. However, as we shall see below, quantum corrections, when computed 
properly, render eq.~(\ref{asymptoticchiral}) false, if the difference of 
winding numbers $n_{+}-n_{-}$  does not vanish. Thus, quantum fluctuations 
introduce a coupling between right-handed and left-handed fermions.

\section{Anomalous $U(1)_A$ currents}

This section is devoted to the computation of the one-loop anomalous 
contributions to the classical conservation equations 
\begin{equation}
\sum_{i}\,(\Dcal_{\mu}\,j_{5}^{(p)\;\mu})_{ii}(x)\,=\,0,\quad \partial_{\mu}\,j_{5}^{(np)\;\mu}(x)\,=\,0,\quad \partial_{\mu}\,j_{5}^{(cn)\,\mu}(x)\,=\,0.
\label{coneq}
\end{equation}
The currents $j_{5}^{(p)\;\mu}$, $j_{5}^{(np)\;\mu}$ and $j_{5}^{(cn)\,\mu}$ 
are given in eqs.~(\ref{nccurrents}) and (\ref{swcurrent}). The anomalous contributions to  the first conservation equation in eq.~(\ref{coneq})
will be computed at any order in $h$, whereas the anomalous contribution to
the remaining equalities in eq.~(\ref{coneq}) will be worked out only up to 
second order in $h^2$. To carry out the  
computations we shall use dimensional regularization and its minimal 
subtraction renormalization  algorithm as defined in 
refs.~\cite{Breitenlohner:1977hr} and \cite{Bonneau:1979jx} 
--see also  ref.~\cite{Martin:1999cc} and references therein.
Hence, our $\gamma_5$ in $D$-dimensions will not anticommute with 
$\gamma^{\mu}$. The dimensionally regularized $\theta^{\mu\nu}$ will be 
defined as an intrinsically ``4-dimensional'' antisymmetric object:
\begin{equation}
 \theta^{\mu\nu}=-\theta^{\nu\mu}, \quad \theta^{\mu\nu}\,\hat{g}_{\nu\rho}=0.
\label{dimtheta}
\end{equation}
Before we plunge into the actual computations, we need some definitions and 
equalities that hold in dimensional regularization. 
Let $ <{\cal O}(a_{\mu},\Psi_{f},\Psib_{f}) >^{(A)}$ be the v.e.v. of the  
operator ${\cal O}(a_{\mu},\Psi_{f},\Psib_{f})$ in the noncommutative 
background $A_{\mu}$ as defined by
\begin{equation}
 <{\cal O}(a_{\mu},\Psi_{f},\Psib_{f})>^{(A)}\,=\,
\frac{1}{Z[A]}\int\prod_{f}\,d\psi_{f} d\psib_{f}\quad 
{\cal O}(a_{\mu},\psi_{f},\psib_{f})\quad 
e^{iS[\Psi_f,\bar{\Psi}_f,A]_{Fermi}^{DR}}.
\label{vevpsi}
\end{equation}
The partition function $Z[A]$ reads
\begin{equation}
Z[A]=\int\prod_{f}d\psi_{f} d\psib_{f}\quad 
e^{iS[\Psi_f,\bar{\Psi}_f,A]_{Fermi}^{DR}}.
\label{partitionfunction}
\end{equation}
In the two previous equations, $S[\Psi_f,\bar{\Psi}_f,A]_{Fermi}^{DR}$ denotes the fermionic part of the action in eq.~(\ref{ncaction}) in the 
``$D$-dimensional'' space-time of dimensional regularization, i.e., 
\begin{equation} 
S[\Psi_f,\bar{\Psi}_f,A]_{Fermi}^{DR}\,=\,\sum_{f=1}^{N_f}\,
{\int\!\! d^D\!x}\Psib_{f}\ST  i\Dirac_{\ST}\Psi_{f}.
\label{draction}
\end{equation}
The noncommutative fields $A_{\mu}$,$\Psi_f$ and  $\bar{\Psi}_f$ are 
given by  the Seiberg-Witten map of eq.~(\ref{SWmap}) with objects defined in 
the ``$D$-dimensional space-time of dimensional regularization. Next, 
by changing variables from $(\psi_f,\psib_f)$ to $(\Psi_f,\Psib_f)$ in the 
path integrals in eqs.~(\ref{vevpsi}) and (\ref{partitionfunction}), we 
conclude that the following string of equalities hold in dimensional 
regularization:
\begin{equation}
\begin{array}{l}
{<{\cal O}(a_{\mu},\Psi_{f},\Psib_{f})>^{(A)}}\\
{\phantom{{\cal O}(a_{\mu},\Psi_{f},\Psib_{f})}
=\frac{1}{Z[A]}\int\prod_{f}\,d\Psi_{f} d\Psib_{f}\;
{\rm det}\big[1+{\rm M}\big]{\rm det}\big[1+\bar{{\rm M}}\big]\,
{\cal O}(a_{\mu},\Psi_{f},\Psib_{f})\, 
e^{iS[\Psi_f,\bar{\Psi}_f,A]_{Fermi}^{DR}}}\\
{\phantom{{\cal O}(a_{\mu},\Psi_{f},\Psib_{f})}=
\frac{1}{Z[A]}\int\prod_{f}\,d\Psi_{f} d\Psib_{f}\;
{\cal O}(a_{\mu},\Psi_{f},\Psib_{f})\;
e^{iS[\Psi_f,\bar{\Psi}_f,A]_{Fermi}^{DR}}.}\\
\end{array}
\label{simplification}
\end{equation} 
The operators ${\rm M}$ and $\bar{{\rm M}}$ are equal, respectively, to the 
formal power expansions in $h$ $\sum_n h^n{\rm M}^{(n)}[\gamma^\rho,\theta^{\rho\lambda},a_{\nu},\partial_{\sigma}]$ and  $\sum_n h^n\bar{{\rm M}}^{(n)}[\gamma^\rho,\theta^{\rho\lambda},a_{\nu},\partial_{\sigma}]$, which are given in 
eq.~(\ref{SWmap}), but with objects defined as ``D-dimensional'' Lorentz 
covariants. Note that the last equality in  eq.~(\ref{simplification}) is a 
consequence of the fact that in dimensional regularization we have 
\begin{displaymath}
{\rm det}\big[1\,+\,{\rm M}\big]={\rm det}\big[1\,+\,\bar{{\rm M}}\big]=1.
\end{displaymath}
Of course, in dimensional regularization, we also have
\begin{equation} 
Z[A]\,=\,\int\prod_{f}\,d\Psi_{f} d\Psib_{f}\quad
e^{iS[\Psi_f,\bar{\Psi}_f,A]_{Fermi}^{DR}},
\label{partitionsimpler}
\end{equation}
if $Z[A]$ is as defined in eq.~(\ref{partitionfunction}). To simplify the 
calculations as much as possible, we shall compute the anomalous contributions 
to the three classical conservation equations in eq.~(\ref{coneq}) keeping 
in the computation the ordering dictated by the latter equation.

\subsection{Anomalous Ward identity for $j_{5}^{(p)\;\mu}$}

The variation of $S[\Psi_f,\bar{\Psi}_f,A]_{Fermi}^{DR}$ 
in eq.~(\ref{draction}) under the chiral transformations
\begin{displaymath}
\quad\quad \delta\Psi_f\,=\,-i\gamma_{5}\alpha\ST \Psi_f\quad
\delta\Psib_f\,=\,-i\Psib_f\ST\alpha\gamma_{5}
\end{displaymath}
reads
\begin{displaymath}
\delta S_{Fermi}^{DR}\,=\,-\iDx\,\big[\sum_{i}\,(\Dcal_{\mu}\,j_{5}^{(p)\;\mu})_{ii}(x)+2\sum_{f}\,(D_{\mu}\Psi_f)_{si}\ST\Psib_{f\;ti}({\hat{\gamma}}^{\mu}\gamma_5)_{ts}\big]
\, \alpha(x)
\end{displaymath}
This result and the invariance of $Z[A]$ in eqs.~(\ref{partitionsimpler}) 
under the previous transformations leads to 
\begin{equation}
\sum_{i}\,(\Dcal_{\mu}\,<j_{5}^{(p)\;\mu}>^{(A)})_{ii}\,=\,
-2\sum_{f}\,<(D_{\mu}\Psi_f)_{si}\ST\Psib_{ti}
({\hat{\gamma}}^{\mu}\gamma_5)_{ts}>^{(A)}.
\label{reganom}
\end{equation}
The v.e.v. in the noncommutative background $A_{\mu}$, $<\cdots>^{(A)}$, 
is defined by the last line of eq.~(\ref{simplification}). 
Always recall that this definition is equivalent 
to the definition in eq.~(\ref{vevpsi}), if dimensional regularization is 
employed. Note that the r.h.s of eq.~(\ref{reganom}) contains an evanescent 
operator --see ref.~\cite{Collins:1984xc}, page 346--, so it will naively 
go to zero as $D\rightarrow 4$, yielding a 
covariant conservation equation. And yet, this evanescent operator will give
a finite contribution when inserted in a divergent loop. This is how the 
anomalies comes about in dimensional regularization.

The minimal subtraction scheme algorithm  
\cite{Breitenlohner:1977hr,Bonneau:1979jx,Collins:1984xc}   
applied to both sides of eq.~(\ref{reganom})  
leads to a renormalized equation in the limit $D\rightarrow 4$: 
\begin{equation}
\sum_{i}\,(\Dcal_{\mu}\,<j_{5}^{(p)\;\mu}>^{(A)}_{\rm MS})_{ii}\,=\,
-2\sum_{f}\,<(D_{\mu}\Psi_f)_{si}\ST\Psib_{ti}({\hat{\gamma}}^{\mu}\gamma_5)_{st}>^{(A)}_{\rm MS}.
\label{msanom} 
\end{equation}
The Feynman diagrams that yield the r.h.s of eq.~(\ref{reganom}) are given in 
Fig. 1.
\vskip 0.5cm 
\begin{figure}[h]
\centering
\subfigure[]{\epsfig{file=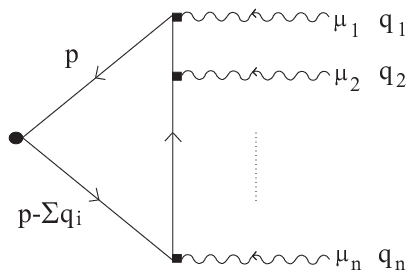}}
\subfigure[]{\epsfig{file=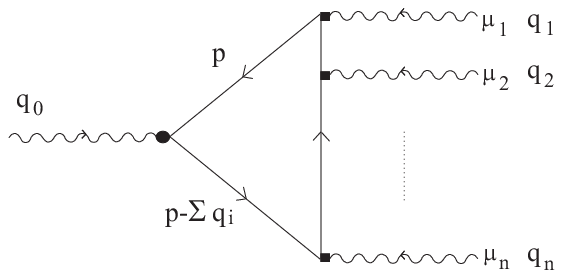}}
\\[5pt]
 \renewcommand{\figurename}{Fig.}
 \renewcommand{\captionlabeldelim}{.}
\caption{Diagrams that give the r.h.s of eq.~(\ref{reganom}).}
\end{figure} 
With the help of the Feynman rules in the Appendix A, we conclude that  
the Feynman diagram in Fig. 1a) represents the following Feynman integral
\begin{equation}
\begin{array}{l}
{{\mathfrak A}_n\,=\,N_f\,\frac{2(-1)^{n}}{n!}\,
       e^{\frac{i}{2}\,h\,\sum_{i>j}{q_i \circ q_j}}\Tr\,
A_{\mu_1}(q_1)A_{\mu_2}(q_2)\dots A_{\mu_n}(q_n)}\\
{\phantom{{\mathfrak A}_n\,=\,\frac{2(-1)^{n}}{n!}
e^{\frac{i}{2}\,h\,\sum_{i>j}{q_i \circ q_j}}A_{\mu_1} }
\iDp
 \tr\frac{\gamma_5 \hat{\pslash} \pslash \gamma^{\mu_1}
        (\pslash-\qslash_1)\gamma^{\mu_2}(\pslash-\qslash_1-\qslash_2)\dots
         \gamma^{\mu_n}\left(\pslash- \sum{ \qslash_i} \right)}{p^2 (p-q_1)^2
        (p-q_q-q_2)^2\dots\left(p-\sum{
         q_i}\right)^2}.}\\
\end{array}
\label{funnya}
\end{equation}
The Feynman diagram in Fig. 1b) yields the following Feynman integral
\begin{equation}
\begin{array}{l}
{{\mathfrak B}_n\,=\,N_f\,\frac{2(-1)^{n}}{n!}\,
         e^{\frac{i}{2}\,h\,\sum_{i>j}{q_i \circ q_j}}\Tr A_{\mu}(q_0)A_{\mu_1}(q_1)
         A_{\mu_2}(q_2)\dots A_{\mu_n}(q_n)}\\
{\phantom{{\mathfrak B}_n\equiv\frac{2(-1)^{n}}{n!}\,
         e^{\frac{i}{2}\,h\,\sum_{i>j}{q_i \circ q_j}}}
\iDp
        \tr\frac{\gamma_5 \hat{\gamma^\mu} \pslash \gamma^{\mu_1}
         (\pslash-\qslash_1)\gamma^{\mu_2}(\pslash-\qslash_1-\qslash_2)\dots
         \gamma^{\mu_n}\left(\pslash-\sum{ \qslash_i} \right)}{p^2 (p-q_1)^2
       (p-q_q-q_2)^2\dots\left(p-\sum{
         q_i}\right)^2}.}\\
\end{array}
\label{bene}
\end{equation}
In eqs.~(\ref{funnya}) and (\ref{bene}), we have used 
$q_i \circ  q_j$ as shorthand for 
$\theta^{\mu\nu}\,q_{\mu\,i} q_{\mu\,j}$. Note that 
from the point of view of its $\theta^{\mu\nu}$-dependence the diagrams in Fig. 1 are planar diagrams. Hence,  no loop momenta is contracted with 
$\theta^{\mu\nu}$ in the corresponding Feynman integrals. This feature of the 
diagrams contributing to the r.h.s. of  eq.~(\ref{reganom}) makes it feasible  
their computation at any order in $h$. Let us remark that in keeping with the 
general strategy adopted in this paper the exponentials involving 
$h\, \theta^{\mu\nu}$ are always understood as given by their expansions in 
powers of $h$.

It turns out that the UV degree of divergence at $D=4$ of the integral that
is obtained from ${\mathfrak A}_n$ by replacing $\hat{\pslash}$ with 
$\pslash$ is negative if $n>4$. Then, for $n>4$, ${\mathfrak A}_n$ vanishes as 
$D\rightarrow 4$. The same type of power-counting arguments can be applied to  
${\mathfrak B}_n$, to conclude that these integrals, if $n>3$, go to zero as 
$D\rightarrow 4$. Now, using the trace identities in 
eq.~(\ref{traza-antic}), one easily shows that 
${\mathfrak A}_1={\mathfrak B}_1=0$. After a little Dirac algebra, one can show
that the contributions to ${\mathfrak B}_2$ and ${\mathfrak B}_3$ that involve
integrals that are not finite by power-counting at $D=4$ are all proportional
to contractions of the type $\hat{g}_{\mu\nu}\epsilon^{\nu\rho\lambda\sigma}$. 
Since these contractions vanish --see eq.~(\ref{gammascom})--, we have
${\mathfrak B}_2={\mathfrak B}_3=0$. In summary, in the limit   
$D \rightarrow 4$, only  
${\mathfrak A}_2$,  ${\mathfrak A}_3$ and ${\mathfrak A}_4$ may give contributions to the r.h.s. of eq.~(\ref{reganom}), and, indeed, they do so. After some 
Dirac algebra --see Appendix B-- and with the help of the integrals in 
Appendix C, one obtains the following results for  
${\mathfrak A}_2$, ${\mathfrak A}_3$ and ${\mathfrak A}_4$ in position space 
and in the dimensional regularization minimal subtraction scheme:
\begin{equation}
\begin{array}{l}
{	{\mathfrak A}_2=-\frac{N_f}{8\pi^2}\epsilon^{\mi\mii\miii\miv}
\Tr\, \Di \Aii\star\Diii \Aiv,}\\
{{\mathfrak A}_3=-i\frac{N_f}{4\pi^2}\epsilon^{\mi\mii\miii\miv}\,\Tr\,\left[\Div\Ai\star\Aii\star\Aiii+
\Ai\star\Aii\star\Div\Aiii\right],}\\
{{\mathfrak A}_4=\frac{N_f}{4\pi^2}\,\epsilon^{\mi\mii\miii\miv}\,\,\Tr\,\Ai\star\Aii\star\Aiii\star\Aiv.}\\
\end{array}
\label{funnyu234}
\end{equation}
Substituting eq.~(\ref{funnyu234}) in the r.h.s. of eq.~(\ref{msanom}), one 
gets
\begin{equation}
\sum_{i}\,(\Dcal_{\mu}\,<j_{5}^{(p)\;\mu}>^{(A)}_{\rm MS})_{ii}\,=\,
\frac{N_f}{16\pi^2}\epsilon^{\mi\mii\miii\miv}\,\Tr\,F_{\mi\mii}\star F_{\miii\miv}.
\label{planeanom}
\end{equation}
This equation looks like the  corresponding equation for $U(N)$ groups 
--see eq.~(9b) in ref.~\cite{Gracia-Bondia:2000pz}. This similarity comes from 
the fact that in both cases no loop momenta is contracted with 
$\theta^{\mu\nu}$, and currents and interaction vertices are  the same type 
of polynomials with respect to the Moyal product. However, there are two 
striking differences. First, the theory in  ref.~\cite{Gracia-Bondia:2000pz} need not be defined by means of the Seiberg-Witten map, but  the theory considered in this paper is unavoidably constructed by using the Seiberg-Witten map. 
Secondly, the object $F_{\mi\mii}$ belongs to the Lie algebra of $U(N)$ in the 
theory of ref.~\cite{Gracia-Bondia:2000pz}, whereas it belongs to the 
enveloping algebra of $SU(N)$, not to its Lie algebra, in the case 
studied here.

Eq.~(\ref{planeanom}) leads to the conclusion that, at least at the one-loop 
level, the classical conservation equation for $j_{5}^{(p)\;\mu}$ in eq.~(\ref{coneq}) 
should be replaced with 
\begin{equation}
\sum_{i}\,(\Dcal_{\mu}\,{\rm N}[j_{5}^{(p)\;\mu}]_{\rm MS})_{ii}\,=\,
\frac{N_f}{8\pi^2}\, \Tr\,F_{\mi\mii}\star \tilde{F}_{\miii\miv},
\label{operatorpanom}
\end{equation}
where ${\rm N}[\phantom{j_{5}}]_{\rm MS}$ denotes normal product of operators in the MS 
scheme~\cite{Bonneau:1979jx,Collins:1984xc}  and  $\tilde{F}^{\mi\mii}=\frac{1}{2}\epsilon^{\mi\mii\miii\miv}F_{\miii\miv}$. Eq.~(\ref{operatorpanom}) tell us that for commutative time, i.e., $\theta^{0i}=0$, the charge $Q_{5}^{(p)}$ is no  longer conserved but verifies the following anomalous equation
\begin{equation}
Q_{5}^{(p)}(t=+\infty)-Q_{5}^{(p)}(t=-\infty)\,=\,
\frac{N_f}{8\pi^2}\idx\, \Tr\,F_{\mi\mii}\star \tilde{F}_{\miii\miv}.
\label{pminf}
\end{equation}
The charge $Q_{5}^{(p)}(t)$ was defined in eq.~(\ref{charge5p}). To obtain  
the l.h.s. of the previous equation, we have integrated the l.h.s of
eq.~(\ref{operatorpanom}) and assumed that the fields vanish fast enough at 
spatial infinity so as to make the following identity
\begin{displaymath}
\int\,d^{3}\vec{x}\;(\Phi_{1}\ST\Phi_2)(t,\vec{x})\,=\,
\int\,d^{3}\vec{x}\; \Phi_1(t,\vec{x})\Phi_2(t,\vec{x})
\end{displaymath}
valid for $\theta^{\mu\nu}$ such that $\theta^{0i}=0$. This choice of 
asymptotic behaviour is standard in noncommutative field theory~ 
\cite{Douglas:2001ba, Szabo:2001kg, Chu:2005ev}
and renders the kinetic terms of the fields in ordinary and
noncommutative space-time equal.

Using the techniques in~\cite{Cerchiai:2002ss}, it is not difficult to 
show that 
\begin{displaymath}
\idx\, \Tr\,F_{\mi\mii}\star \tilde{F}_{\miii\miv}\,=\,
\idx\, \Tr\,f_{\mi\mii}\tilde{f}_{\miii\miv},
\end{displaymath}
at least for the boundary conditions in eq.~(\ref{boundarycond}). 
This equation was obtained for the $U(1)$ gauge group in 
ref.~\cite{Banerjee:2003ce}. 
Now, by combining the previous equation with eq.~(\ref{pminf}), and, then, 
using the temporal gauge and the boundary conditions in 
eq.~(\ref{boundarycond}), one concludes that 
\begin{equation}
Q_{5}^{(p)}(t=+\infty)-Q_{5}^{(p)}(t=-\infty)\,=\,
\frac{N_f}{8\pi^2}\idx\, \Tr\,f_{\mi\mii} \tilde{f}_{\miii\miv}\,=\,
2\,N_f\,(n_{+}-n_{-}).
\label{planeppm}
\end{equation}
The integers $n_{\pm}$ are defined in eq.~(\ref{npm}).

\subsection{Anomalous Ward identity for $j_{5}^{(np)\;\mu}$}

The variation of $S[\Psi_f,\bar{\Psi}_f,A]_{Fermi}^{DR}$ 
in eq.~(\ref{draction}) under the chiral transformations
\begin{displaymath}
\quad\quad \delta\Psi_f\,=\,-i\gamma_{5}\Psi_f\ST\alpha \quad
\delta\Psib_f\,=\,-i\alpha\ST\Psib_f\gamma_{5}
\end{displaymath}
reads
\begin{displaymath}
\delta S_{Fermi}^{DR}\,=\,-\iDx\,\big[
(\partial_{\mu}\,j_{5}^{(np)\;\mu})(x)-2\sum_f\,(\Psib_{f}\ST
{\hat{\gamma}}^{\mu}\gamma_5
\gamma_5\Psi_f)(x)\big]\,\alpha(x).
\end{displaymath}
Now, $Z[A]$ in eqs.~(\ref{partitionsimpler}) is invariant under the previous
chiral transformations. That $\delta Z[A]=0$ and that $\delta S_{Fermi}^{DR}$
be given by the previous expression leads to 
\begin{equation}
\partial_{\mu}<j_{5}^{(np)\;\mu}>^{(A)}(x)=
2\, \sum_f\,<\Psib_{f}\ST
{\hat{\gamma}}^{\mu}\gamma_5
D_{\mu}\Psi_f>^{(A)}(x).
\label{reganomnp}
\end{equation}
The v.e.v. in the noncommutative background $A$,  $<\cdots >^{(A)}$, is 
defined by the last line of  eq.~(\ref{simplification}), which in dimensional 
regularization is equivalent to the original definition in 
eq.~(\ref{vevpsi}). Note that either side of  
eq.~(\ref{reganomnp}) is invariant under $SU(N)$ gauge transformations of 
 $a_{\mu}$, here the MS scheme algorithm of dimensional regularization will 
yield a gauge invariant result when applied to either side of that equation. 

The r.h.s. of  eq.~(\ref{reganomnp}) contains an evanescent 
operator, which  upon  
MS dimensional renormalization will give a finite contribution when inserted 
in UV divergent fermion loops. In this subsection we will compute this finite 
contribution up to second order in $h$. At first order  in $h$, we shall 
work out every Feynman diagram giving, in the $D\rightarrow 4$ limit, a 
nonvanishing contribution to the r.h.s. of eq.~(\ref{reganomnp}). To make this 
computation feasible at order $h^2$, we will take advantage of the gauge 
invariance of the result and compute explicitly only the minimum number 
of Feynman diagrams needed. Let us show next that if we have a gauge invariant 
expression, say ${\cal A}^{(2)}[a_{\mu}]$, that 
matches the contribution obtained by  
explicit computation of the diagrams involving fewer than five fields 
$a_{\mu}$, then there is no room for the Feynman diagrams with five or more 
fields $a_{\mu}$ giving a contribution not included in  ${\cal A}^{(2)}$.  
The standard BRS transformations reads:
\begin{equation}
sa^{a}_{\mu}=s_{0}a^{a}_{\mu}-s_{1}a^{a}_{\mu},\quad s_{0}a^{a}_{\mu}=
\partial_{\mu}c^a,\quad s_{1}a^{a}_{\mu}=-if^{abc}a^{b}_{\mu}c^{c},\quad
\quad sc^a=if^{abc}c^b c^c.
\label{BRS}
\end{equation}
Then, the gauge invariance of ${\cal A}^{(2)}[a_{\mu}]$, $s{\cal A}^{(2)}=0$, 
is equivalent to the following set of equations
\begin{equation}
\begin{array}{l}
{s_{0}{\cal A}^{(2)}_2\,=\,0,}\\
{s_{0}{\cal A}^{(2)}_3\,=\, s_{1}{\cal A}^{(2)}_2,}\\
{s_{0}{\cal A}^{(2)}_4\,=\, s_{1}{\cal A}^{(2)}_3,}\\
{s_{0}{\cal A}^{(2)}_5\,=\, s_{1}{\cal A}^{(2)}_4,}\\
{s_{0}{\cal A}^{(2)}_6\,=\, s_{1}{\cal A}^{(2)}_5,}\\
{s_{0}{\cal A}^{(2)}_7\,=\, s_{1}{\cal A}^{(2)}_6,}\\
{s_{0}{\cal A}^{(2)}_8\,=\, s_{1}{\cal A}^{(2)}_7,}\\
{s_{1}{\cal A}^{(2)_8}\,=\,0.}\\
\end{array}
\label{brseq}
\end{equation}
The symbol ${\cal A}^{(2)}_n$, $n=2,3,4,5,6,7$ and $8$, denotes the 
contribution to ${\cal A}^{(2)}[a_{\mu}]$ involving $n$ fields, and 
its derivatives, $a_{\mu}$:
\begin{displaymath}
{\cal A}^{(2)}[a_{\mu}]\,=\,\sum_{n=2}^{8}\;{\cal A}^{(2)}_{n}[a_{\mu}].
\end{displaymath}
Dimensional analysis shows that $n<9$. Indeed,  ${\cal A}^{(2)}_n$ has 
dimension 4 and   
${\cal A}^{(2)}_n = h^2\,\theta^{\mu_1\mu_2}\theta^{\mu_3\mu_4} 
f_{\mu_1\mu_2\mu_3\mu_4}[a_\mu]$, $f_{\mu_1\mu_2\mu_3\mu_4}[a_\mu]$ being a gauge invariant polynomial of $a_{\mu}$ and its derivatives. The fact that the 
generators of a unitary representation of $SU(N)$ are traceless implies that 
$n>1$. Let ${\cal B}^{(2)}=h^2\,\theta^{\mu_1\mu_2}\theta^{\mu_3\mu_4} 
g_{\mu_1\mu_2\mu_3\mu_4}[a_\mu]$ be a  gauge invariant  
--i.e., $s{\cal B}^{(2)}=0$--  polynomial of $a_{\mu}$ 
and its derivatives which is equal to ${\cal A}^{(2)}$ up to contributions
with more than four $a_{\mu}$, or derivatives of it, and has dimension 4:
\begin{displaymath}
{\cal B}^{(2)}\,=\,\sum_{n=2}^{4}\;{\cal A}^{(2)}_{n}[a_{\mu}]\,+ \,
\sum_{n=5}^{8}\;{\cal B}^{(2)}_{n}[a_{\mu}].
\end{displaymath}
${\cal B}^{(2)}_{n}$ denotes the contribution involving $n$ fields $a_{\mu}$,  
or derivatives of it. Let ${\cal C}^{(2)}_{n}$ stand for the difference 
${\cal A}^{(2)}_{n}-{\cal B}^{(2)}_{n}$, $n=5,6,7$ and $8$. Then, the BRS
invariance of both ${\cal A}^{(2)}$ and ${\cal B}^{(2)}$  
--use eq.~(\ref{brseq})-- leads to
\begin{equation}
\begin{array}{l}
{s_{0}{\cal C}^{(2)}_5\,=\,0,}\\
{s_{0}{\cal C}^{(2)}_6\,=\, s_{1}{\cal C}^{(2)}_5,}\\
{s_{0}{\cal C}^{(2)}_7\,=\, s_{1}{\cal C}^{(2)}_6,}\\
{s_{0}{\cal C}^{(2)}_8\,=\, s_{1}{\cal C}^{(2)}_7,}\\
{s_{1}{\cal C}^{(2)}_8\,=\,0.}\\
\end{array}
\label{brseqdiff}
\end{equation}
Now, the cohomology of the operator $s_{0}$ over the space of polynomials of 
$a^{a}_{\mu}$, $c^a$ and their derivatives has been worked out in 
refs.~\cite{Brandt:1989rd,Brandt:1989gy}. The nontrivial part of this
cohomology is given by polynomials of $f^{a\,(free)}_{\mu\nu}=
\partial_{\mu}a^{a}_{\nu}-\partial_{\nu}a^{a}_{\mu}$ and/or its derivatives 
and/or  $c^{a}$. 
Since ${\cal C}^{(2)}_5$ belongs to the nontrivial part of the cohomology of  
$s_0$ and does not depends on $c^a$, we conclude that it should be either 
zero or a polynomial of  $f^{a\,(free)}_{\mu\nu}$ and its derivatives. 
This last possibility will never be realized in the case under scrutiny since one can show by 
dimensional analysis that ${\cal C}^{(2)}_5$ can contain only two partial 
derivatives, i.e., ${\cal C}^{(2)}_5$ must be a linear combination of monomials of the type 
$\partial_{\mu}a^{a}_{\nu}\partial_{\rho}a^{b}_{\sigma}a^{c}_{\nu_1}
a^{d}_{\nu_2}a^{e}_{\nu_3}$ 
and/or of the form 
$\partial_{\mu}\partial_{\rho}a^{a}_{\nu}a^{b}_{\sigma}a^{c}_{\nu_1}a^{d}_{\nu_2}a^{e}_{\nu_3}$. We have thus shown that ${\cal C}^{(2)}_5$ actually vanishes. 
Substituting, this result in eq.~(\ref{brseqdiff}), one obtains the 
following equation for ${\cal C}^{(2)}_6$: $s_{0}\,{\cal C}^{(2)}_6=0$. The 
same kind of analysis that yielded a vanishing ${\cal C}^{(2)}_5=0$ leads to
the conclusion that ${\cal C}^{(2)}_6=0$. And so on, and so forth. We have 
thus shown that ${\cal C}^{(2)}_n=0$ for all $n$. Hence, 
${\cal A}^{(2)}={\cal B}^{(2)}$. 
Notice that our strategy would have failed if we had decided not to  compute 
diagrams with four gauge fields  (or derivatives of it) $a_{\mu}$. Indeed,  
$s_{0}\,{\cal C}^{(2)}_4=0$, with ${\cal C}^{(2)}_4={\cal A}^{(2)}_4 -{\cal B}^{(2)}_4$, does not imply ${\cal C}^{(2)}_4=0$, since ${\cal C}^{(2)}_4$ may be a
nonvanishing  linear combination of monomials of the type 
$f^{a_1\,(free)}_{\mu_1\nu_1}f^{a_2\,(free)}_{\mu_2\nu_2}
f^{a_3\,(free)}_{\mu_3\nu_3}f^{a_4\,(free)}_{\mu_4\nu_4}$.

The Feynman integrals that yield the r.h.s of eq.~(\ref{reganomnp}) at  order 
$h^n$ can be worked out by extracting the contribution of this order 
coming from the ``master'' Feynman diagrams in Fig. 2. The dimensionally regularized object that these diagrams represent can be obtained by using the 
Feyman rules in Appendix A. In these rules and in all our expressions the 
exponentials $e^{i\frac{h}{2}k_1 \circ k_2}$, with $k_1 \circ k_2=\theta^{\mu\nu}k_{1\,\mu}\,k_{2\,\mu}$, are actually shorthand for their series 
expansions  $\sum_{n=0}^{\infty}\,\frac{i^n h^n}{2^n n!}\,(k_1 \circ k_2)^n$.
\vskip 0.5 cm
\begin{figure}[h]
\centering
\subfigure[]{\epsfig{file=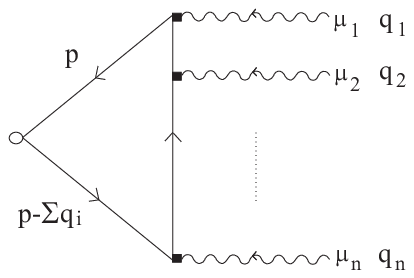}}
\subfigure[]{\epsfig{file=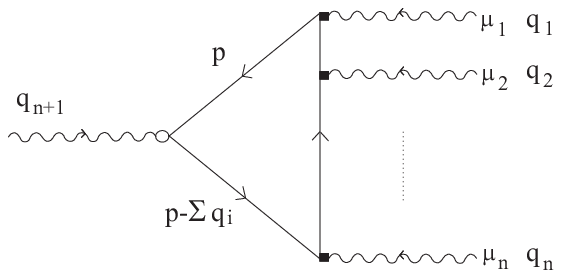}}
\\[5pt]
\renewcommand{\figurename}{Fig.}
\renewcommand{\captionlabeldelim}{.}                                          
\caption{Diagrams that give the r.h.s of eq.~(\ref{reganomnp}).}
\end{figure}
The ``master'' Feynman diagram in Fig. 2a represents the following object:
\begin{equation}
\begin{array}{l}
 {  {\mathfrak F}_n\,=\,N_f\, \frac{2(-1)^{n+1}}{n!}\, 
e^{\frac{i}{2}\,h\,(\sum_{i>j}{q_i \circ q_j})\,}\Tr\,A_{\mu_1}(q_1)A_{\mu_2}(q_2)\dots
         A_{\mu_n}(q_n)}\\
  {\phantom{A_n\,=\, \frac{2(-1)}{n!} }     
\sum_{m=0}^{\infty}
\frac{i^m h^m}{m!}\,\iDp\, (\sum_k p \circ q_k)^m\;
        \tr\frac{\gamma^5 \hat{\pslash} \pslash \gamma^{\mu_1}
(\pslash-\qslash_1)\gamma^{\mu_2}(\pslash-\qslash_1-\qslash_2)\dots
         \gamma^{\mu_n}\left(\pslash- \sum_i \qslash_i \right)}{p^2 (p-q_1)^2
        (p-q_q-q_2)^2\dots\left(p-\sum_{i}
         q_i\right)^2}.}\\
     \end{array}
\label{funnyfn}
\end{equation}
The ``master'' Feynman diagram in Fig. 2b corresponds to the expression that 
follows:
\begin{equation}
\begin{array}{l}
 { {\mathfrak G}_n\,=\,N_f\,      
       \frac{2(-1)^{n+1}}{n!}\,
         e^{\frac{i}{2}\,h\,(\sum_{i>j} q_i \circ q_j )}\;\Tr
A_{\mu}(q_{n+1})A_{\mu_1}(q_1)
         A_{\mu_2}(q_2)\dots A_{\mu_n}(q_n)}\\
{\phantom{ {\mathfrak E}_n\,=\,\frac{2(-1)}{n!}}
\sum_{m=0}^{\infty}\frac{i^m h^m}{m!}\,
\iDp\, (\sum_k p \circ q_k)^m\;
        \tr\frac{\gamma^5 \hat{\gamma^\mu} \pslash \gamma^{\mu_1}
         (\pslash-\qslash_1)\gamma^{\mu_2}(\pslash-\qslash_1-\qslash_2)\dots
         \gamma^{\mu_n}\left(\pslash-\sum_i \qslash_i \right)}{p^2 (p-q_1)^2
       (p-q_q-q_2)^2\dots\left(p-\sum_i 
         q_i \right)^2}.}\\
\end{array}
\label{funnygn}
\end{equation}
Let us see first that the MS dimensional renormalization algorithm 
\cite{Breitenlohner:1977hr, Bonneau:1979jx, Collins:1984xc} sets to zero 
at $D=4$ any contribution coming from  ${\mathfrak G}_n$ in 
eq.~(\ref{funnygn}). 
Using the identities in eqs.~(\ref{gammascom}) and (\ref{traza-antic}), one 
can work out the trace over the the gamma matrices and show that 
\begin{displaymath}
\begin{array}{l}
{\tr\gamma^5 \hat{\gamma^\mu} \pslash \gamma^{\mu_1}
         (\pslash-\qslash_1)\gamma^{\mu2}(\pslash-\qslash_1-\qslash_2)\dots
         \gamma^{\mu_n}\left(\pslash-\sum{ \qslash_i} \right)}\\
   {=\hp^\mu  T_1^{\mu_1\dots\mu_n}+
     \sum_i \hg^{\mu\mu_i} T_{2\,i}^{\mu_1\dots\mu_{i-1}\mu_{i+1}\dots\mu_n}+
         \sum_i \hq_i^\mu T_{3\,i}^{\mu_1\dots\mu_n}.}\\
\end{array}
\end{displaymath}
$T_1$, $T_{2\,i}$ and $T_{3\,i}$ are ``Lorentz covariant tensors'' in the 
$D$-dimensional space-time of dimensional regularization. The expression on the 
r.h.s. of the previous equation shows that any contribution coming from 
${\mathfrak G}_n$ that does not vanish as $D\rightarrow 4$ matches one 
of the following ``tensor'' patterns
\begin{equation}
\begin{array}{l}
{   \frac{1}{D-4}\,
     t_1^{\mu_1\dots\mu_n}\,\Tr\,A_{\mu_1}(q_1)\dots
     \hat{A}_{\mu_k}(q_k)\dots A_{\mu_{n}}(q_{n}),}\\
  {   \frac{1}{D-4}\,
     t_2^{\nu \mu_1\dots\mu_{n}}\,\Tr\, \hat{q}_{i\,\nu} A_{\mu_1}(q_1)\dots
A_{\mu_{i}}(q_{i})\cdots    A_{\mu_{n}}(q_{n}).}
\end{array}
\label{hattedpoles}
\end{equation}
It is important to bear in mind that  $t_1^{\mu_1\dots\mu_n}$ and $t_2^{\nu \mu_1\dots\mu_{n}}$ must be linear combinations of ``Lorentz covariant tensors'' 
with  coefficients that do not depend on $(D-4)$. For instance, a ``tensor'' 
like $t_1^{\mu_1\dots\mu_6}=(D-4)\,\epsilon^{\mu_1\mu_2\mu_3\mu_4}g^{\mu_5\mu_6}$ is not to be admitted, for this type of $t_1$ tensor, when substituted 
back in the first equality  in eq.~(\ref{hattedpoles}),  yields
a contribution that does not go to zero as $D\rightarrow 4$. Now, 
the MS dimensional regularization algorithm removes from ${\mathfrak G}_n$ 
any contribution of the types shown in eq.~(\ref{hattedpoles}). 
Every ${\mathfrak G}_n$ is thus renormalized to zero at $D=4$ in the 
MS renormalization scheme.

The identities in eqs.~(\ref{gammascom}) and (\ref{traza-antic}) can be used 
to remove from ${\mathfrak F}_n$ any  term that upon MS renormalization 
will go away as $D\rightarrow 4$. The trace over the Dirac matrices
of  ${\mathfrak F}_n$ in eq.~(\ref{funnyfn}) is given by 
\begin{equation}
\begin{array}{l} 
{\tr\gamma^5 \hat{\pslash} \pslash \gamma^{\mu_1}
       (\pslash-\qslash_1)\gamma^{\mu2}(\pslash-\qslash_1-\qslash_2)\dots
         \gamma^{\mu_n}\left(\pslash- \sum{ \qslash_i} \right)}\\
{=\hp^2\, R^{\mu_1\dots\mu_n}+\sum_i
         \hp^{\mu_i}\, S_i
        ^{\mu_1\dots\mu_{i-1}\mu_{i+1}\dots\mu_n}+\sum_i \hp\cdot q_i\,
         T^{\mu_1\dots\mu_n}.}\\
\end{array}
\label{workedtrace}
\end{equation}
$R$, $S$ and $T$ are also  ``Lorentz covariant tensors'' in the 
$D$-dimensional space-time of dimensional regularization. Redoing the analysis 
that begun just below eq.~(\ref{funnygn}) for the case at hand 
--{\it mutatis mutandi}--, one shows that the contributions  that go with 
the ``tensors'' $S$ and $T$ in eq.~(\ref{workedtrace}) can be dropped. 
This is so since, after MS renormalization, they will go to zero as  
$D\rightarrow 4$.  
Hence, upon MS renormalization, all nonvanishing contributions at $D=4$ 
coming from ${\mathfrak F}_n$ in eq.~(\ref{funnyfn}) will be furnished by the 
term $\hp^2\, R^{\mu_1\dots\mu_n}$ in eq.~(\ref{workedtrace}). And yet, 
these contributions will also vanish as $D\rightarrow 4$ unless the integration over $p$ yields a pole at $D=4$ when  $\hp^2$ is replaced with 
$p_{\mu}p_{\nu}$. Now, make the latter replacement in the integrals 
of ${\mathfrak F}_n$. Then, some power-counting at $D=4$ tell us that all 
the integrals thus obtained are UV finite if our ${\mathfrak F}_n$ is such that $n>4+m$ --this $m$ indicates that we are dealing with a term of order $h^m$. 
After contraction with $\hg^{\mu\nu}$, these integrals will vanish at $D=4$. 
In summary, to compute, at order $h^m$, the nonzero contribution to the MS 
renormalized r.h.s. of eq.~(\ref{reganomnp}), only the values of the ${\mathfrak F}_n$ objects verifying  
\begin{equation}
 {\mathfrak F}_n\quad\text{such that}\quad n\leq m+4 
\label{nedeedfn}
\end{equation}
are actually needed.  

We shall denote by    
$2\, \sum_f\,<\Psib_{f}\ST{\hat{\gamma}}^{\mu}\gamma_5D_{\mu}\Psi_f>^{(A)}_{MS}$ the 
renormalized object obtained by applying to the r.h.s eq.~(\ref{reganomnp}) 
the minimal subtraction algorithm of dimensional regularization. This object is to be understood as an 
expansion in $h$:
\begin{equation}
 2\, \sum_f\,<\Psib_{f}\ST{\hat{\gamma}}^{\mu}\gamma_5 D_{\mu}\Psi_f>^{(A)}_{MS}\,=\,
{\cal A}^{(0)}\,+\,h\,{\cal A}^{(1)}\,+\,h^2\,{\cal A}^{(2)}\,+\,o(h^3).
\label{severalas}
\end{equation}
${\cal A}^{(0)}$ is given by the well known ordinary $U(1)_{A}$ anomaly:
\begin{equation}
\frac{N_f}{8\pi^2}\idx\, \Tr\,f_{\mi\mii} \tilde{f}_{\miii\miv}.
\label{oranom}
\end{equation}

\subsubsection{The computation of ${\cal A}^{(1)}$}

According to eq.~(\ref{nedeedfn}),   
we shall need the  ${\mathfrak F}_n$'s in eq.~(\ref{funnyfn}) with $n\leq 5$.
We will sort out  the  contributions coming from these ${\mathfrak F}_n$'s  
into two categories. The first type of contributions will be obtained by 
removing from the infinite sum $\sum_{m=0}^{\infty}$ in  ${\mathfrak F}_n$ any 
term with $m>0$. Hence, the first type of contributions will be furnished 
by the terms of order $h$ in $-{\mathfrak A}_n$,   ${\mathfrak A}_n$ being 
given in eq.~(\ref{funnya}). We thus conclude that the   
terms in  ${\cal A}^{(1)}$  that constitute the first  
category can be computed by expanding at first order in $h$ the r.h.s.  
of eq.~(\ref{planeanom}):
\begin{equation}
\begin{array}{l}
{\frac{N_f}{16\pi^2}\,\epsilon^{\mi\mii\miii\miv}\,\Tr\,F_{\mi\mii}
\ST F_{\miii\miv}\mid_{h}=
-\frac{N_f}{4\pi^2}\,\theta^{\rho\sigma}\,\epsilon^{\mu_1\mu_2\mu_3\mu_4}
 \Big\{\Tr\, [\Di a_\rho \partial_\sigma \aii \Diii \aiv 
\!+\!a_\rho \Di 
\partial_\sigma \aii \Diii \aiv }\\
{\phantom{\frac{N_f}{16\pi^2}\,\epsilon^{\mi\mii\miii}}     
+\Di \partial_\sigma \aii a_\rho \Diii \aiv +
        \partial_\sigma \aii \Di a_\rho \Diii \aiv
     -\Di \ar \Dii \as \Diii \aiv +  \Div \ai \Dr \aii \Ds\aiii ]}\\
{\phantom{\frac{N_f}{16\pi^2}\,\epsilon^{\mi\mii\miii}}
-\frac{i}{2}\,\Tr\,[
-\as\ai\Dii\aiii\Div\ar-2\as\ar\Di\aii\Diii\aiv-2\as\aiii\Div\ar\Di\aii}\\
{\phantom{\frac{N_f}{16\pi^2}\,\epsilon^{\mi\mii\miii}+\frac{i}{2}\,\Tr\,[}
-\as\Di\ar\aii\Diii\aiv+\as\Diii\aiv\aii\Di\ar-2\as\Diii\aiv\Di\ar\aii}\\
{\phantom{\frac{N_f}{16\pi^2}\,\epsilon^{\mi\mii\miii}+\frac{i}{2}\,\Tr\,[}  
-\as\Di\ar\Diii\aiv\aii-2\Ds\ai\aii\aiii\Div\ar-2\Ds\aiv\ar\ai\Dii\aiii}\\
{\phantom{\frac{N_f}{16\pi^2}\,\epsilon^{\mi\mii\miii}+\frac{i}{2}\,\Tr\,[}
-2\Ds\aiii\Div\ar\ai\aii-2\Ds\ai\Dii\aiii\aiv\ar+2\Ds\ai\aii\Diii\aiv\ar}\\
{\phantom{\frac{N_f}{16\pi^2}\,\epsilon^{\mi\mii\miii}+\frac{i}{2}\,\Tr\,[}
+2\Ds\aiv\ar\Di\aii\aiii+2\Ds\Di\aii\aiii\aiv\ar+2\Ds\Diii\aiv\ar\ai\aii}\\
{\phantom{\frac{N_f}{16\pi^2}\,\epsilon^{\mi\mii\miii}+\frac{i}{2}\,\Tr\,[} 
+2\Div\ar\Di\as\aii\aiii+\Dii\as\aiii\Div\ai\ar+\aiii\Div\ai\Dii\as\ar}\\
{\phantom{\frac{N_f}{16\pi^2}\,\epsilon^{\mi\mii\miii}+\frac{i}{2}\,\Tr\,[}
+\Diii\as\Div\ai\aii\ar+\Div\ai\aii\Diii\as\ar+2\Ds\ai\Dr\aii\aiii\aiv]}\\
{\phantom{\frac{N_f}{16\pi^2}\,\epsilon^{\mi\mii\miii}}
+\Div\,\Tr\,[
     \ar\as\ai\aii\aiii-\ar\ai\as\aii\aiii]\Big\}.}\\
\end{array}
\label{firstcategory}
\end{equation}
The second type of contributions that make ${\cal A}^{(1)}$ up are obtained 
by setting $h$ to zero everywhere in ${\mathfrak F}_n$, but in the term that
goes with $(\sum_k\,p\circ q_{k})^{m}$, with $m=1$. These substitutions yield 
the following expression: 
\begin{displaymath}
\begin{array}{l}
{{\mathfrak I}_n\,=\,N_f\,\frac{2(-1)^{n+1}}{n!}\,
\Tr\,a_{\mu_1}(q_1)a_{\mu_2}(q_2)\dots a_{\mu_n}(q_n)}\\
 {\phantom{\frac{2(-1)^{n+1}}{n!}\,
a_{\mu_1}(q_1)a_{\mu_2}} \iDp i (\sum_{k}{p \circ q_k})
    \tr\frac{\gamma^5 \hat{\pslash} \pslash \gamma^{\mu_1}
   \nonumber(\pslash-\qslash_1)\gamma^{\mu_2}(\pslash-\qslash_1-\qslash_2)\dots
   \gamma^{\mu_n}\left(\pslash- \sum{ \qslash_i} \right)}{p^2 (p-q_1)^2
   (p-q_q-q_2)^2\dots\left(p-\sum{
   q_i}\right)^2}.}\\
\end{array}
\end{displaymath}
Recall that we saw above that only for $n=2,3,4$ and $5$ may we obtain a 
nonvanishing output. Using the identity in eq.~(\ref{workedtrace}), the 
results in Appendix C and  adding the contributions generated by the 
appropriate permutations of the external momenta, one concludes that
${\mathfrak I}_2$, ${\mathfrak I}_3$, ${\mathfrak I}_4$ and  
${\mathfrak I}_5$ give rise to the following terms in    
${\cal A}^{(1)}$:
\begin{equation}
\begin{array}{l}
{{\mathfrak I}_2\;\rightsquigarrow\; 0,}\\
{{\mathfrak I}_3\;\rightsquigarrow\;+\frac{N_f}{4\pi^2}\,\theta^{\rho\sigma}\epsilon^{\mu_1\mu_2\mu_3\mu_4}
     \Tr\,[\Diii\Ds\aiv\ar\Di\aii+\Ds\Di\aii\Diii\aiv\ar+
\Ds\ar\Di\aii\Diii\aiv],}\\
{{\mathfrak I}_4\;\rightsquigarrow\;-i\frac{N_f}{4\pi^2}\,\theta^{\rho\sigma}\epsilon^{\mu_1\mu_2\mu_3\mu_4}\,\Tr\,
     [\Ds\Di\aii\aiii\aiv\ar+\Ds\Diii\aiv\ar\ai\aii+\Ds\aiii\aiv\ar\Di\aii}\\
{\phantom{{\mathfrak I}_3\;\rightsquigarrow\;-i\frac{N_f}{4\pi^2}\,
\theta^{\rho\sigma}
\epsilon^{\mu_1\mu_2\mu_3\mu_4}\,\Tr\,[}
 +\Ds\ar\ai\aii\Diii\aiv+\Ds\aiv\ar\Di\aii\aiii+\Ds\ai\aii\Diii\aiv\ar}\\
{\phantom{{\mathfrak I}_3\;\rightsquigarrow\;-i\frac{N_f}{4\pi^2}\,
\theta^{\rho\sigma}
\epsilon^{\mu_1\mu_2\mu_3\mu_4}\,\Tr\,[}
    +\Ds\ar\Di\aii\aiii\aiv+\Ds\aii\Diii\aiv\ar\ai],}\\
{{\mathfrak I}_5\;\rightsquigarrow\;-\frac{N_f}{4\pi^2}\,\theta^{\rho\sigma}\epsilon^{\mu_1\mu_2\mu_3\mu_4}\,\Tr\,
\Ds [\ar\ai\aii\aiii\aiv].}\\
\end{array}
\label{secondcategory}
\end{equation}
Before working out the results above, the reader may find it  useful to read 
again the discussion below eq.~(\ref{workedtrace}).
  
Adding the results in eqs.~(\ref{firstcategory}) and (\ref{secondcategory}), 
one obtains that ${\cal A}^{(1)}$ actually vanishes:
\begin{equation}
{\cal A}^{(1)}\,=\,\frac{N_f}{8\pi^2}\,\theta^{\rho\sigma}
\epsilon^{\mu_1\mu_2\mu_3\mu_4}\Tr\,
     [f_{\sigma \mu_1}f_{\mu_2\mu_3}f_{\rho\mu_4}-\frac{1}{4}f_{\mu_1\mu_2}f_{\mu_3\mu_4}
    f_{\sigma\rho}]=0.
\label{zeroaone}
\end{equation}
See eq.~(\ref{D1}).

\subsubsection{The computation of ${\cal A}^{(2)}$}

We saw at the beginning of this subsection --see discussion that begins just 
above eq.~(\ref{brseq})-- that to reconstruct ${\cal A}^{(2)}$ we need gauge invariance and the computation of the values of the Feynman 
diagrams with fewer than   
five $a_{\mu}$. This implies that only the contributions to ${\cal A}^{(2)}$   
coming from ${\mathfrak F}_n$ in eq.~(\ref{funnyfn}) with $2\leq n \leq 4$ will
be worked out by computation of the corresponding dimensionally regularized 
Feynman integrals. This heavy use of the gauge invariance of ${\cal A}^{(2)}$ 
makes the computation feasible: otherwise --see  
eq.~(\ref{nedeedfn})-- one would have to compute the Feynman integrals in 
${\mathfrak F}_5$ and ${\mathfrak F}_6$, which would involve the calculation
of the trace of long strings of gamma matrices. 

The terms in ${\mathfrak F}_n$ in eq.~(\ref{funnyfn}) that will interest us, 
will be distributed in two sets. In the first set, we shall put the 
contributions that have no $(\sum_k\,p\circ q_k)^{m}$, with $m\geq 1$. These 
contributions will be obtained by extracting from $-{\mathfrak A}_n$ every 
term of order $h^2$. ${\mathfrak A}_n$ is in eq.~(\ref{funnya}). We shall 
denote the contributions in the first set by $S^{(1)}_n$, $n$ being the number
of fields $a_{\mu}$ that occur in  it. Since it was the  
${\mathfrak A}_n$'s that  gave the r.h.s. of eq.~(\ref{planeanom}), it is 
clear that
\begin{equation}
\begin{array}{l}
{S^{(1)}_2\,=\,\frac{N_f}{16\pi^2}\epsilon^{\mi\mii\miii\miv}\,\Tr\,F_{\mi\mii}\ST F_{\miii\miv}\big|_{h^2,\,aa},}\\
{S^{(1)}_3\,=\,\frac{N_f}{16\pi^2}\epsilon^{\mi\mii\miii\miv}\,\Tr\,F_{\mi\mii}\ST F_{\miii\miv}\big|_{h^2,\,aaa},}\\
{S^{(1)}_4\,=\,\frac{N_f}{16\pi^2}\epsilon^{\mi\mii\miii\miv}\,\Tr\,F_{\mi\mii}\ST F_{\miii\miv}\big|_{h^2,\,aaaa}.}\\
\end{array}
\label{s1n}
\end{equation}
The subscript  $h^2$ stands for terms of order $h^2$ and the subscripts  
$aa$, $aaa$ and $aaa$ tell us that only contributions with $2$, $3$ and 
$4$ fields $a_{\mu}$ are kept, respectively.

The second set of contributions is made up of the expressions, generically 
denoted by $S^{(2)}_n$ and  $S^{(3)}_n$, given below:
\begin{equation}
\begin{array}{l}
 {S^{(2)}_n \,=\, \frac{(-1)^{n}}{n!}\, 
\Tr\,a_{\mu_1}(q_1)a_{\mu_2}(q_2)\dots
         a_{\mu_n}(q_n)}\\
  {\phantom{S}     
\iDp\,\big\{ (\sum_{i>j} q_i \circ q_j)(\sum_k p \circ q_k)\,+\, (\sum_k p \circ q_k)^2\big\}\;
        \tr\frac{\gamma^5 \hat{\pslash} \pslash \gamma^{\mu_1}
(\pslash-\qslash_1)\gamma^{\mu_2}(\pslash-\qslash_1-\qslash_2)\dots
         \gamma^{\mu_n}\left(\pslash- \sum_i \qslash_i \right)}{p^2 (p-q_1)^2
        (p-q_q-q_2)^2\dots\left(p-\sum_{i}
         q_i\right)^2},}\\
{S^{(3)}_n \,=\,2i \frac{(-1)^{n+1}}{n!}\, 
\Tr\,A_{\mu_1}(q_1)A_{\mu_2}(q_2)\dots
         A_{\mu_n}(q_n)}\mid_{h}\\
  {\phantom{S\big\{ (\sum_{i>j} q_i \circ q_j)\,+\, (\sum_k p \circ q_k)^2\big\}}     
\iDp\,(\sum_k p \circ q_k)\;
        \tr\frac{\gamma^5 \hat{\pslash} \pslash \gamma^{\mu_1}
(\pslash-\qslash_1)\gamma^{\mu_2}(\pslash-\qslash_1-\qslash_2)\dots
         \gamma^{\mu_n}\left(\pslash- \sum_i \qslash_i \right)}{p^2 (p-q_1)^2
        (p-q_q-q_2)^2\dots\left(p-\sum_{i}
         q_i\right)^2},}\\
     \end{array}
\label{s2n}
\end{equation}
Notice that here $n=2,3$ and 4, for $S^{(2)}_n$, and  $n=2$ and $3$, if it 
is $S^{(3)}_n$ that we are talking about.

Let us introduce some more notation. $S_2$,  $S_3$ and  $S_4$ will denote the contributions to ${\cal A}^{(2)}$ carrying $2, 3$ and $4$ fields $a_{\mu}$,  
respectively. Then, 
\begin{equation}
\begin{array}{l}
{S_2\,=\,S^{(1)}_2 + S^{(2),\,MS}_{2},\qquad
S_3\,=\,S^{(1)}_3 + S^{(2),\,MS}_3 + S^{(3),\,MS}_2\big|_{aaa},}\\\
{S_4\,=\,S^{(1)}_4 + S^{(2),\,MS}_4 + S^{(3),\,MS}_2\big|_{aaaa}+
S^{(3),\,MS}_3\big|_{aaaa},}\\
\end{array}
\label{s234}
\end{equation}
where $S^{(1)}_n$, $n=2,3$ and $4$ have been defined in eq.~(\ref{s1n}) and 
$S^{(2),\,MS}_{n}$, $n=2,3$ and $4$, stand for the MS renormalized quantities 
obtained, respectively, from $S^{(2)}_{n}$, $n=2,3$ and $4$ in  
eq.~(\ref{s2n}). After minimal subtraction, $S^{(3)}_2$ yields 
$S^{(3),\,MS}_2\big|_{aaa}$ and $S^{(3),\,MS}_2\big|_{aaaa}$, and 
$S^{(3)}_3$ gives rise to $S^{(3),\,MS}_3\big|_{aaaa}$.

The symbols $S_2^{(inv)}$, $S_3^{(inv)}$ and $S_n^{(inv)}$ will stand for
gauge invariant functions of $a_{\mu}$ that verify the following 
equations
\begin{equation}
S_2^{(inv)}\big|_{aa}=S_2,\;
S_3^{(inv)}\big|_{aaa}=S_3-S_2^{(inv)}\big|_{aaa},\;
S_4^{(inv)}\big|_{aaaa}=S_4-S_2^{(inv)}\big|_{aaaa}-S_3^{(inv)}\big|_{aaaa}.
\label{sinv}
\end{equation}
The subscripts $aa$, $aaa$ and $aaaa$ indicate that a restriction is made to terms with $2$, $3$ and $4$ fields $a_{\mu}$, respectively. 
Besides, we shall assume that the minimum number fields in $S_2^{(inv)}$, 
$S_3^{(inv)}$ and  $S_4^{(inv)}$ is $2$, $3$ and $4$, respectively. 
Furnishing ourselves with these definitions and recalling the discussion 
that begins right above eq.~(\ref{brseq}), one concludes that
\begin{equation}
{\cal A}^{(2)}\,=\,S_2^{(inv)}\,+\,S_3^{(inv)}\,+\,S_4^{(inv)}.
\label{a2=s234}
\end{equation}

We have computed $S_2$,  $S_3$ and $S_4$ by carrying out the lengthy Dirac algebra involved with the help of the identities in Appendix B and using the values of the dimensionally regularized integrals in Appendix C. Many  involved 
algebraic operations that occur in these calculations have been performed with 
the assistance of the algebraic manipulation program {\tt Mathematica}. 
We shall not bother the reader displaying all the intermediate calculations 
since they are not particularly inspiring. $S_2$ defined in eq.~(\ref{s234}) turned out to be given by
\begin{equation}
\begin{array}{l}
{S_2=+\frac{N_f}{96\pi^2}\,\epsilon^{\mu_1\mu_2\mu_3\mu_4}
\theta^{\alpha\beta}\theta^{\rho\sigma}\,\Tr\,\Da\Dr\Di\aiii\Db\Ds\Dii\aiv}\\
{\phantom{S_2=}-\frac{N_f}{24\pi^2}\,\epsilon^{\mu_1\mu_2\mu_3\mu_4}
{\theta_\rho}^{\beta}\theta^{\rho\sigma}g^{\mu\nu}\,\Tr\,\big[\frac{1}{2}\Di\Db\Dm\aiii\Dn\Ds\Dii\aiv
     +\frac{1}{4}\Db\Ds\Di\aiii\Dm\Dn\Dii\aiv}\\
     {\phantom{S_2=}+\frac{1}{4}\Dm\Dn\Di\aiii\Db\Ds\Dii\aiv+\Dm\Dn\Db\Di\aiii\Ds\Dii\aiv
     +\frac{1}{2}\Dm\Db\Ds\Di\aiii\Dn\Dii\aiv}\\
     {\phantom{S_2=}+\frac{1}{2}\Dm\Dn\Db\Ds\Di\aiii\Dii\aiv\big].}\\
\end{array}     
\label{eqS2} 
\end{equation}
Let $[\mu\nu]$ indicate antisymmetrization with respect $\mu$ and $\nu$.
Then, making the following replacements 
\begin{equation}
\begin{array}{l}
{ \partial_{[\mu} a_{\nu]}\rightarrow f_{\mu\nu},\quad 
\partial_{\rho}\partial_{[\mu} a_{\nu]}\rightarrow {\mathfrak D}_{\rho}f_{\mu\nu}}\\
     \partial_\rho\partial_\sigma \partial_{[\mu}a_{\nu]}\rightarrow
     a)\Dcal_\sigma\Dcal_\rho f_{\mu\nu}\quad\text{or}\quad
     b)\Dcal_\rho\Dcal_\sigma f_{\mu\nu}
     \end{array}
 \label{replacements}
\end{equation}
in eq.~(\ref{eqS2}), one obtains a gauge invariant object verifying the first 
equality in eq.~(\ref{sinv}). This object will be our $S_2^{(inv)}$:
\begin{displaymath}
\begin{array}{l}
{S_2^{(inv)}=+\frac{N_f}{384\pi^2}\,
\epsilon^{\mi\mii\miii\miv}\theta^{\alpha\beta}\theta^{\rho\sigma}\,
\Tr\,\Dca\Dcr f_{\mi\miii}\Dcb \Dcs f_{\mii\miv}}\\
     {\phantom{S_2^{(inv)}=}-\frac{N_f}{192\pi^2}\epsilon^{\mu_1\mu_2\mu_3\mu_4}{
\theta_\rho}^\beta \theta^{\rho\sigma}g^{\mu\nu}\Tr\,
\big[\Dcb\Dcm f_{\mi\miii} \Dcs\Dcn f_{\mii\miv}
     +\Dcb\Dcs f_{\mi\miii}\Dcm\Dcn f_{\mii\miv}}\\
    {\phantom{S_2^{(inv)}=-\frac{N_f}{1}}+2\,\Dcb.\Dcm\Dcn f_{\mi\miii}\Dcs f_{\mii\miv}+\Dcb\Dcs\Dcm f_{\mi\miii}\Dcn f_{\mii\miv}+\Dcb\Dcs\Dcm\Dcn f_{\mi\miii} f_{\mii\miv}\big].}\\
\end{array}
\end{displaymath}
All along the computation of the previous result, we have taken advantage of 
the ambiguity that occurs in the replacement in the second line of  
eq.~(\ref{replacements})  and choose in each instance the substitution that  
leads, at the end of the day, to a simpler result. The expression between  
brackets, $\Tr[\cdots]$, on the r.h.s. of the previous equation can be 
expressed as a double total covariant derivative. Hence, 
\begin{equation}
\begin{array}{l}
{S_2^{(inv)}=\frac{N_f}{384\pi^2}\,\epsilon^{\mi\mii\miii\miv}
\theta^{\alpha\beta}\theta^{\rho\sigma}\,\Tr\,\Dca\Dcr 
f_{\mi\miii}\Dcb \Dcs f_{\mii\miv}}\\
 {\phantom{S_2^{(inv)}=} -\frac{1}{384\pi^2}\,\epsilon^{\mu_1\mu_2\mu_3\mu_4}
{\theta_\rho}^\beta \theta^{\rho\sigma}g^{\mu\nu}\,\Tr\,\Dcb\Dcs\big[2\Dcm\Dcn f_{\mi\miii}f_{\mii\miv}+\Dcm f_{\mi\miii}\Dcn f_{\mii\miv}\big].}\\
\end{array}
\label{s2invfinal}
 \end{equation}

To avoid displaying redundant and unnecessarily long expressions we shall 
provide the reader with the value of $S_3-S_2^{(inv)}\big|_{aaa}$ that came 
out of our computations:
\begin{displaymath}
\begin{array}{l}
{S_3-S_2^{(inv)}|_{aaa}=}\\
 {\phantom{S_3-S_2}+\frac{N_f}{8\pi^2}\epsilon^{\mi\mii\miii\miv}
\theta^{\alpha\beta}\theta^{\rho\sigma}\,\Tr\,\Big[+\frac{1}{2}\Da\Dr\aii\Db\Ds\aiii\Di\aiv-\frac{1}{2}\Da\Dr\aii\Db\Diii\as\Di\aiv-}\\
 {\phantom{S_3}-\frac{1}{2}\Da\Dii\ar\Db\Ds\aiii\Di\aiv+
\frac{1}{2}\Da\Dii\ar\Db\Diii\as\Di\aiv
     +\frac{1}{3}\Dr\Di\aiii\Da\Dii\aiv\Db\as}\\
{\phantom{S_3}-\frac{1}{3}\Da\Di\aiii\Dr\Dii\aiv\Db\as+\frac{1}{3}\Da\Di\aiii\Db\Dii\aiv\Dr\as+\frac{1}{3}\Da\Di\aiii\Db\Dr\as\Dii\aiv}\\
{\phantom{S_3-S_2 +\frac{N_f}{8\pi^2}\epsilon^{\mi\mii\miii\miv}
\theta^{\alpha\beta}\theta^{\rho\sigma}\,\Tr\,\big[}
+\frac{1}{3}\Da\Dr\as\Db\Di\Di\aiii\Dii\aiv\Big]}\\
{\phantom{S_3-S_2}+\frac{N_f}{48\pi^2}\epsilon^{\mi\mii\miii\miv}
{\theta_\rho}^{\beta}\theta^{\rho\sigma}\,\Tr\,\Big[+
     \Db\Dm\Ds\aii\partial^{\mu}\aiii\Di\aiv-\Db\Ds\Dii\am\partial^{\mu}
\aiii\Di\aiv}\\
{\phantom{S_3}-\Db\Ds\Dm\aii\Diii a^{\mu}\Di\aiv+
\Db\Ds\Dii\am\Diii a^{\mu}\Di\aiv+
    2\Db\Dm\aii\Ds\partial^{\mu}\aiii\Di\aiv}\\
  {\phantom{S_3}-2\Db\Dii\am\Ds\partial^{\mu}\aiii\Di\aiv-2\Db\Dm\aii\Ds\Diii a^{\mu}\Di\aiv
     +2\Db\Dii\am\Ds\Diii a^{\mu}\Di\aiv}\\
{\phantom{S_3}+\Db\Ds\Dm\aii\Di\aiii\partial^{\mu}\aiv-\Db\Ds\Dii\am\Di\aiii\partial^{\mu}\aiv
-\Db\Ds\Dm\aii\Di\aiii\Div a^{\mu}}\\
{\phantom{S_3} + \Db\Ds\Dii\am\Ds\Di\aiii\Div a^{\mu}
     +\Db\Ds\Di\aii\Dm\aiii\partial^{\mu}\aiv-\Db\Ds\Di\aii\Diii\am\partial^{\mu} \aiv}\\
 {\phantom{S_3}-\Db\Ds\Di\aii\Dm\aiii\Div a^{\mu}+\Db\Ds\Di\aii\Diii\am\Div a^{\mu}
     +2\Db\Dm\aii\Ds\Di\aiii\partial^{\mu}\aiv}\\
 {\phantom{S_3}   -2\Db\Dii\am\Ds\Di\aiii\partial^{\mu}\aiv-2\Db\Dm\aii\Ds\Di\aiii\Div a^{\mu}+2
     \Db\Dii\am\Ds\Di\aiii\Div a^{\mu}}\\
{\phantom{S_3}+ 2\Db\Di\aii\Ds\Dm\aiii\partial^{\mu}\aiv-2\Ds\ai\aii\Ds\Diii\am\partial^{\mu}\aiv
-2\Db\Di\aii\Ds\Dm\aiii\Div a^{\mu}}\\
 {\phantom{S_3-S_2+2\frac{N_f}{48\pi^2}\epsilon^{\mi\mii\miii\miv}{\theta_\rho}^{\beta}\theta^{\rho\sigma}\,\Tr\,\big[}
 +2\Db\Di\aii\Ds\Diii\am\Div a^{\mu}\Big].}\\
 \end{array}
\end{displaymath}
Applying to this result the substitutions in eq.~(\ref{replacements}), one 
obtains 
\begin{displaymath}
\begin{array}{l}
{S_3^{(inv)}=+i\frac{N_f}{32\pi^2}\epsilon^{\mi\mii\miii\miv}\theta^{\alpha\beta}\theta^{\rho\sigma}\,\Tr\,\Big[+\frac{1}{3}\Dcr f_{\mi\miii}\Dca f_{\mii\miv} f_{\b\s}+\frac{1}{6}\Dca f_{\mi\miii}\Dcb f_{\mii\miv} f_{\r\s}}\\
     {\phantom{S_3}+\frac{1}{6}\Dca f_{\mi\miii}\Dcb f_{\r\s} f_{\mii\miv}+
     \frac{1}{6}\Dca f_{\r\s}\Dcb f_{\mi\miii} f_{\mii\miv}+\Dca f_{\r\mii} \Dcb f_{\s\miii} f_{\mi\miv}\Big]}\\
 {\phantom{S_3^{(inv)}=}  +i\frac{N_f}{48\pi^2}\epsilon^{\mi\mii\miii\miv}
{\theta_\rho}^{\beta}\theta^{\rho\sigma}g^{\mu\nu}\,\Tr\,
\Big[\frac{1}{2}\Dcb\Dcs f_{\mu\mii} f_{\nu\miii} f_{\mi\miv}+\Dcb f_{\mu\mii}\Dcs f_{\nu\miii} f_{\mi\miv}}\\
 {\phantom{S_3}  + \frac{1}{2}\Dcb\Dcs f_{\mu\mii} f_{\mi\miii} f_{\nu\miv}+\frac{1}{2}\Dcb\Dcs f_{\mi\mii} f_{\mu\miii} f_{\nu\miv}+\Dcb f_{\mu\mii} \Dcs f_{\mi\miii} f_{\nu\miv}}\\
{\phantom{S_3^{(inv)}=  +\frac{i}{48\pi^2}\epsilon^{\mi\mii\miii\miv}
{\theta_\rho}^{\beta}\theta^{\rho\sigma}g^{\mu\nu}\,\Tr\, 
\big[}   +\Dcb f_{\mi\mii} \Dcs f_{\mu \miii} f_{\nu\miv}\Big].}\\
\end{array}
\end{displaymath}
Using the cyclicity of the trace and the antisymmetric character of some of the
objects in the previous expression, one may express the term that goes with 
${\theta_\rho}^{\beta}\theta^{\rho\sigma}g^{\mu\nu}$ as a double covariant 
derivative. Thus, we have
\begin{equation}
\begin{array}{l}
{S_3^{(inv)}=+i\frac{N_f}{32\pi^2}\epsilon^{\mi\mii\miii\miv}\theta^{\alpha\beta}\theta^{\rho\sigma}\,\Tr\,\Big[
\frac{1}{3}\Dcr f_{\mi\miii}\Dca f_{\mii\miv} f_{\b\s}+\frac{1}{6}\Dca f_{\mi\miii}\Dcb f_{\mii\miv} f_{\r\s}}\\
    {\phantom{S_3^{(inv)}=+\frac{i}{\pi^2}}+\frac{1}{6}\Dca f_{\mi\miii}\Dcb f_{\r\s} f_{\mii\miv}+
     \frac{1}{6}\Dca f_{\r\s}\Dcb f_{\mi\miii} f_{\mii\miv}+\Dca f_{\r\mii} \Dcb f_{\s\miii} f_{\mi\miv}\Big]}\\
{\phantom{S_3^{(inv)}=}+i\frac{N_f}{96\pi^2}\,\epsilon^{\mi\mii\miii\miv}{\theta_\rho}^{\beta}\theta^{\rho\sigma}g^{\mu\nu}\,\Tr\,\Dcb\Dcs\Big[f_{\mu\mi} f_{\nu\mii} f_{\miii\miv}
\Big].}\\
\end{array}
\label{s3invfinal}
\end{equation}
Note that the minimum number of fields in $S_3^{(inv)}$ is $3$, as we had 
assumed when writing eq.~(\ref{a2=s234}).

Using the Feynman integrals in Appendix C, we have computed $S_{4}$ and 
obtained the following result:
\begin{displaymath}
\begin{array}{l}
  {S_4-S_3^{(inv)}|_{aaaa}-S_2^{(inv)}|_{aaaa}=\frac{N_f}{16\pi^2}\,\epsilon^{\mi\mii\miii\miv}
\theta^{\a\b}\theta^{\r\s}\,\Tr\,\Big[
 \partial_{[\r}a_{\mi]}\partial_{[\a}a_{\mii]}\partial_{[\b}a_{\miii]}
\partial_{[\s}a_{\miv]}}\\
{+\partial_{[\r}a_{\mii]}\partial_{[\mi}a_{\miii]}
 \partial_{[\a}a_{\miv]} \partial_{[\b}a_{\s]}
+\frac{1}{2}\partial_{[\mi}a_{\mii]}\partial_{[\r}a_{\miii]}
\partial_{[\b}a_{\miv]}\partial_{[\a}a_{\s]}
+\frac{1}{2}\partial_{[\mi}a_{\mii]}\partial_{[\b}a_{\miii]}
\partial_{[\a}a_{\miv]}\partial_{[\r}a_{\s]}}\\
{+\frac{1}{2}\partial_{[\r}a_{\mii]}\partial_{[\mi}a_{\miii]}
\partial_{[\b}a_{\s]}\partial_{[\a}a_{\miv]}
+\frac{1}{2}\partial_{[\a}a_{\mii]}\partial_{[\mi}a_{\miii]}\partial_{[\r}a_{\s]}\partial_{[\b}a_{\miv]}
-\frac{1}{3}\partial_{[\mii}a_{\miii]}
\partial_{[\mi}a_{\miv]}\partial_{[\r}a_{\b]}\partial_{[\a}a_{\s]}}\\
{-\frac{1}{12}\partial_{[\mi}a_{\miii]}\partial_{[\mii}a_{\miv]}
\partial_{[\a}a_{\b]}\partial_{[\r}a_{\s]}+\frac{1}{12}
\partial_{[\mi}a_{\miii]}\partial_{[\r}a_{\a]}\partial_{[\mii}a_{\miv]}
\partial_{[\b}a_{\s]}-\frac{1}
{24}
\partial_{[\mi}a_{\miii]}\partial_{[\a}a_{\b]}
\partial_{[\mii}a_{\miv]}\partial_{[\r}a_{\s]}\Big].}\\
\end{array}
\end{displaymath}
The substitutions in eq.~(\ref{replacements}) applied to the previous equation
yield an object that verifies by construction 
the last equality in  eq.~(\ref{sinv}) and has four or more gauge fields 
$a_{\mu}$. This object is our  $S_4^{(inv)}$:
\begin{equation}
\begin{array}{l}
  {S_4^{(inv)}=\frac{N_f}{16\pi^2}\,\epsilon^{\mi\mii\miii\miv}
\theta^{\a\b}\theta^{\r\s}\,\Tr\,\Big[
 f_{\r\mi}f_{\a\mii}f_{\b\miii}f_{\s\miv}+
  f_{\r\mii}f_{\mi\miii}
 f_{\a\miv} f_{\b\s}+\frac{1}{2}f_{\mi\mii}f_{\r\miii}f_{\b\miv}f_{\a\s}}\\
{\phantom{S_4}+\frac{1}{2}f_{\mi\mii}f_{\b\miii}f_{\a\miv}f_{\r\s}+\frac{1}{2}f_{\r\mii}f_{\mi\miii}f_{\b\s}
f_{\a\miv}+\frac{1}{2}f_{\a\mii}f_{\mi\miii}f_{\r\s}f_{\b\miv}-\frac{1}{3}f_{\mii\miii}
f_{\mi\miv}f_{\r\b}f_{\a\s}}\\
{\phantom{S_4}-\frac{1}{12}f_{\mi\miii}f_{\mii\miv}
f_{\a\b}f_{\r\s}+\frac{1}{12}f_{\mi\miii}f_{\r\a}f_{\mii\miv}f_{\b\s}-\frac{1}
{24}
f_{\mi\miii}f_{\a\b}f_{\mii\miv}f_{\r\s}\Big].}\\
\end{array}
\label{s4inv}
 \end{equation}
Substituting the r.h.s. of eqs.~(\ref{s2invfinal}),  (\ref{s3invfinal})  and 
(\ref{s4inv}) in eq.~(\ref{a2=s234}), one obtains
\begin{equation}
\begin{array}{l}
{{\cal A}^{(2)}\,=\,
    \frac{N_f}{\pi^2}\,\epsilon^{\mi\mii\miii\miv}\theta^{\alpha\beta}\theta^{\rho\sigma}\,\Tr\,\Big[+
\frac{1}{384}\Dca\Dcr f_{\mi\miii}\Dcb \Dcs f_{\mii\miv}}\\
{\phantom{{\cal A}} +\frac{i}{96}\Dcr f_{\mi\miii}\Dca f_{\mii\miv}
f_{\b\s}+\frac{i}{192}\Dca f_{\mi\miii}\Dcb f_{\mii\miv} f_{\r\s}
+\frac{i}{192}\Dca f_{\mi\miii}\Dcb f_{\r\s}
f_{\mii\miv}}\\
 {\phantom{{\cal A}}+\frac{i}{192}\Dca
f_{\r\s}\Dcb f_{\mi\miii} f_{\mii\miv}+\frac{i}{32}\Dca f_{\r\mii}
\Dcb f_{\s\miii} f_{\mi\miv} 
+\frac{1}{16}f_{\r\mi}f_{\a\mii}f_{\b\miii}f_{\s\miv}}\\
{\phantom{{\cal A}}+\frac{1}{16}f_{\r\mii}f_{\mi\miii}
 f_{\a\miv} f_{\b\s}+\frac{1}{32}f_{\mi\mii}f_{\r\miii}f_{\b\miv}f_{\a\s}
 +\frac{1}{32}f_{\mi\mii}f_{\b\miii}f_{\a\miv}f_{\r\s}
+\frac{1}{32}f_{\r\mii}f_{\mi\miii}f_{\b\s}f_{\a\miv}}\\
{\phantom{{\cal A}}
+\frac{1}{32}f_{\a\mii}f_{\mi\miii}f_{\r\s}f_{\b\miv}-\frac{1}{48}f_{\mii\miii}
f_{\mi\miv}f_{\r\b}f_{\a\s}-\frac{1}{192}f_{\mi\miii}f_{\mii\miv}
f_{\a\b}f_{\r\s}+\frac{1}{192}f_{\mi\miii}f_{\r\a}f_{\mii\miv}f_{\b\s}
}\\
{\phantom{{\cal A}^{(2)}\,=\,
    \frac{1}{\pi^2}\,\epsilon^{\mi\mii\miii\miv}\theta^{\alpha\beta}\theta^{\rho\sigma}\,\Tr\,\Big[+}-\frac{1}{384} f_{\mi\miii}f_{\a\b}f_{\mii\miv}f_{\r\s}
\Big]}\\
 {-\frac{N_f}{\pi^2}\,\epsilon^{\mi\mii\miii\miv}{\theta_\rho}^\beta\theta^{\rho\sigma}\Tr\,
\Dcb\Dcs\Big[\frac{1}{384}\big(2\Dcm\Dcal^{\mu} f_{\mi\miii}f_{\mii\miv}+
\Dcm f_{\mi\miii}\Dcal^{\mu} f_{\mii\miv}\big)-\frac{i}{96}f_{\mu\mi} 
f^{\mu}_{\phantom{\mu}\mii} f_{\miii\miv}\Big].}\\
\end{array}
\label{A2}
\end{equation}
In Appendix D, we shall show that the previous result can be simplified 
to 
\begin{equation}
\begin{array}{l}
{{\cal A}^{(2)}\,=\,
    \frac{N_f}{96\pi^2}\,\epsilon^{\mi\mii\miii\miv}\theta^{\alpha\beta}
\theta^{\rho\sigma}\,\Tr\,\Big[+
\frac{1}{4}\Dca(\Dcr f_{\mi\miii}\Dcb \Dcs f_{\mii\miv})+
i\,\Dcr ( f_{\mi\miii}\Dca f_{\mii\miv}
f_{\b\s})}\\
{\phantom{{\cal A}^{(2)}\,=\,
    \frac{1}{\pi^2}\,\epsilon^{\mi\mii\miii\miv}\theta^{\alpha\beta}
\theta^{\rho\sigma}\,\Tr\,\Big[+}+\frac{i}{4}\Dcb(\Dca f_{\r\s}f_{\mi\mii}f_{\miii\miv})+i\,
\Dcb(\Dca f_{\mi\mii}f_{\r\s}f_{\miii\miv})\Big]}\\
{-\frac{N_f}{96\pi^2}\,\epsilon^{\mi\mii\miii\miv}{\theta_\rho}^\beta\theta^{\rho\sigma}\,\Tr\,
\Dcb\Dcs\Big[\frac{1}{4}\left(2\Dcm\Dcal^{\mu} f_{\mi\miii}f_{\mii\miv}+
\Dcm f_{\mi\miii}\Dcal^{\mu} f_{\mii\miv}\right)-i 
f_{\mu\mi} f^{\mu}_{\phantom{\mu}\mii} f_{\miii\miv}\Big].}\\
\end{array}
\label{a2final}
\end{equation}
Hence, 
\begin{equation}
{\cal A}^{(2)}\,=\,\partial_{\lambda}\,{\cal X}^{\lambda},
\label{a2divergence}
\end{equation}
where 
\begin{equation}
\begin{array}{l}
{{\cal X}^{\lambda}\,=\,
    \frac{1}{96\pi^2}\,\epsilon^{\mi\mii\miii\miv}\theta^{\lambda\alpha}
\theta^{\rho\sigma}\,\Tr\,\Big[+
\frac{1}{4} \Dcr f_{\mi\miii}\Dca \Dcs f_{\mii\miv}+
i\,f_{\mi\miii}\Dcr f_{\mii\miv}
f_{\s\alpha} }\\
{\phantom{{\cal A}^{(2)}\,=\,
    \frac{1}{\pi^2}\,\epsilon^{\mi\mii\miii\miv}\theta^{\alpha\beta}
\theta^{\rho\sigma}\,\Tr\,\Big[+}-\frac{i}{4}
\Dca f_{\r\s}f_{\mi\mii}f_{\miii\miv}-i\,
\Dca f_{\mi\mii}f_{\r\s}f_{\miii\miv})\Big]}\\
{+\frac{1}{96\pi^2}\,\epsilon^{\mi\mii\miii\miv}{\theta_\rho}^\lambda\theta^{\rho\sigma}g^{\mu\nu}\,\Tr\,
\Dcs\Big[-\frac{1}{4}\left(2\Dcm\Dcn f_{\mi\miii}f_{\mii\miv}+
\Dcm f_{\mi\miii}\Dcn f_{\mii\miv}\right)+i f_{\mu\mi} f_{\nu\mii} f_{\miii\miv}\Big].}\\
\end{array}
\label{xlambda}
\end{equation}
Let us remark that ${\cal X}^{\lambda}$ is a gauge invariant quantity. 

Taking into account eqs.~(\ref{reganomnp}), (\ref{severalas}), (\ref{oranom}),
(\ref{zeroaone}), (\ref{a2divergence}), one concludes that 
\begin{equation}
\partial_{\mu}<j_{5}^{(np)\;\mu}>^{(A)}_{MS}(x)\,=\,
\frac{N_f}{8\pi^2}\, \Tr\,f_{\mi\mii}(x) \tilde{f}_{\miii\miv}(x)\,+\,
h^2\,\partial_{\lambda}\,{\cal X}^{\lambda}(x).
\label{finalresultjnp}
\end{equation}
The subscript $MS$ signals the fact that the  previous equation has been computed by applying the minimal subtraction algorithm of dimensional regularization \cite{Breitenlohner:1977hr, Bonneau:1979jx, Collins:1984xc} to both 
sides of eq.~(\ref{reganomnp}). 
${\cal X}^{\lambda}(x)$ is given in eq.~(\ref{xlambda}).  
Eq.~(\ref{finalresultjnp}) shows that the classical conservation equation in  
eq.~(\ref{conservation}) no longer holds at the quantum level and should be
replaced with 
\begin{equation}
\partial_{\mu}N[j_{5}^{(np)\;\mu}]_{MS}(x)\,=\, 
\frac{N_f}{8\pi^2}\, \Tr\,f_{\mi\mii}(x) \tilde{f}_{\miii\miv}(x)\,+\,h^2\,
\partial_{\lambda}\,{\cal X}^{\lambda}(x).
\label{opanomeq}
\end{equation}
Where $N[j_{5}^{(np)\;\mu}]_{MS}(x)$ is the normal product operator 
 --see~\cite{Bonneau:1979jx,Collins:1984xc}-- obtained from the regularized  
current $j_{5}^{(np)\;\mu}(x)$ by MS renormalization. 
However, the term $\partial_{\lambda}\,{\cal X}^{\lambda}(x)$ is not an 
anomalous contribution since, as a consequence of the gauge invariance of 
${\cal X}^{\lambda}(x)$, we may introduce a new renormalized gauge invariant 
current
\begin{equation}
{\cal J}_{5}^{(np)\;\mu}(x)\,=\,N[j_{5}^{(np)\;\mu}]_{MS}(x)\,-\,h^2\,
{\cal X}^{\lambda}(x),
\label{newcalnp}
\end{equation}
which verifies the standard $U(1)_A$ anomaly equation. Note that, for 
$\theta^{0i}=0$, $N[j_{5}^{(np)\;\mu}]_{MS}(x)$ and 
${\cal J}_{5}^{(np)\;\mu}(x)$ lead to the same renormalized charge 
$Q_{5}^{(np)}$, at least up to order $h^2$. Indeed, if time commutes 
${\cal X}^{0}(x)=0$. By employing the temporal gauge $a_{0}(x,t)=0$, 
integrating both sides of eq.~(\ref{opanomeq}) over all
values of $x$ and taking into account the boundary conditions in 
eq.~(\ref{boundarycond}), one gets
\begin{equation}
Q_{5}^{(np)}(t=+\infty)-Q_{5}^{(np)}(t=-\infty)\,=\,
\frac{N_f}{8\pi^2}\idx\, \Tr\,f_{\mi\mii}(x) \tilde{f}_{\miii\miv}(x)
\,= \,2\,N_f\,(n_{+}-n_{-}).
\label{qnpanom}
\end{equation}
$n_{\pm}$ are defined in eq.~(\ref{npm}). To obtain the l.h.s of the previous
equation, we have assumed that the fields go to zero fast enough as 
$|\vec{x}|\rightarrow \infty$ so as to make sure that the are no surface  
contributions at spatial infinity. Note that 
\begin{displaymath}
\int d^{3}{\vec x}\quad \partial_{i}{\cal X}^{i}(x)\,=\,0,
\end{displaymath}
even for gauge fields that  vanish as $1/|\vec{x}|$  when 
$|\vec{x}|\rightarrow \infty$.

Eq.~(\ref{qnpanom}) looks suspiciously similar to eq.~(\ref{planeppm}). They 
are actually the same equation. Indeed, in the MS scheme, as we shall show 
below, the quantum charges $Q_{5}^{(p)}(t)$ and $Q_{5}^{(np)}(t)$  
are equal if $\theta^{0i}=0$.  
To show that $Q_{5}^{(p)}(t)=Q_{5}^{(np)}(t)$, we shall need some properties 
of the MS normal product operation 
--see ref.~\cite{Bonneau:1979jx, Collins:1984xc}-- that we recall next.
Let $N[\phantom{a}]_{MS}$ denote the MS normal product operation acting on 
monomials of the fields and their derivatives, then 
\begin{equation}
\begin{array}{l}
{N[\,c_1\, O_1+c_2\,O_2]_{MS}=c_1\,N[O_1]_{MS}\,+\,c_2\,N[O_2]_{MS},}\\  
\;\, N[\partial_{\mu}\,O^{\mu}]_{MS}=\partial_{\mu}\,N[O^{\mu}]_{MS},\;\,
{N[\theta^{\mu\nu}\,O_{\nu\rho}]_{MS}\,=\,
\theta^{\mu\nu}\,N[O_{\nu\rho}]_{MS},}\\
\end{array}
\label{properties}
\end{equation}
where $c_1$, and $c_2$ are numbers which do not depend on $D$ and 
$O_1$, $O_2$ $O^{\mu}$ and $O_{\nu\rho}$ are monomials of the fields and their 
derivatives. It is clear that in dimensional regularization 
\begin{displaymath} 
j_{5}^{(np)\;\mu}\,=\,j_{5}^{(p)\;\mu}\,+\,
\sum_{f}\,\sum_{i}[\Psi_{f\, s i},\bar{\Psi}_{f\,t i}]_{\ST}
(\gamma^{\mu}\gamma_5)_{st},
\end{displaymath}
and that  
\begin{displaymath}
\sum_{f}\sum_{i}\;[\Psi_{f\,s i} ,\bar{\Psi}_{f\,t i}]_{\ST}\,=\,
\partial_{\mu}\,\Big[\frac{i}{2}\;\theta^{\mu\beta}\,\sum_{i}
\,\int_{0}^{h}\,dt\;\big(\{\Psi_{f\, si},\partial_{\beta}\bar{\Psi}_{f\,t i}\}_{\ST_t}\big)_{ii}\Big].
\end{displaymath}
Now, upon using the Seiberg-Witten map, the r.h.s of this equation is an 
infinite sum of monomials of the ordinary 
fields and their derivatives with coefficients not depending on $D$. Then, 
taking into account eq.~(\ref{properties}) and the equations below it, 
one concludes that  
\begin{equation}
N[j_{5}^{(np)\;\mu}]_{MS}\,=\,N[j_{5}^{(p)\;\mu}]_{MS}\,+\,
\partial_{i}\,\theta^{i\rho}\sum_{m}\,
N[{\cal O}_{\rho\,s t}^{(m)}(\gamma^{\mu}\gamma_5)_{st}]_{MS}.
\label{opop}
\end{equation}
${\cal O}_{\rho\,s t}^{(m)}$ are the monomials of 
the ordinary fields and their   
derivatives we have just mentioned and $m$ collects all the indices needed to 
label them. In the previous equation we have already used the equality 
$\theta^{0i}=0$. Setting $\mu=0$ and integrating over all values of 
$\vec{x}$, leads to 
\begin{displaymath}
Q_{5}^{(np)}(t)=\int d^3\vec{x}N[j_{5}^{(np)\;0}]_{MS}(\vec{x},t)=
\int d^3\vec{x}N[j_{5}^{(p)\;0}]_{MS}(\vec{x},t)=Q_{5}^{(p)}(t).
\end{displaymath}
Note that the integral of the second term on the r.h.s of eq.~(\ref{opop}) 
vanishes for fields that decrease sufficiently rapidly as 
$|\vec{x}|\rightarrow\infty$.

\subsection{Anomalous Ward identity for $j_{5}^{(cn)\,\mu}$}

In this subsection we shall compute 
$\partial_{\mu}\,<j_{5}^{(cn)\,\mu}>^{(A)}$  
in the MS scheme of dimensional regularization at second order in $h^2$.
To carry out this calculation we shall employ the results obtained for
$j_{5}^{(np)\,\mu}$ in the previous subsection. To do so, let us find first  
the relation between the two currents at hand in the dimensionally regularized 
theory. For the time being, $j_{5}^{(cn)\,\mu}$ will 
denote  the natural dimensionally regularized current 
obtained from its 4-dimensional counterpart in eq.~(\ref{swcurrent}). This 
$j_{5}^{(cn)\,\mu}$ is given by an expression which is exactly the expression 
displayed in eq~(\ref{swcurrent}) provided the objects that make it up live 
in the ``D-dimensional space-time'' of dimensional regularization. The object 
$\theta^{\mu\nu}$ in dimensional regularization  was defined 
in eq.~(\ref{dimtheta}) as an intrinsically ``four dimensional'' object. We
 shall use the same symbol for the current  
$j_{5}^{(np)\,\mu}$ and for its dimensionally regularized counterpart, 
the context will tell us clearly what the symbol stands for. The difference  
between the dimensionally regularized currents  $j_{5}^{(cn)\,\mu}$ and $j_{5}^{(np)\,\mu}$ is given by the following equations:
\begin{equation}
j_{5}^{(cn)\,\mu}=j_{5}^{(np)\,\mu}\,+\,{\cal Y}^{\mu},
\label{ddimdiff}
\end{equation}
where
\begin{equation} 
\begin{array}{l}
{{\cal Y}^{\mu}=\sum_{f}\Big[-i\frac{h}{2}\Da(\theta^{\alpha\beta}\bar{\psi_{f}}\g^\mu\g^5 
D_\beta\psi_{f})+i\frac{h}{2}\Da(\theta^{\mu\beta}\bar{\psi_{f}}\g^\alpha
\g^5 D_\beta\psi_{f})}\\
{\phantom{{\cal Y}^{\mu}=\sum_{f}\Big[}+i\frac{h}{2}\theta^{\alpha\mu}(\overline{D_\nu\psi_{f}}\g^\nu\g^5 D_\alpha\psi_{f}+\bar{\psi_{f}}\g^\nu\g^5 
D_\alpha D_\nu\psi_{f})}\\
    {\phantom{{\cal Y}^{\mu}=\sum_{f}\Big[}+h^2\theta^{\alpha\beta}\theta^{\rho\sigma}\big[+\frac{1}{8}
\Da\Dr(\bar{\psi_{f}}\g^\mu\g^5D_\beta D_\sigma\psi_{f})-\frac{1}{8}
\Dr(\bar{\psi_{f}}\g^\mu\g^5D_\alpha\{D_\beta,D_\sigma\}\psi_{f})}\\
{\phantom{{\cal Y}^{\mu}=\sum_{f}\Big[+h^2\theta^{\alpha\beta}
\theta^{\rho\sigma}\big[}
-\frac{1}{4}\Da(\bar{\psi_{f}}\g^{\mu}\g^5D_\beta D_\rho D_\sigma \psi_{f})
 +\frac{1}{4}\Da(\bar{\psi_{f}}\g^\mu\g^5D_\rho D_\beta D_\sigma \psi_{f})\big]
}\\
{\phantom{{\cal Y}^{\mu}=\sum_{f}\Big[}
+ h^2  \theta^{\alpha\beta}\theta^{\mu\rho}\big[\frac{i}{8}
\Dn(\bar{\psi_{f}}\g^\nu\g^5\Dcal_\alpha f_{\beta\rho}\psi_{f})+
\frac{i}{8}\Db(\bar{\psi_{f}}\g^\nu\g^5\Dcal_{\alpha}f_{\rho\nu}\psi_{f})\big]}\\
{ \phantom{{\cal Y}^{\mu}=\sum_{f}\Big[}-h^2\theta^{\alpha\beta}
\theta^{\mu\rho}
\bar{\psi_{f}}\g^\nu\g^5\big[\frac{i}{8}\Dcal_\alpha f_{\beta\rho}D_\nu+
\frac{i}{8}\Dcal_\rho f_{\alpha\nu}D_\beta+\frac{i}{8}
\Dcal_\alpha f_{\rho\nu}D_\beta\big]\psi_{f}}\\
{\phantom{{\cal Y}^{\mu}=\sum_{f}\Big[}+h^2\theta^{\alpha\beta}
\theta^{\mu\rho}\bar{\psi_{f}}\g^\nu\g^5\big[
+\frac{1}{8}f_{\rho\alpha}f_{\beta\nu}+\frac{1}{4}
f_{\alpha\nu}f_{\beta\rho}
    +\frac{1}{8}f_{\alpha\beta}f_{\rho\nu}\big]\psi_{f}\Big]+o(h^3).}\\
\end{array}
\label{dcaligy}
\end{equation}
In the previous equation all objects live in the ``D-dimensional'' space-time 
of dimensional regularization. It was shown long ago~\cite{Breitenlohner:1977hr} that the equations of  motion holds in the dimensionally regularized theory. 
Using the equations of motion and eqs.~(\ref{dcaligy}), one gets that 
\begin{displaymath}
\partial_{\mu}{\cal Y}^{\mu}\,=\,\partial_{\sigma}\hat{{\cal X}}^{\sigma},
\end{displaymath}
where
\begin{equation}
\begin{array}{l}
{\hat{{\cal X}}^{\sigma}=\sum_{f}\Big[+
ih\,\theta^{\rho\sigma}(\bar{\psi_{f}}
\hat{\gamma}^\nu\g_5 D_\rho D_\nu \psi_{f})}\\
{\phantom{\hat{{\cal X}}}+h^2\,\theta^{\rho\sigma}\theta^{\alpha\beta}\big[-\frac{i}{2}
\bar{\psi_{f}}\hat{\gamma}^\nu \g_5 D_\rho(f_{\nu\alpha}D_\beta\psi_{f})+
\frac{1}{4}\Db(\bar{\psi_{f}}\hat{\g}^\nu\g_5 D_\alpha D_\rho D_\nu \psi_{f})}\\
{\phantom{\hat{{\cal X}}=+\,\,\theta^{\rho\sigma}\theta^{\alpha\beta}\Ds\big[}+\frac{i}{4}\bar{\psi_{f}}\hat{\g}^\nu\g_5 
 f_{\alpha\rho}D_\beta D_\nu\psi_{f}+\frac{i}{4}\bar{\psi_{f}}\hat{\g}^\nu\g_5 
f_{\beta\alpha}D_\rho D_\nu\psi_{f}+\frac{i}{8}\bar{\psi_{f}}\hat{\g}^\nu\g_5 \
\Dcal_\rho f_{\beta\alpha} D_\nu\psi_{f}\big]\Big].}\\
\end{array}
\label{hatx}
\end{equation}
Note that at variance with the result for the classical theory, the dimensionally regularized difference $\partial_{\mu}j_{5}^{(cn)\,\mu}-
\partial_{\mu}j_{5}^{(np)\,\mu}=\partial_{\mu}{\cal Y}^{\mu}$ does not vanish
upon imposing the equation of motion. The operator 
$\partial_{\sigma}\,\hat{{\cal X}}^{\sigma}$ is an evanescent operator --it 
vanishes as $D\rightarrow 4$-- so it may yield --and, indeed, it will-- a 
$D\rightarrow 4$ finite contribution when inserted into an UV 
divergent fermion loop. In summary, quantum corrections will make 
$\partial_{\mu}<j_{5}^{(cn)\,\mu}>^{(A)}_{MS}$ different from the renormalized 
$\partial_{\mu}<j_{5}^{(np)\,\mu}>^{(A)}_{MS}$. Let us work out this 
difference .

Since $\hat{{\cal X}}^{\sigma}$ is an invariant quantity under $SU(N)$
gauge transformations, it so happens that the MS renormalized 
$\partial_{\sigma}<\hat{{\cal X}}^{\sigma}>^{(A)}_{MS}$ is equal to 
a $\partial_{\sigma}{\cal A}^{(cn)\,\sigma}$, with ${\cal A}^{(cn)\,\sigma}$ 
being a gauge invariant function of $a_{\mu}$ and its derivatives. 
${\cal A}^{(cn)\,\mu}$ has mass dimension equal to 3.  To compute 
$\partial_{\sigma}{\cal A}^{(cn)\,\sigma}$, we shall follow the strategy used 
in the computation of ${\cal A}^{(2)}$. We shall thus use gauge invariance and
the result obtained by explicit computation of appropriate Feynman diagrams 
to reconstruct $\partial_{\sigma}{\cal A}^{(cn)\,\sigma}$. If we adjust to 
the case at hand  the analysis that begins just above eq.~(\ref{BRS}), we 
will conclude that the Feynman diagrams that must be unavoidably computed
have 2 gauge fields, in the case of the contribution of order $h$, and 
2 and 3 gauge fields in the case of the contribution of order $h^2$. The  
terms with $4,5,$ {\it etc}... gauge fields are obtained by using locality, 
gauge invariance, the replacements in eq.~(\ref{replacements}) and the results 
concerning the cohomology of $s_0$ quoted right below eq.~(\ref{brseqdiff}). 

Let us introduce some more notation and denote by  ${\cal A}^{(cn)\,(1)}$ and 
${\cal A}^{(cn)\,(2)}$ the $o(h)$ and $o(h^2)$ contributions to 
$\partial_{\sigma}{\cal A}^{(cn)\,\sigma}$. Then,
\begin{equation}
\partial_{\sigma}<\hat{{\cal X}}^{\sigma}>^{(A)}_{MS}\,=\,
\partial_{\sigma}{\cal A}^{(cn)\,\sigma}\,=\,h\, {\cal A}^{(cn)\,(1)}\,+\,
h^2\,{\cal A}^{(cn)\,(2)}.
\label{asw1asw2}
\end{equation}
The diagram with two gauge fields that give  the two-field terms in 
${\cal A}^{(cn)\,(1)}$ is depicted in  Fig. 3. 
\vskip 0.5 cm
\begin{figure}[h]
\centering
\epsfig{file=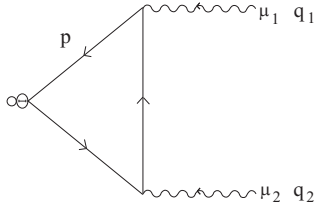}\\[15pt]
 \renewcommand{\figurename}{Fig.}
 \renewcommand{\captionlabeldelim}{.}
\caption{Diagram that contributes ${\cal A}^{(cn)\,(1)}$ in eq.~(\ref{asw1asw2}).}
\label{diagramasjsw1}
\end{figure}
With the help of the Feynman rules in Appendix A and the Feynman 
integrals in Appendix C, one shows that this two-field contribution vanishes
in the limit $D\rightarrow 4$ in the MS scheme. Hence, gauge invariance leads
to the conclusion that in this renormalization scheme:   
\begin{equation} 
{\cal A}^{(cn)\,(1)}=0.
\label{asw1}
\end{equation}

Let $\partial_{\b}{\cal T}_2^{\b}$ and $\partial_{\b}{\cal T}_3^{\b}$ 
be  the contributions to ${\cal A}^{(cn)\,(2)}$ carrying 2  and 3 
gauge fields, respectively. Let us introduce local gauge invariant functions 
${\cal T}_2^{(inv)\,\b}$ and ${\cal T}_3^{(inv)\,\b}$ such that
$\partial_{\b}{\cal T}_2^{(inv)\,\b}$ and 
$\partial_{\b}{\cal T}_3^{(inv)\,\b}$ verify
\begin{equation}
\partial_{\b}{\cal T}_2^{(inv)\,\b}|_{aa}=\partial_{\b}{\cal T}_2^{\b},\quad
\partial_{\b}{\cal T}_3^{(inv)\,\b}|_{aaa}=\partial_{\b}{\cal T}_3^{\b}-
\partial_{\b}{\cal T}_2^{(inv)\,\b}|_{aaa}.
\label{tinvdef}
\end{equation}
Let us further assume that the minimum number of fields in  
${\cal T}_3^{(inv)\,\b}$ is 3. Then, one can show that 
\begin{equation}
{\cal A}^{(cn)\,(2)}\,=\,\partial_{\b}{\cal T}_2^{(inv)\,\b}\,+\,
\partial_{\b}{\cal T}_3^{(inv)\,\b}.
\label{asw2}
\end{equation}
The Feynman diagrams that give $\partial_{\b}{\cal T}_2^{\b}$ are the diagrams 
with two wavy lines depicted in Fig. 4. Some Dirac algebra and the integrals 
in Appendix C leads to 
\begin{displaymath}
\begin{array}{l}
{\partial_{\b}{\cal T}_2^{\b}=\partial_{\beta}\Big\{+
    \frac{N_f}{96\pi^2}\,\theta^{\alpha\beta}
\theta^{\rho\sigma}\epsilon^{\mi\mii\miii\miv}\,
   g_{\alpha\rho} g^{\mu\nu}\Tr\,\big[\Dm\Dn\Ds\Di\aiii\Dii\aiv
+\Dm\Ds\Di\aiii\Dn\Dii\aiv}\\
{\phantom{\partial_{\mu}{\cal T}_2^{\mu}=\partial_{\beta}\Big\{\Tr\Big[}
+\Dm\Dn\Di\aiii\Ds\Dii\aiv+\Ds\big(\frac{1}{2}
\Dm\Dn\Di\aiii\Dii\aiv+\frac{1}{4}\Dm
\Di\aiii\Dn\Dii\aiv\big)\big]}\\
 {\phantom{\partial_{\mu}{\cal T}_2^{\mu}=\partial_{\beta}\Big\{}
+\frac{N_f}{96\pi^2}\,\theta^{\alpha\beta}\epsilon^{\mi\mii\miii\miv}\Tr\,
\big[-\Da\Ds\Di\aiii\Dr\Dii\aiv-\Ds\big(\frac{3}{2}\Da\Di\aiii\Dr\Dii\aiv}\\
{\phantom{\partial_{\mu}{\cal T}_2^{\mu}=\partial_{\beta}\Big\{
+\frac{N_f}{96\pi^2}\,\theta^{\alpha\beta}\epsilon^{\mi\mii\miii\miv}\Tr\,
\big[}
+2\Dr\Da\Di\aiii\Dii\aiv\big)\big]\Big\}.}\\
\end{array}
\end{displaymath}
The replacements in eq.~(\ref{replacements}) turn the previous equation into
the following identity:
\begin{equation}
\begin{array}{l}
    {\partial_{\b}{\cal T}_2^{(inv)\,\b}=\Db\Big\{\frac{N_f}{384\pi^2}\,
\theta^{\alpha\beta}\theta^{\rho\sigma}\epsilon^{\mi\mii\miii\miv}\,
    \Tr\Big[\Dcal_\alpha \Dcal_\rho f_{\mi\miii}\Dcal_\sigma f_{\mii\miv}-
\Ds\big[\frac{3}{2}\Dcal_\alpha f_{\mi\miii}\Dcal_\rho f_{\mii\miv}}\\
   {\phantom{\partial_{\b}{\cal T}_2^{(inv)\,\b}=}
 +2\Dcal_\rho\Dcal_\alpha f_{\mi\miii}f_{\mii\miv}\big]\Big]
    +\frac{N_f}{512\pi^2}\,\theta^{\alpha\beta}\theta^{\rho\sigma}\epsilon^{\mi\mii\miii\miv}g_{\alpha\rho}g^{\mu\nu}\Ds\Tr\big[2\,\Dcal_\mu\Dcal_\nu f_{\mi\miii}f_{\mii\miv}}\\
    {\phantom{\partial_{\b}{\cal T}_2^{(inv)\,\b}=}
+\Dcal_\mu f_{\mi\miii}\Dcal_\nu f_{\mii\miv}\big]\Big\}.}\\
\end{array}
 \label{t2inv}
\end{equation}
\begin{figure}[h]
\centering
\epsfig{file=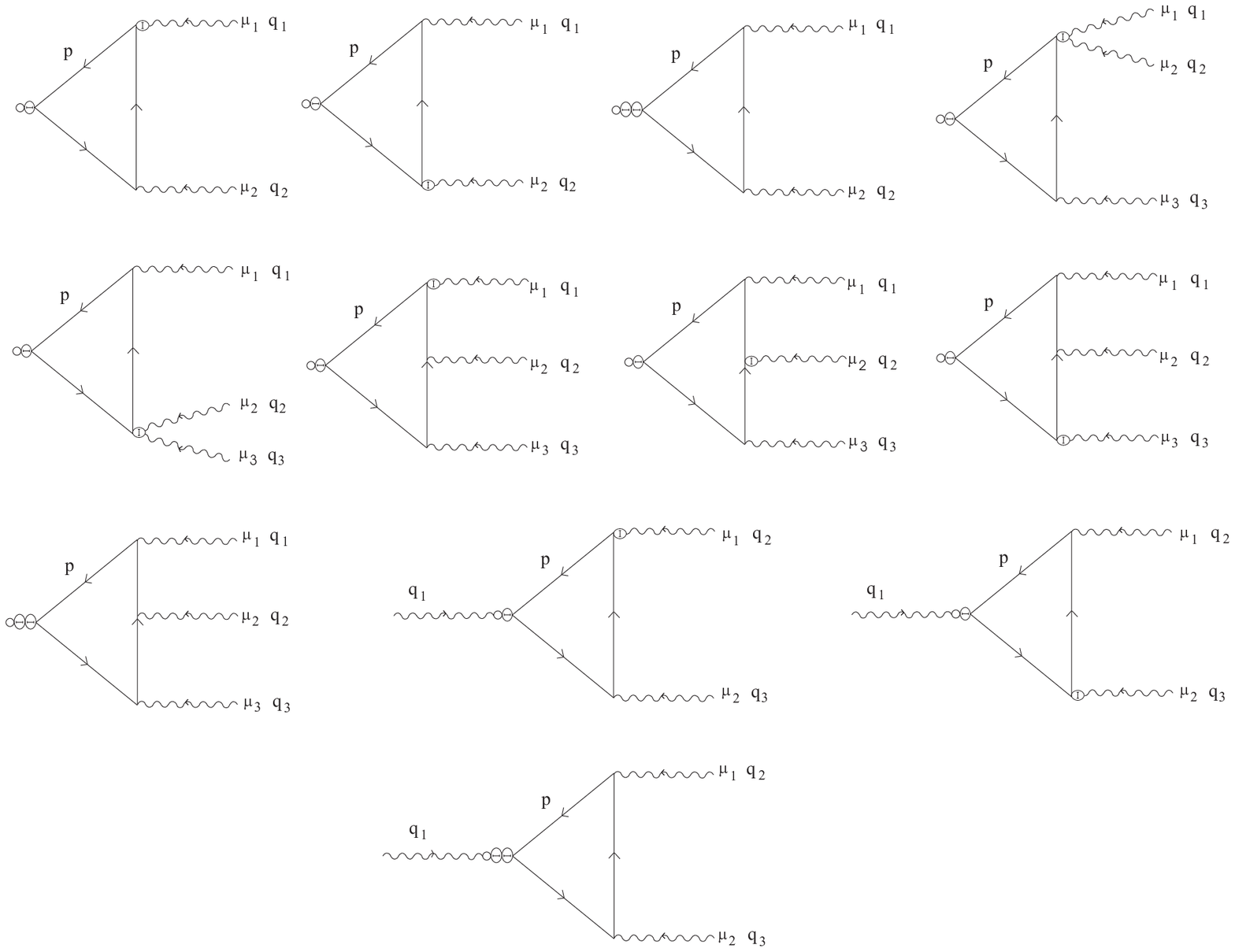,height=12cm}
 \renewcommand{\figurename}{Fig.}
 \renewcommand{\captionlabeldelim}{.}
\caption{Diagrams that yield ${\cal A}^{(cn)\,(2)}$ in eq.~(\ref{asw1asw2}).}
\label{diagramasjsw2}
\end{figure}
The computation of the Feynman diagrams in Fig.~4 with three gauge fields 
$a_{\mu}$ gives $\partial_{\b}{\cal T}_3^{\b}$. By performing that
computation, we have obtained the following result:
\begin{displaymath}
\begin{array}{l}
{\partial_{\b}{\cal T}_3^{\b}-\partial_{\b}{\cal T}_2^{(inv)\,\b}|_{aaa}=
\Db\Big\{i\frac{N_f}{96\pi^2}\theta^{\alpha\beta}\theta^{\rho\sigma}
    \epsilon^{\mi\mii\miii\miv}\Tr\big[6\,\Da\partial_{[\,\rho}a_{\mii]}\partial_{[\,\sigma}a_{\miii]}\Di\aiv}\\
{\phantom{\partial_{\b}{\cal T}_3^{\b}}
-6\,\Da\partial_{[\,\rho}a_{\mii]}\Di\aiii\partial_{[\,\sigma}a_{\miv]}
+6\,\Da\Di\aii\partial_{[\,\rho}a_{\miii]}\partial_{[\,\sigma}a_{\miv]}+
   7\,\Da\Di\aiii\Dii\aiv\Dr\as}\\
{\phantom{\partial_{\b}{\cal T}_3^{\b}}
+3\,\Dr\Di\aiii\partial_{[\,\sigma}a_{\alpha]}\Dii\aiv
  +5\,\Da\Di\aiii\Dr\as\Dii\aiv +6\,\Da\Dr\as\Di\aiii\Dii\aiv\big]}\\
{\phantom{\partial_{\b}{\cal T}_3^{\b}}
-i\frac{N_f}{64\pi^2}\theta^{\alpha\beta}\theta^{\rho\sigma}
  \epsilon^{\mi\mii\miii\miv}\,g_{\alpha\rho}g^{\mu\nu}\Ds\Tr\big[\partial_{\,[\mu}a_{\mii]}
    \partial_{\,[\nu}a_{\miii]}\Di\aiv\big]\Big\}.}\\
\end{array}
\end{displaymath}
Applying to the r.h.s. of the previous equation the replacements in  
eq.~(\ref{replacements}), one gets that 
\begin{equation}
\begin{array}{l}
{\partial_{\b}{\cal T}_3^{(inv)\,\b}=\Db\Big\{
i\frac{N_f}{96\pi^2}\theta^{\alpha\beta}\theta^{\rho\sigma}
    \epsilon^{\mi\mii\miii\miv}\Tr\big[3\,\Dcal_\alpha f_{\rho\mii}f_{\sigma\miii}f_{\mi\miv}-
    3\,\Dcal_\alpha f_{\rho\mii}f_{\mi\miii}f_{\sigma\miv}}\\
{\phantom{\partial_{\b}{\cal T}_2^{(inv)\,\b}=\Db\Big\{}
+3\,\Dcal_\alpha f_{\mi\mii}f_{\rho\miii}f_{\sigma\miv}+
   \frac{7}{8}\,\Dcal_\alpha f_{\mi\miii}f_{\mii\miv}f_{\rho\sigma}+
   \frac{3}{4}\,\Dcal_\rho f_{\mi\miii}f_{\sigma\alpha}f_{\mii\miv}}\\
   {
+\frac{5}{8}\,\Dcal_\alpha f_{\mi\miii}f_{\rho\sigma}f_{\mii\miv}
+\!\frac{3}{4}\Dcal_\alpha f_{\rho\sigma}f_{\mi\miii}f_{\mii\miv}
\big]-\!i\frac{N_f}{128\pi^2}\theta^{\alpha\beta}\theta^{\rho\sigma}
    \epsilon^{\mi\mii\miii\miv}g_{\alpha\rho}g^{\mu\nu}\Ds\Tr\big[f_{\mu\mii}f_{\nu\miii}f_{\mi\miv}\big]\Big\}.}\\
\end{array}
\label{t3inv}
\end{equation}
It is clear that $\partial_{\b}{\cal T}_3^{(inv)\,\b}$ verifies 
eq.~(\ref{tinvdef}).  

Substituting eqs.~(\ref{t3inv}) and (\ref{t2inv}) in 
eq~(\ref{asw2}), one obtains the following result:  
\begin{displaymath}
\begin{array}{l}
 { {\cal A}^{(cn)\,(2)}=\Db\Big\{-\frac{N_f}{384\pi^2}\,\theta^{\alpha\beta}
\theta^{\rho\sigma}\epsilon^{\mi\mii\miii\miv}\,
\big(\Tr\,\Dcal_\alpha \Dcal_\sigma f_{\mi\miii}\Dcal_\rho f_{\mii\miv}
+\Ds\Tr\big[\frac{3}{2}\Dcal_\alpha f_{\mi\miii}\Dcal_\rho f_{\mii\miv}}\\
{\phantom{{\cal A}^{(cn)\,(2)}=\Db\Big\{}
+2\Dcal_\rho\Dcal_\alpha f_{\mi\miii}f_{\mii\miv}\big]\big)
+\frac{N_f}{512\pi^2}\,\theta^{\alpha\beta}\theta^{\rho\sigma}
\epsilon^{\mi\mii\miii\miv}g_{\alpha\rho}g^{\mu\nu}\Ds
\Tr\big[2\,\Dcal_\mu\Dcal_\nu f_{\mi\miii}f_{\mii\miv} }\\
   {\phantom{{\cal A}^{(cn)\,(2)}=\Db\Big\{}+\Dcal_\mu 
f_{\mi\miii}\Dcal_\nu f_{\mii\miv}\big]}\\
{\phantom{{\cal A}^{(cn)\,(2)}=\Db\Big\{}+i\frac{N_f}{96\pi^2}\theta^{\alpha\beta}\theta^{\rho\sigma}
\epsilon^{\mi\mii\miii\miv}\Tr[3\,\Dcal_\alpha f_{\rho\mii}
f_{\sigma\miii}f_{\mi\miv}-
    3\,\Dcal_\alpha f_{\rho\mii}f_{\mi\miii}f_{\sigma\miv}}\\
{\phantom{{\cal A}^{(cn)\,(2)}=\Db\Big\{}
+3\,\Dcal_\alpha f_{\mi\mii}f_{\rho\miii}f_{\sigma\miv}+
   \frac{7}{8}\,\Dcal_\alpha f_{\mi\miii}f_{\mii\miv}f_{\rho\sigma}+
   \frac{3}{4}\,\Dcal_\rho f_{\mi\miii}f_{\sigma\alpha}f_{\mii\miv}}\\
{\phantom{{\cal A}^{(cn)\,(2)}=\Db\Big\{}+\frac{5}{8}\,
\Dcal_\alpha f_{\mi\miii}f_{\rho\sigma}f_{\mii\miv}+
\frac{3}{4}\,\Dcal_\alpha f_{\rho\sigma}f_{\mi\miii}f_{\mii\miv}
\big]}\\
{\phantom{{\cal A}^{(cn)\,(2)}=\Db\Big\{}
-i\frac{N_f}{128\pi^2}\theta^{\alpha\beta}\theta^{\rho\sigma}
\epsilon^{\mi\mii\miii\miv}\,g_{\alpha\rho}g^{\mu\nu}\Ds\Tr f_{\mu\mii}f_{\nu\miii}f_{\mi\miv}\Big\}.}\\
\end{array}
\end{displaymath}
Using linear equations that can be derived quite easily from eqs.~(\ref{D3o}) 
and~(\ref{D3}), one may  simplify the previous equation and obtain:
\begin{equation}
 {\cal A}^{(cn)\,(2)}= \partial_{\beta}{\cal Z}^{\beta},
\label{asw2partial}
\end{equation}
where
\begin{equation}
\begin{array}{l}
 {{\cal Z}^{\beta}= +\frac{N_f}{1536\pi^2}\,
\theta^{\alpha\beta}\theta^{\rho\sigma}\epsilon^{\mi\mii\miii\miv}\,\Tr\,
  \big[2\,\Dcal_\alpha \Dcal_\rho f_{\mi\miii}\Dcal_\sigma f_{\mii\miv}+
10\,i\,\Dcal_\rho f_{\mi\miii}f_{\sigma\alpha}
    f_{\mii\miv}}\\
{\phantom{{\cal Z}^{\beta}=+\frac{N_f}{1536\pi^2}\,\theta}
+2\,i\,\Dcal_\rho f_{\mi\miii}f_{\mii\miv}f_{\sigma\alpha}+i
\Dcal_\alpha f_{\mi\mii}f_{\rho\sigma}f_{\miii\miv}-i\,\Dcal_{\alpha}f_{\mi\mii}f_{\miii\miv}
     f_{\rho\sigma}\big]}\\
  {\phantom{{\cal Z}^{\beta}=}+\frac{N_f}{512\pi^2}\,\theta^{\alpha\beta}\theta^{\rho\sigma}\epsilon^{\mi\mii\miii\miv}g_{\alpha\rho}g^{\mu\nu}\,\Tr\,\Ds
     \big[2\,\Dcal_\mu \Dcal_\nu f_{\mi\miii}f_{\mii\miv}+\Dcal_\mu f_{\mi\miii}\Dcal_\nu f_{\mii\miv}}\\
{\phantom{{\cal Z}^{\beta}=+\frac{N_f}{1536\pi^2}\,\theta }     
-4\,i\,f_{\mu\mii}f_{\nu\miii}f_{\mi\miv}\big].}\\
\end{array}
\label{zbeta}
\end{equation}
Finally, taking into account eqs~(\ref{zbeta}), (\ref{asw2partial}), 
(\ref{asw1}), (\ref{asw1asw2}) and (\ref{ddimdiff}), one comes to the 
conclusion that
\begin{equation} 
\partial_{\mu}<j_{5}^{(cn)\,\mu}(x)>^{(A)}_{MS}-
\partial_{\mu}<j_{5}^{(np)\,\mu}(x)>^{(A)}_{MS}\,=\,h^2\,
{\cal A}^{(cn)\,(2)}\neq 0.
\label{finaldiffswnp}
\end{equation}  
Let $N[j_{5}^{(cn)\,\mu}]_{MS}(x)$ and $N[j_{5}^{(np)\,\mu}]_{MS}(x)$ be  
renormalized operators --called normal products~\cite{Bonneau:1979jx,Collins:1984xc}-- constructed  by MS renormalization from the dimensionally regularized 
currents $j_{5}^{(cn)\,\mu}(x)$ and $j_{5}^{(np)\,\mu}(x)$, respectively.
Then, the previous  equation shows that the difference between 
$N[j_{5}^{(cn)\,\mu}]_{MS}(x)$ and $N[j_{5}^{(np)\,\mu}]_{MS}(x)$ 
is an operator, say $N[{\cal Y}^{\mu}]$, which does not verify 
$\partial_{\mu} N[{\cal Y}^{\mu}]_{MS}(x)=0$, even upon imposing  
the equations of motion. This is in contradiction to the classical case.   
And yet, as we shall see below, both currents, defined in terms of  normal 
products, yield, if $\theta^{0i}=0$, the same chiral charge  
up to order $h^2$.  But first, let us see that   
the $\theta^{\mu\nu}$-dependent  contributions in eq.~(\ref{finaldiffswnp}) 
are not anomalous contributions, but finite renormalizations of the current  
 $N[j_{5}^{(cn)\,\mu}(x)]$. Indeed, eqs.~(\ref{finaldiffswnp}) and  
(\ref{opanomeq}) lead to  
\begin{displaymath}
\partial_{\mu}N[j_{5}^{(cn)\;\mu}]_{MS}(x)\,=\, 
\frac{N_f}{8\pi^2} \Tr\,f_{\mi\mii}(x) \tilde{f}_{\miii\miv}(x)\,+\,h^2\,
\partial_{\lambda}\,{\cal X}^{\lambda}(x)\,+\,h^2\,\partial_{\beta}{\cal Z}^{\beta},
\end{displaymath}
where ${\cal X}^{\lambda}(x)$ and ${\cal Z}^{\beta}$ are the gauge invariant 
vector fields in  eqs.~(\ref{xlambda}) and (\ref{zbeta}), respectively. Then, 
we introduce a new  current, say ${\cal J}_5^{(cn)\,\mu}$, defined as follows
\begin{equation} 
{\cal J}_5^{(cn)\,\mu}(x)\,=\,N[j_{5}^{(cn)\;\mu}]_{MS}(x)\,-\,h^2\,
{\cal X}^{\mu}(x)\,-\,h^2\,{\cal Z}^{\mu}(x).
\label{newcn}
\end{equation}
This new current is to be understood as a finite renormalization of $N[j_{5}^{(cn)\;\mu}]_{MS}(x)$, and satisfies the ordinary $U(1)_A$ anomaly equation:
\begin{equation}
\partial_{\mu}{\cal J}_5^{(cn)\,\mu}(x)\,=\,\
\frac{N_f}{8\pi^2}\,\Tr\,f_{\mi\mii}(x) \tilde{f}_{\miii\miv}(x).
\label{anomcaljcn}
\end{equation}
It is plain that ${\cal X}^{0}=0={\cal Z}^{0}$, if $\theta^{0i}=0$. Hence,  
if time is commutative both ${\cal J}_5^{(cn)\,\mu}(x)$ and 
$N[j_{5}^{(cn)\;\mu}]_{MS}(x)$ give rise to the same chiral charge
$Q_{5}^{(cn)}$. Integrating both sides of 
eq.~(\ref{anomcaljcn}) over all values of $x$, one concludes that, unlike
its ordinary counterpart, the quantum $Q_{5}^{(cn)}$ is no longer conserved 
in the presence of topologically nontrivial field configurations:
\begin{equation}
Q_{5}^{(cn)}(t=+\infty)-Q_{5}^{(cn)}(t=-\infty)\,=\,
\frac{N_f}{8\pi^2}\idx\, \Tr\,f_{\mi\mii} \tilde{f}_{\miii\miv}\,=\,
2\,N_f\,(n_{+}-n_{-}).
\label{planenpm}
\end{equation}
The integers $n_{\pm}$ are defined in eq.~(\ref{npm}). To obtain 
the result in the far right of the previous equation we have used the temporal 
gauge $a_{0}(x)=0$ and the boundary conditions in eq.~(\ref{boundarycond}).
Again, eq.~(\ref{planenpm}) is the spitting image of eq.~(\ref{qnpanom}). This
is no wonder since, as we shall see next, the MS renormalized $Q_{5}^{(cn)}(t)$
and $Q_{5}^{(np)}(t)$ agree up to second order in $h$, at the least. 
Using the identities in eq.~(\ref{properties}), one can show that 
the  following equations hold for $\theta^{0i}=0$:
\begin{displaymath}
N[j_{5}^{(cn)\,0}]_{MS}-N[j_{5}^{(np)\,0}]_{MS}=\partial_{i}\,R^{i}.
\end{displaymath} 
$R^i$ denotes the operator 
\begin{displaymath} 
\begin{array}{l}
{-i\frac{h}{2}\,\theta^{i j}N[\bar{\psi}\g^0 \g_5 D_j \psi]_{MS}
+h^2\,\theta^{i j}\theta^{i' j'}\Big[+\frac{1}{8}
\partial_{i'}N[\bar{\psi}\g^0 \g_5 D_j D_{j'}\psi]_{MS}
\phantom{aaaaaaaaa}}\\ 
{\phantom{aaaaaaaaaaaaaaa}-\frac{1}{8}N[\bar{\psi}\g^0 \g_5 D_{i'}\{D_{j'},D_j \}\psi]_{MS}
-\frac{1}{4}N[\bar{\psi}
\g^0 \g_5 D_j D_{i'} D_{j'} \psi]_{MS}\phantom{aaaaaaaaa}}\\
{\phantom{aaaaaaaaaaaaaaaaaaaaaa}
+\frac{1}{4}
N[\bar{\psi}
\g^o \g_5 D_{i'} D_j D_{j'}\psi]_{MS}\,\Big]+o(h^3).}\\
\end{array}
\end{displaymath}
Then, 
\begin{displaymath}
Q_{5}^{(cn)}=\int d^{3}\vec{x}\;N[j_{5}^{(cn)\,0}]_{MS}=
\int d^{3}\vec{x}\;N[j_{5}^{(np)\,0}]_{MS}+
\int d^{3}\vec{x}\; \partial_{i}\,R^{i}=Q_{5}^{(np)}\,+\,o(h^3).
\end{displaymath}
We have assumed that the fields go sufficiently rapidly to zero at spatial 
infinity so that the last integral vanishes.

\section{Nonsinglet chiral currents are anomaly free}

The $SU(N_f)_A$ canonical Noether current, i.e., the canonical nonsinglet 
chiral current, reads
\begin{displaymath}
j_{5}^{(cn)\,a\mu}\,=\,\sum_{ff'}\bar{\psi}_{f}\,{\cal H}\,T^{a}_{ff'}
\psi_{f'}. 
\end{displaymath}
Where  ${\cal H}$ is the object that is left after removing from 
$j^\mu_{(cn)}$ in eq.~(\ref{swcurrent}) the fields $\bar{\psi}_{f}$ and 
$\psi_{f}$. 
We also have the nonsinglet current, $j_{5}^{(np)\,a\mu}$, which is the analog of the singlet current $j_{5}^{(np)\,\mu}$ in eq.~(\ref{nccurrents}):
\begin{displaymath}
j_{5}^{(np)\,a\mu}=\sum_{ff'}\bar{\Psi}_{f}T^{a}_{ff'}\gamma_5\ST\Psi_{f'}.
\end{displaymath}
These two nonsinglet currents are divergenceless classically since the 
classical theory has the $SU(N_f)_A$ symmetry in eq.~(\ref{rigidtrans}). 
The dimensionally regularized currents constructed from 
$j_{5}^{(cn)\,a\mu}$ and $j_{5}^{(np)\,a\mu}$ above verify the 
following equations
\begin{equation}
\begin{array}{l}
{\partial_{\mu} <j_{5}^{(cn)\,a\mu}>^{(A)}=\partial_{\mu}<j_{5}^{(np)\,a\mu}>^
{(A)}\,+\, \partial_{\mu}\hat{{\cal X}}^{a\,\mu}_{(ns)}},\\
{\partial_{\mu}<j_{5}^{(np)\,a\mu}>^{(A)}=2\, \sum_{ff'}\,
<\Psib_{f}\ST T^a_{ff'}{\hat{\gamma}}^{a\,\mu}\gamma_5\Psi_{f'}>^{(A)}.}\\
\end{array}
\label{nonsinglet}
\end{equation}
Here, 
$\hat{{\cal X}}^{a\,\mu}_{(ns)}=\sum_{ff'}\bar{\psi}_{f}\,{\cal K}\,T^{a}_{ff'}
\psi_{f'}$. ${\cal K}$ is obtained by stripping $\bar{\psi}_{f}$ and 
$\psi_{f}$ off the r.h.s. of eq.~(\ref{hatx}).  Now, since the 
kinetic terms and vertices of our noncommutative theory are in flavour space  
proportional to the identity, it is clear that the contributions to the 
r.h.s. of the equalities  in eq.~(\ref{nonsinglet}) can be  
obtained from the corresponding  
singlet contributions by multiplying them by $\Tr\,T^{a}$ --see eqs.~(\ref{severalas}), (\ref{zeroaone}), (\ref{a2divergence}), (\ref{xlambda}) and  
(\ref{hatx}), and diagrams in Figs. 1, 2 and 3. But,  $\Tr\,T^{a}=0$, so that
\begin{displaymath}
\partial_{\mu} <j_{5}^{(cn)\,a\mu}>^{(A)}=0,\qquad
 \partial_{\mu}<j_{5}^{(np)\,a\mu}>^{(A)}=0.
\end{displaymath} 
We have thus shown that, at least at the one-loop level and second order in 
$h$, the quantum nonsinglet currents of the $SU(N_f)_A$ classical symmetry 
of the theory are anomaly free.

\section{Summary and conclusions}

In this paper we have obtained, at the one-loop level and second order 
in $\theta^{\mu\nu}$, the anomaly equation for the canonical 
Noether current --$j_{5}^{(cn)\,\mu}$ in eq.~(\ref{swcurrent})-- of the 
classical $U(1)_A$ symmetry of noncommutative 
$SU(N)$ gauge theory with massless fermions. All along this paper 
the physical  $\theta^{\mu\nu}$ has been considered to be of ``magnetic'' 
type: $\theta^{i0}=0$. We have shown that the current $j_{5}^{(cn)\,\mu}$ can 
be renormalized to a current --${\cal J}_5^{(cn)\,\mu}$ in 
eq.~(\ref{newcn})-- such that the anomalous contribution
to the fourdivergence of the latter is just the ordinary anomaly. This is a
highly nontrivial result since, {\it a priori}, there are 
$\theta^{\mu\nu}-$dependent  candidates to the $U(1)_A$ anomaly such as
\begin{displaymath}
\theta^{\rho\sigma}
\epsilon^{\mu_1\mu_2\mu_3\mu_4}\Tr\,
     [f_{\sigma \mu_1}f_{\mu_2\mu_3}f_{\rho\mu_4}], 
\quad 
\theta^{\a\b}\theta^{\r\s}\,\epsilon^{\mi\mii\miii\miv}\,\Tr\,\Big[
 f_{\mi\mii}f_{\miii\miv}f_{\a\r}f_{\b\s}\Big],\quad{\it etc}...
\end{displaymath}
We have shown that all these would-be anomalous contributions neatly cancel 
among themselves --see eqs.~(\ref{zeroaone}), (\ref{A2}) and (\ref{a2final}).
We have also studied the anomaly equation 
for other noncommutative currents that are classically  (covariantly) conserved
as a consequence of the $U(1)_A$ invariance of the classical action. 
These currents go under the names of $j_5^{(p)\,\mu}$ and  
$j_5^{(np)\,\mu}$ and their (covariant) fourdivergences in the MS scheme 
are given in eqs.~(\ref{operatorpanom}) and (\ref{opanomeq}). Classically, 
the current $j_5^{(np)\,\mu}$ is also a Noether current, for it is related 
with the canonical Noether current $j_5^{(cn)\,\mu}$ by 
eq.~(\ref{diffnpcn}) --see also eq.~(\ref{ambi})--. This relationship  
does not hold for the MS renormalized currents. However, at the one-loop 
level, we have been able to introduce a current --${\cal J}_5^{(np)\,\mu}$ in eq.~(\ref{newcalnp})-- which is obtained by nonminimal renormalization of    
$j_5^{(np)\,\mu}$ and whose difference with ${\cal J}_5^{(cn)\,\mu}$ is 
a certain ${\cal Y}^{\mu}$ satisfying the criteria in eq.~(\ref{zerocalgy}).

We have also shown that, at least up to second order in $\theta^{\mu\nu}$, all 
the $U(1)_A$  currents considered above yield the same chiral charge, 
say $Q^{(cn)}_5(t)$, if $\theta^{0i}=0$. Of course, this classically 
conserved charge is not conserved at the quantum level, but verifies the 
following equation:
\begin{equation}
Q^{(cn)}_{5} (t=+\infty)-Q^{(cn)}_{5}(t=-\infty)\,=\,
\frac{N_f}{8\pi^2}\idx\, \Tr\,f_{\mi\mii} \tilde{f}_{\miii\miv}\,=\,
2\,N_f\,(n_{+}-n_{-}).
\label{compulsory}
\end{equation}
To obtain the result in the far right of the previous equation, the temporal
gauge, $a_{o}(x)=0$, has been used and the boundary conditions in 
eq.~(\ref{boundarycond}) have been imposed. The integers $n_{\pm}$ are
defined in eq.~(\ref{npm}).  The  identity on the  
far right of eq.~(\ref{asymptoticchiral}) puts us in the position of 
giving to eq.~(\ref{compulsory}) a clear physical meaning. What 
eq.~(\ref{compulsory})
shows is that in any quantum transition from $t=-\infty$ to $t=+\infty$ that
involve  a change in the topological properties of the asymptotic gauge 
fields  --i.e., $n_{+}-n_{-}=n\neq 0$--, there is, for $(n<0) n>0$, a 
transmutation of the (right-) left-handed fermionic degrees of 
freedom at $t=-\infty$ into (left-) 
right-handed degrees at $t=\infty$. For instance, take $n>0$, then, if in that transition the fermionic part of the physical state at 
$t=-\infty$ is constituted  by $n N_f$ left-handed fermions, then, 
the fermionic part of the physical state at $t=+\infty$ will be made of  
$n N_f$ right-handed fermions. Of course, there will be  ``compulsory'' creation of fermion-antifermion  pairs at $t=+\infty$, if there are no fermionic 
degrees of freedom at $t=-\infty$. It is well known that these phenomena also 
occur in ordinary space-time, so introducing noncommutative space-time does 
not change the qualitative picture; it does change, however, the quantitative 
analysis of these phenomena. For instance, upon Wick rotation the dominant 
contribution to the path integral coming from the gauge fields is a certain 
$\theta^{\mu\nu}-$deformation of the ordinary BPST instanton. This deformation,
 in turn, gives rise to a $\theta^{\mu\nu}-$dependent effective 't Hooft 
vertex. We shall report on these findings elsewhere~\cite{Ctamarit}.

Next, taking into account that we have shown that  
$Q^{(cn)}_{5} (t)=Q^{(pn)}_{5} (t)=Q^{(p )}_{5} (t)$ is verified at least up 
to second order in $\theta^{\mu\nu}$ and the fact that eq.~(\ref{planeppm})  
is valid  at the one-loop level and any order in $\theta^{\mu\nu}$, it is not  
foolish to conjecture that eq.~(\ref{compulsory}) will  hold at any order 
in $\theta^{\mu\nu}$.

Now, since in our computations the actual properties of $T^a$ --the 
generators of $SU(N)$-- have played no role, barring its hermiticity and the 
cyclicity of trace of any product of them, we conclude that all our 
expressions  
are valid for $SO(N)$ groups. Our expressions are also valid for  
$U(1)$ provided we replace $a_{\mu}$ with $Q\,a_{\mu}$. $Q$ being the charge 
of the fermion coupled to the $U(1)$ field $a_{\mu}$.
 
Finally, it is quite obvious how to generalize our expressions to encompass 
the situation where several representations --labeled by {\cal R}-- of the 
gauge group are at work in the fermionic action. Let us give just one 
instance. Assume that we have $N^{({\cal R})}_{f}$ fermions which couple 
to the gauge field, $a^{({\cal R})}_{\mu}$, in the ${\cal R}$ representation 
of the gauge group. Then, eq.~(\ref{anomcaljcn}) will read: 
\begin{displaymath}
\partial_{\mu}{\cal J}_5^{(cn)\,\mu}(x)\,=\,\
\sum_{{\cal R}}\frac{N^{({\cal R})}_f}{8\pi^2}\,\Tr\,f^{({\cal R})}_{\mi\mii}(x) \tilde{f}^{({\cal R})}_{\miii\miv}(x),
\end{displaymath}
with
\begin{displaymath}
{\cal J}_5^{(cn)\,\mu}(x)\,=\,\sum_{{\cal R}}\,N[j_{5}^{({\cal R})\,(cn)\;\mu}]_{MS}(x)\,-\,h^2\,
\sum_{{\cal R}}\,{\cal X}^{({\cal R})\,\mu}(x)\,-\,h^2\,
\sum_{{\cal R}}\,{\cal Z}^{({\cal R})\,\mu}(x).
\end{displaymath}
The gauge fields in $j_{5}^{({\cal R})\,(cn)\;\mu}$, 
${\cal X}^{({\cal R})\,\mu}$ and ${\cal Z}^{({\cal R})\,\mu}$ are all in the 
${\cal R}$ representation of the gauge group.

\section{Acknowledgments}

We thank  L. Moeller for fruitful correspondence. This work has been 
financially supported in part by MEC through grant BFM2002-00950. The work 
of C. Tamarit has  also received financial support from MEC trough FPU grant   
AP2003-4034.

\newpage

\setcounter{section}{0}

\section*{Appendix A}
In this Appendix we give the Feynman rules needed to turn into
mathematical objects the Feynman diagrams displayed in this paper. These
Feynman rules are in Fig.~5.
\begin{figure}[h]
\flushleft
\begin{minipage}[c]{0.12\textwidth}
\epsfig{file=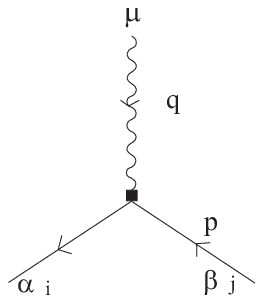,height=2cm}
\end{minipage}%
\begin{minipage}[c]{0.23\textwidth}
\small
\begin{equation*}
    \leftrightarrow i \gamma^\mu_{\alpha\beta}\,\,e^{-i\frac{h}{2}q\circ
         p}A_{\mu,ij}
\end{equation*}
\end{minipage}%
\begin{minipage}[c]{0.11\textwidth}
\epsfig{file=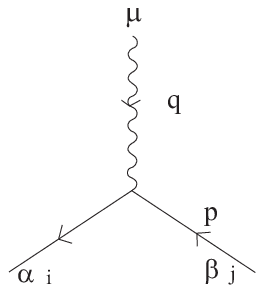,height=2cm}
\end{minipage}%
\begin{minipage}[c]{0.12\textwidth}
\small
\begin{equation*}
        \leftrightarrow i \gamma^\mu_{\alpha\beta}
     \end{equation*}
\end{minipage}%
\begin{minipage}[c]{0.12\textwidth}
\epsfig{file=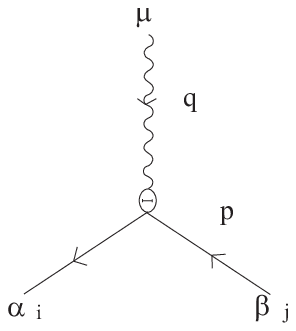,height=2cm}
\end{minipage}%
\begin{minipage}[c]{0.30\textwidth}
\small
 \begin{equation*}
 \begin{array}{l}
\leftrightarrow-\frac{h}{2}\,\theta^{\rho\sigma}\gamma^\mu_{\alpha\beta}\left\{q_\rho a_\sigma(q)
     p_\mu+\right.\\
    \phantom{\leftrightarrow}
    \left.+p_\sigma[q_\mu a_\rho(q)-q_\rho a_\mu(q)]\right\}
\end{array}
\end{equation*}
\end{minipage}
\flushleft
\begin{minipage}[c]{0.15\textwidth}
  \epsfig{file=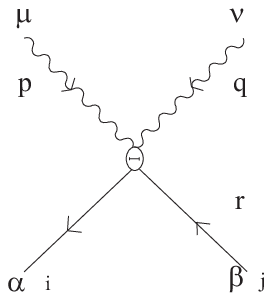,height=1.8cm}
\end{minipage}%
\begin{minipage}[c]{0.8\textwidth}
\small
 \begin{equation*}
 \begin{array}{l}
\leftrightarrow -\frac{h}{2}\,\theta^{\rho\sigma}\gamma^\mu_{\alpha\beta}\left\{{p}_\rho a_\sigma (p)a_\mu (q)
     +\frac{1}{2}\,r_\mu[a_\rho(p)a_\sigma(q)+a_\rho(q)a_\sigma(p)]+
     {p}_\mu a_\rho(p)a_\sigma(q)-\right.\\
     \phantom{\leftrightarrow}\left.-{p}_\rho a_\mu(p)a_\sigma(q)+r_\sigma[a_\mu(p)a_\rho(q)-a_\rho(p)a_\mu(q)]\right\}_{ij}
\end{array}
     \end{equation*}
\end{minipage}\\[15pt]
\begin{minipage}[c]{0.14\textwidth}
  \epsfig{file=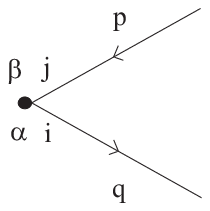,height=1.8cm}
\end{minipage}%
\begin{minipage}[c]{0.31\textwidth}
\small
 \begin{equation*}\leftrightarrow-2ip_\mu\,(
         \gamma^5\hgamma^\mu)_{\alpha\beta}\,\delta_{ij}\,e^{i\frac{h}{2}p\circ q}
    \end{equation*}
\end{minipage}%
\begin{minipage}[c]{0.25\textwidth}
\centering
  \epsfig{file=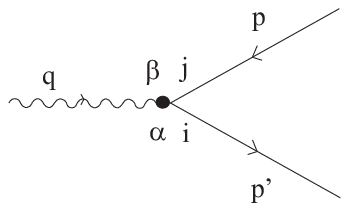,height=1.8cm}
\end{minipage}%
\begin{minipage}[c]{0.3\textwidth}
\small
 \begin{equation*}
 \begin{array}{c}
 \leftrightarrow -2i\, (
        \gamma^5\hgamma^\mu)_{\alpha\beta}\,A_{\mu,ij}(q)\\
    \phantom{\leftrightarrow -2i\,} e^{i\frac{h}{2}(q+p)\circ
         p'}\,e^{-i\frac{h}{2}q\circ p}
    \end{array}
    \end{equation*}
 \end{minipage}\\[15pt]
\begin{minipage}[c]{0.15\textwidth}
 \epsfig{file=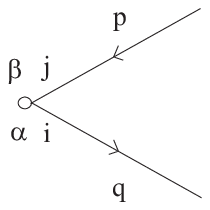,height=1.8cm}
\end{minipage}%
\begin{minipage}[c]{0.3\textwidth}
\small
 \begin{equation*}
 \leftrightarrow-2ip_\mu\,(
 \gamma^5\hgamma^\mu)_{\alpha\beta}\,\delta_{ij}\,e^{-i\frac{h}{2}p\circ q}
 \end{equation*}
\end{minipage}%
\begin{minipage}[c]{0.25\textwidth}
\centering
 \epsfig{file=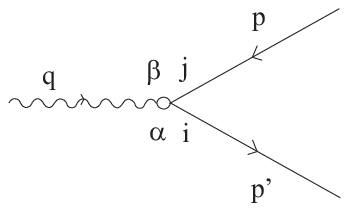,height=1.8cm}
\end{minipage}%
\begin{minipage}[c]{0.3\textwidth}\small
 \begin{equation*}
 \begin{array}{l}
 \leftrightarrow -2i\, (
        \gamma^5\hgamma^\mu)_{\alpha\beta}\,A_{\mu,ij}(q)\\
  \phantom{\leftrightarrow -2i}e^{i\frac{h}{2}p\circ
         (p'+q)}\,e^{-i\frac{h}{2}q\circ p}
 \end{array}
 \end{equation*}
\end{minipage}\\[13pt]
\begin{minipage}[c]{0.12\textwidth}
\epsfig{file=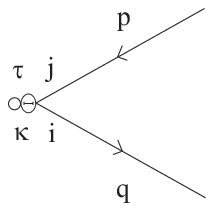,height=1.8cm}
\end{minipage}%
\begin{minipage}[c]{0.36\textwidth}
\small
\begin{equation*}
\begin{array}{l}
        \leftrightarrow -\,h\,\theta^{\alpha\beta}(p-q)_\beta\, p_\alpha p_\nu\, \delta_{ij}\\
    \phantom{\leftrightarrow}{(\g^5\hat{\g}^\nu)_{\kappa
         \tau}}
\end{array}
     \end{equation*}
\end{minipage}%
\begin{minipage}[c]{0.22\textwidth}
\epsfig{file=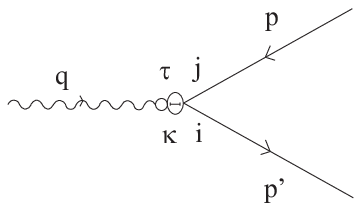,height=1.8cm}
\end{minipage}%
\begin{minipage}[c]{0.27\textwidth}
\small
\begin{equation*}
\begin{array}{l}
         \leftrightarrow -\,h\,\theta^{\alpha\beta} (p+q-p')_\beta\\
          (p_\nu \,{a_\alpha}+a_\nu(q_\alpha+p_\alpha))_{ij}\\(\g^\nu \g^5)_{\kappa
         \tau}
\end{array}
     \end{equation*}
\end{minipage}
\begin{minipage}[c]{0.3\textwidth}
\epsfig{file=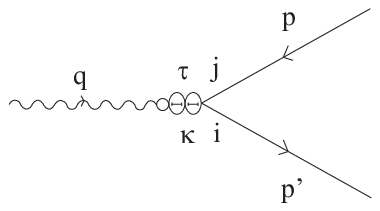,height=1.8cm}
\end{minipage}%
\begin{minipage}[c]{0.7\textwidth}
\flushleft
\small
\begin{equation*}
\begin{array}{l}
        \leftrightarrow \frac{i\,h^2}{4}\,\theta^{\alpha\beta}\theta^{\rho\sigma} (p+q-p')_\beta\,(\g^\nu \g^5)_{\kappa\tau}
    \left\{({q}_\rho a_{\alpha}-{q}_\alpha a_\rho)p_\sigma p_\nu \right.\\
         \left.-({q}_\rho a_{\sigma}-{q}_\sigma a_\rho)p_\alpha p_\nu -{q}_\alpha{q}_\rho a_\sigma p_\nu-2(p_\alpha+q_\alpha)p_\sigma(q_\nu a_\rho-q_\rho a_\nu)\right.\\\left.
         -(p+q-p')_\sigma[(q_\alpha q_\rho+2q_\rho p_\alpha+p_\alpha p_\rho)a_\nu+(q_\alpha a_\rho+2 a_\rho p_\alpha)p_\nu]\right\}_{ij}
\end{array}
\end{equation*}
\end{minipage}\\[5pt]
\begin{minipage}[c]{0.16\textwidth}
\epsfig{file=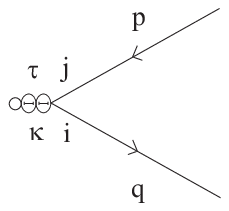,height=1.8cm}
\end{minipage}%
\begin{minipage}[c]{0.50\textwidth}
\small
\begin{equation*}
        \leftrightarrow  -\frac{i\,h^2}{4}\,\theta^{\alpha\beta}\theta^{\rho\sigma}(p-q)_\beta(p-q)_\sigma\, p_\alpha p_\rho p_\nu\, \delta_{ij}\,(\g^\nu \g^5)_{\kappa\tau}
     \end{equation*}
\end{minipage}\\[13pt]
 \renewcommand{\figurename}{Fig.}
 \renewcommand{\captionlabeldelim}{.}
\caption{Feynman rules}
\end{figure}

\section*{Appendix B}
\renewcommand{\theequation}{B.\arabic{equation}}

In this Appendix we display some equalities verified by the 
``$D$-dimensional''  Lorentz covariants introduced in ref.~\cite{Breitenlohner:1977hr}, which are used in our computations. 
\begin{equation}
\begin{array}{l}
 {\,\,\phantom{\hg}g^{\mu\nu}\hg_{\nu\rho}=\hg^\mu _\rho,\quad\quad\quad\quad\quad\quad\quad\quad\quad
\hg^{\mu}_\mu = 2\epsilon,\quad\quad\quad\quad\quad\quad\quad\quad
     \epsilon^{\mu\nu\rho\sigma}\hg_{\sigma\eta}=0,}\\
{  \{\gamma^\mu,\gamma^\nu \}= 2\g^{\mu\nu}\mathbb{I},\quad\quad\quad
     \{\gamma_5,\gamma^\mu\}=\{\gamma_5,\hgamma^\mu\}=2 \gamma_5
\hgamma^\mu,\quad\quad\quad\quad\,
     [\gamma_5,\hgamma^\mu]\,=\,0.}\\
\end{array}
\label{gammascom}
\end{equation}
\begin{equation}
\begin{array}{l}
{\tr \gamma_5
   \hat{\gamma}^{\mu_1}\gamma^{\mu_2}\dots\gamma^{\mu_2k}=}\\
   {\frac{1}{2}
         \left[\tr \gamma_5\{\hat{\gamma}^{\mu_1},\gamma^{\mu_2}\}\gamma^{\mu_3}\dots
         \gamma^{\mu_{2k}}+\tr \gamma_5\gamma^{\mu_2}\left\{\hat{\gamma}^{\mu_1}
         ,\gamma^{\mu_3}\right\}\dots\gamma^{\mu_{2k}}+\dots+
         \tr \gamma_5\gamma^{\mu_2}\dots \{\hat{\gamma}^{\mu_1},\gamma^{\mu_2}\}\right].}
\end{array}
\label{traza-antic}
 \end{equation}
 \begin{displaymath}
     \tr\,\g_5 \g^{\mu_1}\g^{\mu_2}\,=\,0,\quad\quad
     \tr\,\g_5\g^{\mu_1}\g^{\mu_2}\g^{\mu_3}\,=\,0,\quad\quad
         \tr\,
         \gamma_5\gamma^{\mu_1}\dots\gamma^{\mu_4}\,=\,i\;\tr\,\mathbb{I}\;\epsilon^{\mu_1
         \dots\mu_4}.
 \end{displaymath}
 \begin{displaymath}
     \tr\,\g_5\g^{\mu_1}\dots\g^{\mu_6}\,=\,\;\tr\,\mathbb{I}\;\sum_{p<q}{(-1)^{q-p}
    \epsilon^{\mu_1\dots\mu_{p-1}\mu_{p+1}\dots\mu_{q-1}\mu_{q+1}\dots\mu_6}}
\,g^{\mu_p\mu_q}.
  \end{displaymath}

\section*{Appendix C}
\renewcommand{\theequation}{C.\arabic{equation}}

Here we include the list of the dimensionally regularized integrals that
are need to work out the would-be anomalous contributions. The dimensional regularization regulator $\epsilon$ is equal to $\frac{D-4}{2}$. Contributions that vanish as
$\epsilon\rightarrow 0$ are never included.
The symbol ``$\sim$'' shows that we have dropped contributions 
of the type $\frac{1}{\epsilon}\hat{O}$, where $\hat{O}$ is an evanescent 
tensor, for they are not actually needed: they are subtracted by the 
renormalization algorithm. 
\begin{displaymath}
 \begin{array}{l}  
  {\iDp\frac{\hp^2}{(p-a)^2(p-b)^2(p-c)^2}=\iDp\frac{\hp^2 p^2}{(p-a)^2(p-b)^2(p-c)^2(p-d)^2}=\frac{-i}{32\pi^2},}\\
  {\iDp\frac{\hp_\a p_\b}{p^2(p-a)^2(p-b)^2}=\frac{-i}{64\pi^2\epsilon}\hg_{\a\b},}\\
  {\iDp\frac{\hp^2
     p_\alpha p_\beta}{(p-a)^2(p-b)^2(p-c)^2(p-d)^2}=\frac{-i}{4!8\pi^2}\Big\{
     g_{\alpha\beta}+\frac{1}{\epsilon}\hg_{\alpha\beta}\Big\},}\\
 {\iDp\frac{\hp^2 p^2
     p_\alpha}{(p-a)^2(p-b)^2(p-c)^2(p-d)^2}=\frac{-i}{6(4\pi)^2}\Big\{
     (a+\dots+d)_\alpha+\frac{1}{\epsilon}\,\hg^\nu _\alpha
     (a+\dots+d)_\nu \Big\},}\\
  {\iDp\frac{\hp^2 p^2 p_\alpha
     p_\beta}{(p-a)^2(p-b)^2(p-c)^2(p-d)^2(p-e)^2}=\frac{-i}{4!8\pi^2}\Big\{g_{\alpha\beta}+
     \frac{1}{\epsilon}\hg_{\alpha\beta}\Big\},}\\
 {\iDp\frac{\hp^2 p_\alpha p_\beta}{(p-a)^2(p-b)^2(p-c)^2}=-\frac{i}{2(4\pi)^2}
\Big\{
     2\Delta_{\alpha\beta}+\frac{4}{\epsilon}\hat{\Delta}_{\alpha\beta}+\frac{1}{\epsilon}g_{\alpha\beta}
\hat{\Delta}^\mu _\mu\Big\}}\\
 {\hspace{3cm}+\frac{i}{24(4\pi)^2}(a^2+b^2+c^2-a\cdot b-a\cdot c-b\cdot c)\Big\{g_{\alpha\beta}+\frac{1}{\epsilon}\hat{g}_{\alpha\beta}\Big\},}\\
   {\hspace{3cm}\Delta_{\alpha\beta}\equiv\frac{1}{12}(a_\alpha a_\beta+b_\alpha b_\beta +c_\alpha c_\beta)
     +\frac{1}{24}(a_\alpha b_\beta+a_\alpha c_\beta+b_\alpha c_\beta+\alpha\leftrightarrow \beta),}\\
 {\iDp\frac{\hp^2 p_\alpha p_\beta p_\rho}{(p-a)^2(p-b)^2(p-c)^2(p-d)^2}\sim}\\
       { \hspace{2cm}\sim\frac{-i}{4!32\pi^2}\Big\{
     g_{\alpha\beta}(a+b+c+d)_\rho+g_{\alpha\rho}(a+b+c+d)_\beta+g_{\beta\rho}(a+b+c+d)_\alpha\Big\},}\\
\end{array}
\end{displaymath}
\begin{displaymath}
\begin{array}{l}
   {\iDp\frac{\hp^2 p^2 p_\alpha p_\beta}{(p-a)^2(p-b)^2(p-c)^2(p-d)^2}\sim}\\
     {\sim\frac{i}{384\pi^2}\,g_{\alpha\beta}\Big\{a^2+b^2+c^2+d^2-a\cdot(b+c+d)-b\cdot(c+d)-c\cdot d
     \Big\},}\\
     {-\frac{i}{384\pi^2}\Big\{2a_\a a_\b +2b_\a b_\b +2 c_\a c_\b+2d_\a d_\b
    +\big[a_\a (b+c+d)_\b+b_\a (c+d)_\b+c_\a d_\b+\alpha\leftrightarrow\beta\big]\Big\},}\\
     {\iDp\frac{\hp^2 p_\alpha p_\beta p_\r p_\s}{p^2(p-a)^2(p-b)^2(p-c)^2(p-d)^2}\sim
    \frac{-i}{1536\pi^2}\Big\{g_{\a\b}g_{\r\s}+g_{\a\r}g_{\b\s}+g_{\a\s}g_{\b\r}\Big\}.}\\
    \end{array}
\end{displaymath}

\section*{Appendix D}
\renewcommand{\theequation}{D.\arabic{equation}}

In this Appendix  we shall work out a number of identities among the terms on the r.h.s of eq.~(\ref{A2}) and explain how to use them to obtain our final 
answer for ${\cal A}^{(2)}$ given in eq.~(\ref{a2final}).

Let $t_{\mu_1\mu_2\cdots\mu_n}$ be an object with indices $\mu_{i}$,  
$i=1\cdots n$, where $\mu_{i}=0,1,2,3$ $\forall i$ and $n>4$. Then, 
if $[\mu_1\mu_2\cdots\mu_n]$ stands for antisymetrization of the indices, we
have 
\begin{displaymath}
t_{[\mu_1\mu_2\cdots\mu_n]}\,=\,0.
\end{displaymath}
Taking into account the previous identity, the cyclicity of $\Tr$ and the 
antisymmetry properties of $\epsilon^{\mu_1\mu_2\mu_3\mu_4}$, one obtains the collection of beautiful identities displayed below:
\begin{equation}
 \theta^{\rho\sigma}\epsilon^{\mu_1\mu_2\mu_3\mu_4}\,\Tr\,f_{\sigma[\mu_1}f_{\mu_2\mu_3}
     f_{\mu_4\rho ]}=0\Rightarrow
 \theta^{\rho\sigma}\epsilon^{\mu_1\mu_2\mu_3\mu_4}\,\Tr\,
 \big[f_{\sigma \mu_1}f_{\mu_2\mu_3}f_{\rho\mu_4}-\frac{1}{4}
f_{\mu_1\mu_2}f_{\mu_3\mu_4}f_{\sigma\rho}\big]=0.
 \label{D1}
\end{equation}
\begin{equation}
\begin{array}{l}
{\epsilon^{\mi\mii\miii\miv}\,\Tr\,\Dca f_{\r[\s}\Dcb f_{\mi\mii} 
f_{\miii\miv]}=0 \Rightarrow}\\
   {\phantom{\epsilon^{\mi\mii\miii\miv}}\epsilon^{\mi\mii\miii\miv}\,\Tr\,[\Dca f_{\r\s} \Dcb f_{\mi\mii} f_{\miii\miv}
    +2\Dca f_{\r\mi} \Dcb f_{\mii\miii} f_{\miv\s}
    +2\Dca f_{\r\miii} \Dcb f_{\miv\s} f_{\mi\mii}]=0,}\\
{\epsilon^{\mi\mii\miii\miv}\,\Tr\,\Dca f_{[\r\mi}\Dcb f_{\mii\miii} 
f_{\miv]\s}=0\Rightarrow }\\
{\phantom{\epsilon^{\mi\mii\miii\miv}}\epsilon^{\mi\mii\miii\miv}\,\Tr\,[2\Dca f_{\r\mi} \Dcb 
f_{\mii\miii} f_{\miv\s}
    +2\Dca f_{\mii\miii} \Dcb f_{\miv\r} f_{\mi\s}
    +\Dca f_{\mi\mii} \Dcb f_{\miii\miv} f_{\r\s}]=0,}\\
        {\epsilon^{\mi\mii\miii\miv}\,\Tr\,\Dca f_{[\s\mi}\Dcb f_{|\r|\mii} f_{\miii\miv]}=0\Rightarrow}\\
    {\phantom{\epsilon^{\mi\mii\miii\miv}}\epsilon^{\mi\mii\miii\miv}\,\Tr\,[2\Dca f_{\s\mi} \Dcb f_{\r\mii} f_{\miii\miv}
    +2\Dca f_{\mii\miii} \Dcb f_{\r\miv} f_{\s\mi}
    +\Dca f_{\miii\miv} \Dcb f_{\r\s} f_{\mi\mii}]=0.}\\
\end{array}
\label{D2}
\end{equation}
\begin{equation}
\begin{array}{l}
{\epsilon^{\mi\mii\miii\miv}\,\Tr\,f_{\a\b} f_{\r[\s} f_{\mi\mii} f_{\miii\miv]}=0 \Rightarrow}\\ 
{\phantom{\epsilon^{\mi\mii\miii\miv}}
\epsilon^{\mi\mii\miii\miv}\,\Tr\,[f_{\a\b} f_{\r\s} f_{\mi\mii} f_{\miii\miv}
    +2f_{\a\b} f_{\r\mi} f_{\mii\miii} f_{\miv\s}
    +2f_{\a\b} f_{\r\miii} f_{\miv\s} f_{\mi\mii}]=0,}\\
 {\epsilon^{\mi\mii\miii\miv}\,\Tr\,f_{\r\a} f_{\b[\s} f_{\mi\mii} f_{\miii\miv]}=0\Rightarrow }\\
{\phantom{\epsilon^{\mi\mii\miii\miv}}
\epsilon^{\mi\mii\miii\miv}\,\Tr\,[f_{\r\a} f_{\b\s} f_{\mi\mii} f_{\miii\miv}
    +2f_{\r\a} f_{\b\mi} f_{\mii\miii} f_{\miv\s}
    +2f_{\r\a} f_{\b\miii} f_{\miv\s} f_{\mi\mii}]=0,}\\
 {\epsilon^{\mi\mii\miii\miv}\,\Tr\,f_{\r[\s} f_{|\a\b|} f_{\mi\mii} f_{\miii\miv]}=0 \Rightarrow}\\ 
{\phantom{\epsilon^{\mi\mii\miii\miv}}
\epsilon^{\mi\mii\miii\miv}\,\Tr\,[f_{\r\s} f_{\a\b} f_{\mi\mii} f_{\miii\miv}
    +2f_{\r\mi} f_{\a\b} f_{\mii\miii} f_{\miv\s}
    +2f_{\r\miii} f_{\a\b} f_{\miv\s} f_{\mi\mii}]=0.}\\
\end{array}
\label{D3o}
\end{equation}
 \begin{equation}
\begin{array}{l}
 {\epsilon^{\mi\mii\miii\miv}\,\Tr\,f_{\a\b} f_{[\s\mi} f_{|\r|\mii} f_{\miii\miv]}=0\Rightarrow}\\ 
{\phantom{\epsilon^{\mi\mii\miii\miv}}\epsilon^{\mi\mii\miii\miv}\,\Tr\,
[2f_{\a\b} f_{\s\mi} f_{\r\mii} f_{\miii\miv}
    +2f_{\a\b} f_{\mi\mii} f_{\r\miii} f_{\miv\s}
    +f_{\a\b} f_{\miii\miv} f_{\r\s} f_{\mi\mii}]=0,}\\
 {\epsilon^{\mi\mii\miii\miv}\,\Tr\,f_{\a[\s} f_{|\b\r|} f_{\mi\mii} f_{\miii\miv]}=0\Rightarrow }\\
{\phantom{\epsilon^{\mi\mii\miii\miv}}
\epsilon^{\mi\mii\miii\miv}\,\Tr\,[f_{\a\s} f_{\b\r} f_{\mi\mii} f_{\miii\miv}
    +2f_{\a\mi} f_{\b\r} f_{\mii\miii} f_{\miv\s}
    +2f_{\a\miii} f_{\b\r} f_{\miv\s} f_{\mi\mii}]=0,}\\
 {\epsilon^{\mi\mii\miii\miv}\,\Tr\,f_{\r\a} f_{[\s\mi} f_{|\b|\mii} f_{\miii\miv]}=0\Rightarrow}\\ 
{\phantom{\epsilon^{\mi\mii\miii\miv}}
\epsilon^{\mi\mii\miii\miv}\,\Tr\,[2f_{\r\a} f_{\s\mi} f_{\b\mii} f_{\miii\miv}
    +2f_{\r\a} f_{\mi\mii} f_{\b\miii} f_{\miv\s}
    +2f_{\r\a} f_{\miii\miv} f_{\b\s} f_{\mi\mii}]=0,}\\
{\epsilon^{\mi\mii\miii\miv}\,\Tr\,f_{\r[\s} f_{|\a|\mi} f_{|\b|\mii}f_{\miii\miv]}=0\Rightarrow}\\ 
{\phantom{\epsilon^{\mi\mii\miii\miv}}
\epsilon^{\mi\mii\miii\miv}\,\Tr\,[f_{\r\s} f_{\a\mi} f_{\b\mii} f_{\miii\miv}
    +2f_{\r\mi} f_{\a\mii} f_{\b\miii} f_{\miv\s}
    +2f_{\r\miii} f_{\a\miv} f_{\b\s} f_{\mi\mii}]=0,}\\
{\epsilon^{\mi\mii\miii\miv}\,\Tr\,f_{\a[\s} f_{|\r|\mi} f_{|\b|\mii} f_{\miii\miv]}=0\Rightarrow}\\
{\epsilon^{\mi\mii\miii\miv}\,\Tr\,[f_{\a\s} f_{\r\mi} f_{\b\mii} f_{\miii\miv}
    +2f_{\a\mi} f_{\r\mii} f_{\b\miii} f_{\miv\s}+f_{\a\miii} f_{\r\miv} f_{\b\s} f_{\mi\mii}
    +f_{\a\miv} f_{\r\s} f_{\b\mi} f_{\mii\miii}]=0.}\\
\end{array} 
   \label{D3}
\end{equation}
We shall also need the following identities:
\begin{displaymath}
\begin{array}{l}
{\epsilon^{\mi\mii\miii\miv}\theta^{\a\b}\theta^{\r\s}\,\Tr\,
\Dca\Dcr f_{\mi\miii}\Dcb\Dcs
f_{\mii\miv}}\\
{\phantom{\epsilon^{\mi\mii\miii\miv}\theta^{\a\b}\theta^{\r\s}\,\Tr\,
\Dca}
=\epsilon^{\mi\mii\miii\miv}\theta^{\a\b}\theta^{\r\s}\,\Tr\,
\big[\Dca(\Dcr f_{\mi\miii}\Dcb\Dcs f_{\mii\miv})+i\Dca
f_{\mi\miii}f_{\r\s}\Dcb f_{\mii\miv}\big],}\\
{\epsilon^{\mi\mii\miii\miv}\theta^{\a\b}\theta^{\r\s}\,\Tr\,\Dcr f_{\mi\miii} \Dca f_{\mii\miv} f_{\b\s}=
 \epsilon^{\mi\mii\miii\miv}\theta^{\a\b}\theta^{\r\s}\,\Tr\,\big[\Dcr\big(f_{\mi\miii}\Dca f_{\mii\miv} f_{\b\s}big)}\\
  {\phantom{\epsilon^{\mi\mii\miii\miv}\theta^{\a\b}\theta^{\r\s}\,\Tr\,
\Dcr f_{\mi\miii}  }
-\frac{1}{2}\Dca f_{\mii\miv} \Dcb f_{\r\s} f_{\mi\miii}+\frac{i}{2} f_{\mi\miii} 
    f_{\r\a} f_{\mii\miv} f_{\b\s}-\frac{i}{2}f_{\mi\miii}f_{\mii\miv}f_{\r\a}f_{\b\s}\big].}\\
\end{array}
\end{displaymath}
Substituting the two previous equations in eq.~(\ref{A2}), one gets 
\begin{equation}
\begin{array}{l}
{{\cal A}^{(2)}=
  \frac{1}{\pi^2}\,\epsilon^{\mi\mii\miii\miv}
\theta^{\alpha\beta}\theta^{\rho\sigma}\,\Tr\,\big[
\frac{1}{384}\Dca(\Dcr f_{\mi\miii}\Dcb \Dcs f_{\mii\miv})+
\frac{i}{96}\Dcr ( f_{\mi\miii}\Dca f_{\mii\miv}f_{\b\s})}\\
{\phantom{{\cal A}^{(2)}=}+\frac{i}{384}\Dca f_{\mi\miii}\Dcb 
f_{\mii\miv} f_{\r\s}+\frac{i}{192}\Dca
f_{\r\s}\Dcb f_{\mi\miii} f_{\mii\miv}+\frac{i}{32}\Dca f_{\r\mii}
\Dcb f_{\s\miii} f_{\mi\miv}}\\
{\phantom{{\cal A}^{(2)}=}
+\frac{1}{16}f_{\r\mi}f_{\a\mii}f_{\b\miii}f_{\s\miv}+\frac{1}{16}f_{\r\mii}f_{\mi\miii}
 f_{\a\miv} f_{\b\s}+\frac{1}{32}f_{\mi\mii}f_{\r\miii}f_{\b\miv}f_{\a\s}
 +\frac{1}{32}f_{\mi\mii}f_{\b\miii}f_{\a\miv}f_{\r\s}}\\
{\phantom{{\cal A}^{(2)}=}+\frac{1}{32}f_{\r\mii}f_{\mi\miii}f_{\b\s}f_{\a\miv}
+\frac{1}{32}f_{\a\mii}f_{\mi\miii}f_{\r\s}f_{\b\miv}
-\frac{1}{64}f_{\mii\miii}f_{\mi\miv}f_{\r\b}f_{\a\s}
-\frac{1}{192}f_{\mi\miii}f_{\mii\miv}f_{\a\b}f_{\r\s}}\\
{\phantom{{\cal A}^{(2)}=}-\frac{1}{384} f_{\mi\miii}f_{\a\b}f_{\mii\miv}f_{\r\s}
\big]}\\
{ -\frac{1}{\pi^2}\epsilon^{\mi\mii\miii\miv}
{\theta_\rho}^\beta\theta^{\rho\sigma}g^{\mu\nu}\Tr
\Dcb\Dcs\big[\frac{1}{384}\big(2\Dcm\Dcn f_{\mi\miii}f_{\mii\miv}+
\Dcm f_{\mi\miii}\Dcn f_{\mii\miv}\big)-\frac{i}{96}f_{\mu\mi} 
f_{\nu\mii} f_{\miii\miv}\big].}\\
\end{array}
\label{A22}
\end{equation}
Let us introduce next the following shorthand
\begin{displaymath}
\begin{array}{l}  
{x_1=\epsilon^{\mi\mii\miii\miv}
\theta^{\alpha\beta}\theta^{\rho\sigma}\Tr
\big[\Dca f_{\r\s}\Dcb f_{\mi\mii} f_{\miii\miv}\big],\quad
x_2= \epsilon^{\mi\mii\miii\miv}
\theta^{\alpha\beta}\theta^{\rho\sigma}\Tr\big[\Dca f_{\mi\mii}\Dcb 
f_{\r\s} f_{\miii\miv}\big],}\\
{x_3=\epsilon^{\mi\mii\miii\miv}\theta^{\alpha\beta}\theta^{\rho\sigma}\Tr
\big[
\Dca f_{\mi\mii}\Dcb f_{\mii\miv} f_{\r\s}\big],\quad 
 x_4= \epsilon^{\mi\mii\miii\miv}\theta^{\alpha\beta}\theta^{\rho\sigma}\Tr
\big[\Dca f_{\r\mi}\Dcb f_{\s\mii} f_{\miii\miv}\big],}\\
 {x_5=\epsilon^{\mi\mii\miii\miv}\theta^{\alpha\beta}\theta^{\rho\sigma}\Tr
\big[
\Dca f_{\r\mi}\Dcb f_{\mii\miii} f_{\s\miv}\big],\quad
x_6=\epsilon^{\mi\mii\miii\miv}\theta^{\alpha\beta}\theta^{\rho\sigma}\Tr
\big[
\Dca f_{\mi\mii}\Dcb f_{\r\miii} f_{\s\miv}\big].}\\
\end{array}
\end{displaymath}
The objects $x_{i}$, $i=1,\cdots,6$ are not linearly independent. They are 
related by the three identities in eq.~({\ref{D2}). These identities read 
\begin{displaymath}
 x_1-2x_5-2x_4=0,\quad
    -2x_5-2x_6+x_3=0,\quad
    -2x_4-2x_6+x_2=0.\quad
\end{displaymath}
This liner system can be solved yielding the following result:
\begin{displaymath}
x_4=\frac{1}{4}(x_1+x_2-x_3),\quad x_5=\frac{1}{4}(x_1-x_2+x_3),\quad 
x_6=\frac{1}{4}(-x_1+x_2+x_3).
\end{displaymath}
It is not difficult to convince oneself that $x_1$, $x_2$ and $x_3$ are 
linearly   independent. We shall employ the previous result to express the sum 
of the terms on the r.h.s of eq.~(\ref{A22}) with only two covariant 
derivatives as follows:
\begin{equation}
\begin{array}{l}
{\frac{i}{384}\Dca f_{\mi\miii}\Dcb f_{\mii\miv} f_{\r\s}
+\frac{i}{192}\Dca
f_{\r\s}\Dcb f_{\mi\miii} f_{\mii\miv}+\frac{i}{32}\Dca f_{\r\mii}
\Dcb f_{\s\miii} f_{\mi\miv}=}\\
{-\frac{i}{384}x_3-\frac{i}{192}x_1+\frac{i}{32}x_4=\frac{i}{384}x_1+\frac{i}{128}x_2
-\frac{i}{96}x_3=\frac{i}{384}(x_1-x_2)+\frac{i}{96}(x_2-x_3).}\\
\end{array}
\label{newexpres}
\end{equation}
Using eq.~(\ref{D2}) and the cyclicity of the trace, one can show that
\begin{equation}
\begin{array}{l}
{x_1-x_2=\epsilon^{\mi\mii\miii\miv}\theta^{\alpha\beta}\theta^{\rho\sigma}\,
\Tr\,
\big[\Dcb(\Dca f_{\r\s}f_{\mi\mii}f_{\miii\miv})\big]}\\
{ x_2-x_3=
\epsilon^{\mi\mii\miii\miv}\theta^{\alpha\beta}\theta^{\rho\sigma}
\Tr
\big[\Dcb(\Dca f_{\mi\mii} f_{\r\s} f_{\miii\miv})-\frac{i}{2}f_{\a\b}
f_{\mi\mii}
 f_{\r\s}f_{\miii\miv}+\frac{i}{2}f_{\mi\mii}f_{\a\b}f_{\r\s}
f_{\miii\miv}\big].}\\
\end{array}
\label{x1-x2}
\end{equation}
Substituting eq.~(\ref{newexpres}) in eq.~(\ref{A22}), and then using the 
identities in eq.~(\ref{x1-x2}), one obtains the following intermediate 
expression for ${\cal A}^{(2)}$:
\begin{equation}
\begin{array}{l}
 {{\cal A}^{(2)}=
    \frac{1}{\pi^2}\,\epsilon^{\mi\mii\miii\miv}\theta^{\alpha\beta}
\theta^{\rho\sigma}\Tr\big[\frac{1}{384}\Dca(\Dcr f_{\mi\miii}\Dcb \Dcs f_{\mii\miv})+
\frac{i}{96}\Dcr ( f_{\mi\miii}\Dca f_{\mii\miv}
f_{\b\s})}\\
{\phantom{{\cal A}=}+\frac{i}{384}\Dcb(\Dca f_{\r\s}f_{\mi\mii}f_{\miii\miv})+\frac{i}{96}
\Dcb(\Dca f_{\mi\mii}f_{\r\s}f_{\miii\miv})}\\
{\phantom{{\cal A}=}+
\frac{1}{16}f_{\r\mi}f_{\a\mii}f_{\b\miii}f_{\s\miv}+\frac{1}{16}f_{\r\mii}f_{\mi\miii}
 f_{\a\miv} f_{\b\s}+\frac{1}{32}f_{\mi\mii}f_{\r\miii}f_{\b\miv}f_{\a\s}
 +\frac{1}{32}f_{\mi\mii}f_{\b\miii}f_{\a\miv}f_{\r\s}}\\
{\phantom{{\cal A}=}
+\frac{1}{32}f_{\r\mii}f_{\mi\miii}f_{\b\s}f_{\a\miv}
+\frac{1}{32}f_{\a\mii}f_{\mi\miii}f_{\r\s}f_{\b\miv}
-\frac{1}{64}f_{\mii\miii}f_{\mi\miv}f_{\r\b}f_{\a\s}
-\frac{1}{128} f_{\mi\miii}f_{\a\b}f_{\mii\miv}f_{\r\s}
\big]}\\
 {
-\frac{1}{\pi^2}\epsilon^{\mi\mii\miii\miv}
{\theta_\rho}^\beta\theta^{\rho\sigma}g^{\mu\nu}\Tr
\Dcb\Dcs\big[\frac{1}{384}\big(2\Dcm\Dcn f_{\mi\miii}f_{\mii\miv}+
\Dcm f_{\mi\miii}\Dcn f_{\mii\miv}\big)-\frac{i}{96}f_{\mu\mi} f_{\nu\mii} f_{\miii\miv}\big].}\\
\end{array}
\label{A23}
\end{equation}
Let us finally show that the contributions on the r.h.s. of the previous 
identity which are of the type $\epsilon\theta\theta\Tr ffff$, with obvious 
notation, add up to zero. To make the discussion as clear as possible, we 
shall introduce the following notation:
\begin{displaymath}
\begin{array}{l}
{  y_1=
\epsilon^{\mi\mii\miii\miv}\theta^{\alpha\beta}\theta^{\rho\sigma}\Tr
\big[f_{\a\b}f_{\r\s}f_{\mi\mii}f_{\miii\miv}\big],\quad
 y_2 = \epsilon^{\mi\mii\miii\miv}\theta^{\alpha\beta}\theta^{\rho\sigma}\Tr\big[f_{\a\b}f_{\mi\mii}f_{\r\s}f_{\miii\miv}\big],}\\
{y_3=\epsilon^{\mi\mii\miii\miv}\theta^{\alpha\beta}\theta^{\rho\sigma}\Tr
\big[f_{\a\b}f_{\mi\mii}f_{\miii\miv}f_{\r\s}],\quad

y_4= \epsilon^{\mi\mii\miii\miv}\theta^{\alpha\beta}\theta^{\rho\sigma}\Tr
\big[  f_{\a\b}f_{\r\mi}f_{\s\mii}f_{\miii\miv}\big],}\\ 
{y_5=\epsilon^{\mi\mii\miii\miv}\theta^{\alpha\beta}\theta^{\rho\sigma}\Tr
\big[
f_{\a\b}f_{\mi\mii}f_{\r\miii}f_{\s\miv}\big],\quad
y_6=\epsilon^{\mi\mii\miii\miv}\theta^{\alpha\beta}\theta^{\rho\sigma}\Tr
\big[
f_{\a\r}f_{\b\s}f_{\mi\mii}f_{\miii\miv}],}\\
{y_7=\epsilon^{\mi\mii\miii\miv}\theta^{\alpha\beta}\theta^{\rho\sigma}\Tr
\big[
f_{\a\r}f_{\mi\mii}f_{\b\s}f_{\miii\miv} \big],\quad
y_8=\epsilon^{\mi\mii\miii\miv}\theta^{\alpha\beta}\theta^{\rho\sigma}\Tr
\big[ 
f_{\a\r}f_{\b\mi}f_{\s\mii}f_{\miii\miv}],}\\
{y_9=\epsilon^{\mi\mii\miii\miv}\theta^{\alpha\beta}\theta^{\rho\sigma}\Tr
\big[
f_{\a\r}f_{\b\mi}f_{\s\mii}f_{\miii\miv}],\quad 
y_{10}=\epsilon^{\mi\mii\miii\miv}\theta^{\alpha\beta}\theta^{\rho\sigma}\Tr
\big[
 f_{\a\r}f_{\mi\mii}f_{\b\miii}f_{\s\miv}],}\\
{y_{11}= \epsilon^{\mi\mii\miii\miv}\theta^{\alpha\beta}\theta^{\rho\sigma}\Tr
\big[ 
  f_{\a\mi}f_{\b\mii}f_{\r\miii}f_{\s\miv}\big],\quad
y_{12}=\epsilon^{\mi\mii\miii\miv}\theta^{\alpha\beta}\theta^{\rho\sigma}\Tr
\big[  f_{\a\mi}f_{\r\mii}f_{\b\miii}f_{\s\miv}].}\\
\end{array}
\end{displaymath}
These objects are not linearly independent since they verify the linear equations in eq.~(\ref{D3o}) and ~(\ref{D3}). These linear equations read  
\begin{displaymath}
\begin{array}{l}
  { y_1-2y_4-2y_3=0,\phantom{y_9=0}\qquad  
   y_6-2y_9-2y_8=0,\phantom{y_9=0}\qquad y_1-2y_5-2y_4=0,}\\
  {y_2-2y_3-2y_5=0,\phantom{y_9=0}\qquad y_6-2y_{10}-2y_9=0,\phantom{y_9=\,\,}\qquad y_7-2y_8-2y_{10}=0,}\\
  {y_3-2y_{11}-y_{10}-y_9=0,\qquad\; y_8-2y_{12}+y_{10}-y_4=0,}\\
\end{array}
\end{displaymath}
where the symbols $y_i$ have been introduced above. The previous linear system 
can be solved in terms of, say, $y_1$, $y_2$, $y_6$ and $y_7$. The solution 
is the following:
\begin{displaymath}
\begin{array}{l}
 {y_3=\frac{1}{4}y_2,\quad y_4=\frac{1}{2}y_1-\frac{1}{4}y_2,\quad y_5=\frac{1}{4}y_2
    y_8=\frac{1}{4}y_7,\quad y_9=\frac{1}{2}y_6-\frac{1}{4}y_7,\quad y_{10}=\frac{1}{4}y_7}\\
   { y_{11}=\frac{1}{8}y_2-\frac{1}{4}y_6,\quad y_{12}=-\frac{1}{4}y_1+\frac{1}{8}y_2+\frac{1}{4}y_7.}\\
\end{array}
\end{displaymath}
Using this result, one can easily show that  the following equation holds
\begin{displaymath}
\begin{array}{l}
{\epsilon^{\mi\mii\miii\miv}\theta^{\a\b}\theta^{\r\s}\Tr
\big[\frac{1}{16}f_{\r\mi}f_{\a\mii}f_{\b\miii}f_{\s\miv}+
\frac{1}{16}f_{\r\mii}f_{\mi\miii}f_{\a\miv} f_{\b\s}+
 \frac{1}{32}f_{\mi\mii}f_{\r\miii}f_{\b\miv}f_{\a\s}}\\
{\phantom{\epsilon^{\mi\mii\miii}}
 +\frac{1}{32}f_{\mi\mii}f_{\b\miii}f_{\a\miv}f_{\r\s}
+\frac{1}{32}f_{\r\mii}f_{\mi\miii}f_{\b\s}f_{\a\miv}
+\frac{1}{32}f_{\a\mii}f_{\mi\miii}f_{\r\s}f_{\b\miv}-
\frac{1}{64}f_{\mii\miii}f_{\mi\miv}f_{\r\b}f_{\a\s}}\\
{\phantom{\epsilon^{\mi\mii\miii}}
-\frac{1}{128} f_{\mi\miii}f_{\a\b}f_{\mii\miv}f_{\r\s}\big]}\\
{\phantom{\epsilon}
=\frac{1}{16}y_{11}+\frac{1}{16}y_9+\frac{1}{32}y_{10}
-\frac{1}{32}y_5+\frac{1}{32}y_8-\frac{1}{32}y_3-\frac{1}{64}y_6
+\frac{1}{128}y_2=0.}\\
\end{array}
\end{displaymath}
By substituting this result in eq.~(\ref{A23}), one obtains
\begin{displaymath}
\begin{array}{l}
 {{\cal A}^{(2)}=
    \frac{1}{\pi^2}\,\epsilon^{\mi\mii\miii\miv}\theta^{\alpha\beta}
\theta^{\rho\sigma}\Tr\big[+\frac{1}{384}\Dca(\Dcr f_{\mi\miii}\Dcb \Dcs f_{\mii\miv})+
\frac{i}{96}\Dcr ( f_{\mi\miii}\Dca f_{\mii\miv}
f_{\b\s})}\\
{\phantom{{\cal A}^{(2)}=\epsilon^{\mi\mii\miii\miv}\theta^{\alpha\beta}
\theta^{\rho\sigma}\Tr\big[aa\,}+\frac{i}{384}\Dcb(\Dca f_{\r\s}
f_{\mi\mii}f_{\miii\miv})+\frac{i}{96}
\Dcb(\Dca f_{\mi\mii}f_{\r\s}f_{\miii\miv})\big]}\\
 {
-\frac{1}{\pi^2}\epsilon^{\mi\mii\miii\miv}{\theta_\rho}^\beta\theta^{\rho\sigma}g^{\mu\nu}\Tr
\Dcb\Dcs\big[\frac{1}{384}\big(2\Dcm\Dcn f_{\mi\miii}f_{\mii\miv}+
\Dcm f_{\mi\miii}\Dcn f_{\mii\miv}\big)-\frac{i}{96}f_{\mu\mi} f_{\nu\mii} f_{\miii\miv}\big].}\\
\end{array}
\end{displaymath}
This is eq.~(\ref{a2final}).

\end{document}